\documentclass[prd,amsmath,aps,floats,amssymb, floatfix, superscriptaddress,nofootinbib,twocolumn,preprintnumbers]{revtex4-2}  

\setcitestyle{authoryear,round}

\usepackage[T1]{fontenc}
\usepackage{aecompl}

\usepackage{newtxtext,newtxmath}
\usepackage{mdframed,bm}
\usepackage[T1]{fontenc}
\usepackage{ae,aecompl}
\usepackage{graphicx}	
\usepackage{amsmath}	
\usepackage{amssymb}	
\usepackage{booktabs}
\usepackage{color}
\usepackage{enumitem}
\usepackage{xspace}
\usepackage{tabularx}

\usepackage[colorlinks=true]{hyperref}

\hypersetup{
    urlcolor=black,
    citecolor=blue,
    linkcolor=blue,
  }

\usepackage[]{lineno}
\newcommand\dnf{\textsc{dnf}}
\newcommand\dz{{\rm d}  z}
\newcommand{\diff}{{\rm d}}
\newcommand\dthet{{\rm d} \theta}
\newcommand\wur{w_{\rm ur}}
\newcommand\rmu{{\rm u}}
\newcommand\rmr{{\rm r}}
\newcommand\rmur{{\rm ur}}
\newcommand\rp{r_{\rm p}}
\newcommand{\vecb}{{\bf b}}
\newcommand{\vecu}{{\bf u}}
\newcommand{\vecq}{{\bf q}}
\newcommand{\vecw}{{\bf w}}
\newcommand{\vecn}{{\bf n}}

\newcommand{\aap}{A\&A}
\newcommand{\apjs}{ApJS}
\newcommand{\aj}{A.J.}
\newcommand{\mnras}{MNRAS}

\newcommand{\jcap}{JCAP}
\newcommand{\apjl}{ApJ}

\usepackage{verbatim}
\usepackage[usenames,dvipsnames]{xcolor}
\usepackage{tabulary}
\usepackage{tabularx}
\usepackage{todonotes}
\usepackage{hyperref}
\usepackage{ulem}

\newcommand\be{\begin{equation}}
\newcommand\ee{\end{equation}}
\def\bea{\begin{eqnarray}}
\def\eea{\end{eqnarray}}

\newcommand\red[1]{{#1}}

\allowdisplaybreaks

\begin{document}

\title{Dark Energy Survey Year 6 Results: Clustering-redshifts and importance sampling of Self-Organised-Maps $n(z)$ realizations for $3\times2$pt samples.}

\author{W.~d'Assignies}\email{E-mail:wdoumerg@ifae.es}
\affiliation{Institut de F\'{\i}sica d'Altes Energies (IFAE), The Barcelona Institute of Science and Technology, Campus UAB, 08193 Bellaterra (Barcelona) Spain}

\author{G.~M.~Bernstein}
\affiliation{Department of Physics and Astronomy, University of Pennsylvania, Philadelphia, PA 19104, USA}
\author{B.~Yin}
\affiliation{Department of Physics, Duke University Durham, NC 27708, USA}
\author{G.~Giannini}
\affiliation{Department of Astronomy and Astrophysics, University of Chicago, Chicago, IL 60637, USA}
\affiliation{Kavli Institute for Cosmological Physics, University of Chicago, Chicago, IL 60637, USA}

\author{A.~Alarcon}
\affiliation{Institute of Space Sciences (ICE, CSIC),  Campus UAB, Carrer de Can Magrans, s/n,  08193 Barcelona, Spain}
\author{M.~Manera}
\affiliation{Institut de F\'{\i}sica d'Altes Energies (IFAE), The Barcelona Institute of Science and Technology, Campus UAB, 08193 Bellaterra (Barcelona) Spain}
\affiliation{Barcelona, 
Edifici C Facultat de Ciències, 08193 Bellaterra, Spain }
\author{C.~To}
\affiliation{Department of Astronomy and Astrophysics, University of Chicago, Chicago, IL 60637, USA}
\author{M.~Yamamoto}
\affiliation{Department of Physics, Duke University Durham, NC 27708, USA}
\affiliation{Department of Astrophysical Sciences, Princeton University, Peyton Hall, Princeton, NJ 08544, USA}
 \author{N.~Weaverdyck}
\affiliation{Lawrence Berkeley National Laboratory, 1 Cyclotron Road, Berkeley, CA 94720, USA}
\affiliation{Berkeley Center for Cosmological Physics, Berkeley, CA 94720, USA}

\author{R.~Cawthon}
\affiliation{Oxford College of Emory University, Oxford, GA 30054, USA}
\author{M.~Gatti}
\affiliation{Kavli Institute for Cosmological Physics, University of Chicago, Chicago, IL 60637, USA}

\author{A.~Amon}
\affiliation{Department of Astrophysical Sciences, Princeton University, Peyton Hall, Princeton, NJ 08544, USA}
\author{D.~Anbajagane}
\affiliation{Kavli Institute for Cosmological Physics, University of Chicago, Chicago, IL 60637, USA}
\author{S.~Avila}
\affiliation{Centro de Investigaciones Energ\'eticas, Medioambientales y Tecnol\'ogicas (CIEMAT), Madrid, Spain}
\author{M.~R.~Becker}
\affiliation{ Argonne National Laboratory, 9700 South Cass Avenue, Lemont, IL 60439, USA}
\author{K.~Bechtol}
\affiliation{ Physics Department, 2320 Chamberlin Hall, University of Wisconsin-Madison, 1150 University Avenue Madison, WI  53706-1390}

\author{C.~Chang}
\affiliation{Department of Astronomy and Astrophysics, University of Chicago, Chicago, IL 60637, USA}
\affiliation{Kavli Institute for Cosmological Physics, University of Chicago, Chicago, IL 60637, USA}
\author{M.~Crocce}
\affiliation{Institute of Space Sciences (ICE, CSIC),  Campus UAB, Carrer de Can Magrans, s/n,  08193 Barcelona, Spain}
\affiliation{Institut d'Estudis Espacials de Catalunya (IEEC), 08034 Barcelona, Spain}
\author{J.~De Vicente}
\affiliation{Centro de Investigaciones Energ\'eticas, Medioambientales y Tecnol\'ogicas (CIEMAT), Madrid, Spain}
\author{S.~Dodelson}
\affiliation{Department of Astronomy and Astrophysics, University of Chicago, Chicago, IL 60637, USA}
\affiliation{Kavli Institute for Cosmological Physics, University of Chicago, Chicago, IL 60637, USA}
\affiliation{Fermi National Accelerator Laboratory, P. O. Box 500, Batavia, IL 60510, USA}
\author{J.~Fang}
\affiliation{Physics Department, William Jewell College, Liberty, MO 64068, USA}
\author{A.~Ferté}
\affiliation{SLAC National Accelerator Laboratory, Menlo Park, CA 94025, USA}
\author{D.~Gruen}
\affiliation{University Observatory, LMU Faculty of Physics, Scheinerstr. 1, 81679 Munich, Germany}

\author{E.~Legnani}
\affiliation{Institut de F\'{\i}sica d'Altes Energies (IFAE), The Barcelona Institute of Science and Technology, Campus UAB, 08193 Bellaterra (Barcelona) Spain}
\author{A.~Porredon}
\affiliation{ Ruhr University Bochum, Faculty of Physics and Astronomy, Astronomical Institute, German Centre for Cosmological Lensing, 44780 Bochum, Germany}
\affiliation{ Centro de Investigaciones Energ\'eticas, Medioambientales y Tecnol\'ogicas (CIEMAT), Madrid, Spain}

\author{J. Prat}
\affiliation{Department of Astronomy and Astrophysics, University of Chicago, Chicago, IL 60637, USA}
\affiliation{Nordita, KTH Royal Institute of Technology and Stockholm University, Hannes Alfv\'ens v\"ag 12, SE-10691 Stockholm, Sweden}
\author{M.~Rodriguez-Monroy}
\affiliation{ Instituto de Física Teórica UAM/CSIC, Universidad Autónoma de Madrid, 28049 Madrid, Spain }
\affiliation{Laboratoire de physique des 2 infinis Irène Joliot-Curie, CNRS Université Paris-Saclay, Bât. 100, F-91405 Orsay Cedex, France}

\author{C.~S{\'a}nchez}
\affiliation{Institut de F\'{\i}sica d'Altes Energies (IFAE), The Barcelona Institute of Science and Technology, Campus UAB, 08193 Bellaterra (Barcelona) Spain}
\affiliation{Departament de F\'{\i}sica, Universitat Aut\`{o}noma de Barcelona (UAB), 08193 Bellaterra (Barcelona), Spain}
\author{T.~Schutt}
\affiliation{SLAC National Accelerator Laboratory, Menlo Park, CA 94025, USA}
\affiliation{Kavli Institute for Particle Astrophysics \& Cosmology, P. O. Box 2450, Stanford University, Stanford, CA 94305, USA}
\affiliation{Department of Physics, Stanford University, 382 Via Pueblo Mall, Stanford, CA 94305, USA}
\author{I.~Sevilla-Noarbe}
\affiliation{ Centro de Investigaciones Energ\'eticas, Medioambientales y Tecnol\'ogicas (CIEMAT), Madrid, Spain}
\author{D.~Sanchez Cid}
\affiliation{ Centro de Investigaciones Energ\'eticas, Medioambientales y Tecnol\'ogicas (CIEMAT), Madrid, Spain}
\affiliation{ Physik-Institut, University of Zürich, Winterthurerstrasse 190, CH-8057 Zürich, Switzerland}

\author{M.~A.~Troxel}
\affiliation{Department of Physics, Duke University Durham, NC 27708, USA}

\author{T.~M.~C.~Abbott}
\affiliation{Cerro Tololo Inter-American Observatory, NSF's National Optical-Infrared Astronomy Research Laboratory, Casilla 603, La Serena, Chile}
\author{F.~Andrade-Oliveira}
\affiliation{ Physik-Institut, University of Zürich, Winterthurerstrasse 190, CH-8057 Zürich, Switzerland}
\author{M.~Aguena}
\affiliation{Laborat\'orio Interinstitucional de e-Astronomia - LIneA, Av. Pastor Martin Luther King Jr, 126 Del Castilho, Nova Am\'erica Offices, Torre 3000/sala 817 CEP: 20765-000, Brazil}
\affiliation{INAF-Osservatorio Astronomico di Trieste, via G. B. Tiepolo 11, I-34143 Trieste, Italy}
\author{O.~Alves}
\affiliation{Department of Physics, University of Michigan, Ann Arbor, MI 48109, USA}
\author{D.~Bacon}
\affiliation{Institute of Cosmology and Gravitation, University of Portsmouth, Portsmouth, PO1 3FX, UK}
\author{J. Blazek},
\affiliation{Department of Physics, Northeastern University, Boston, MA 02115, USA}

\author{S.~Bocquet}
\affiliation{University Observatory, LMU Faculty of Physics, Scheinerstr. 1, 81679 Munich, Germany}
\author{D.~Brooks}
\affiliation{Department of Physics \& Astronomy, University College London, Gower Street, London, WC1E 6BT, UK}
\author{R.~Camilleri}
\affiliation{School of Mathematics and Physics, University of Queensland,  Brisbane, QLD 4072, Australia}
\author{A.~Carnero~Rosell}
\affiliation{Laborat\'orio Interinstitucional de e-Astronomia - LIneA, Av. Pastor Martin Luther King Jr, 126 Del Castilho, Nova Am\'erica Offices, Torre 3000/sala 817 CEP: 20765-000, Brazil}
\affiliation{Universidad de La Laguna, Dpto. Astrofísica, E-38206 La Laguna, Tenerife, Spain}
\affiliation{Instituto de Astrofisica de Canarias, E-38205 La Laguna, Tenerife, Spain}
\author{M.~Carrasco~Kind}
\affiliation{Center for Astrophysical Surveys, National Center for Supercomputing Applications, 1205 West Clark St., Urbana, IL 61801, USA}
\affiliation{Department of Astronomy, University of Illinois at Urbana-Champaign, 1002 W. Green Street, Urbana, IL 61801, USA}

\author{J.~Carretero}
\affiliation{Institut de F\'{\i}sica d'Altes Energies (IFAE), The Barcelona Institute of Science and Technology, Campus UAB, 08193 Bellaterra (Barcelona) Spain}
\author{F.~J.~Castander}
\affiliation{Kavli Institute for Cosmological Physics, University of Chicago, Chicago, IL 60637, USA}
\affiliation{Institute of Space Sciences (ICE, CSIC),  Campus UAB, Carrer de Can Magrans, s/n,  08193 Barcelona, Spain}
\affiliation{Institut d'Estudis Espacials de Catalunya (IEEC), 08034 Barcelona, Spain}

\author{L.~N.~da Costa}
\affiliation{Laborat\'orio Interinstitucional de e-Astronomia - LIneA, Av. Pastor Martin Luther King Jr, 126 Del Castilho, Nova Am\'erica Offices, Torre 3000/sala 817 CEP: 20765-000, Brazil}
\author{M.~E.~da Silva Pereira}
\affiliation{Hamburger Sternwarte, Universit\"{a}t Hamburg, Gojenbergsweg 112, 21029 Hamburg, Germany}
\author{T.~M.~Davis}
\affiliation{School of Mathematics and Physics, University of Queensland,  Brisbane, QLD 4072, Australia}
\author{S.~Desai}
\affiliation{Department of Physics, IIT Hyderabad, Kandi, Telangana 502285, India}
\author{P.~Doel}
\affiliation{Department of Physics \& Astronomy, University College London, Gower Street, London, WC1E 6BT, UK}
\author{C.~Doux}
\affiliation{Universit\'e Grenoble Alpes, CNRS, LPSC-IN2P3, 38000 Grenoble, France}
\affiliation{Department of Physics and Astronomy, University of Pennsylvania, Philadelphia, PA 19104, USA}
\author{A.~Drlica-Wagner}

\affiliation{Department of Astronomy and Astrophysics, University of Chicago, Chicago, IL 60637, USA}
\affiliation{Kavli Institute for Cosmological Physics, University of Chicago, Chicago, IL 60637, USA}
\affiliation{Fermi National Accelerator Laboratory, P. O. Box 500, Batavia, IL 60510, USA}
\author{T.~Eifler}
\affiliation{Department of Astronomy/Steward Observatory, University of Arizona, 933 North Cherry Avenue, Tucson, AZ 85721-0065, USA}
\affiliation{Jet Propulsion Laboratory, California Institute of Technology, 4800 Oak Grove Dr., Pasadena, CA 91109, USA}

\author{J.~Elvin-Poole}
\affiliation{Department of Physics and Astronomy, University of Waterloo, 200 University Ave W, Waterloo, ON N2L 3G1, Canada}
\author{S.~Everett}
\affiliation{California Institute of Technology, 1200 East California Blvd, MC 249-17, Pasadena, CA 91125, USA}
\author{B.~Flaugher}
\affiliation{Fermi National Accelerator Laboratory, P. O. Box 500, Batavia, IL 60510, USA}
\author{P.~Fosalba}
\affiliation{Institute of Space Sciences (ICE, CSIC),  Campus UAB, Carrer de Can Magrans, s/n,  08193 Barcelona, Spain}
\affiliation{Institut d'Estudis Espacials de Catalunya (IEEC), 08034 Barcelona, Spain}

\author{J.~Frieman}
\affiliation{Department of Astronomy and Astrophysics, University of Chicago, Chicago, IL 60637, USA}
\affiliation{Kavli Institute for Cosmological Physics, University of Chicago, Chicago, IL 60637, USA}
\affiliation{Fermi National Accelerator Laboratory, P. O. Box 500, Batavia, IL 60510, USA}

\author{J.~Garc\'ia-Bellido}
\affiliation{Instituto de Fisica Teorica UAM/CSIC, Universidad Autonoma de Madrid, 28049 Madrid, Spain}
\author{E.~Gaztanaga}
\affiliation{Institute of Space Sciences (ICE, CSIC),  Campus UAB, Carrer de Can Magrans, s/n,  08193 Barcelona, Spain}
\affiliation{Institut d'Estudis Espacials de Catalunya (IEEC), 08034 Barcelona, Spain}
\affiliation{Institute of Cosmology and Gravitation, University of Portsmouth, Portsmouth, PO1 3FX, UK}

\author{P.~Giles}
\affiliation{Department of Physics and Astronomy, Pevensey Building, University of Sussex, Brighton, BN1 9QH, UK}

\author{G.~Gutierrez}
\affiliation{Fermi National Accelerator Laboratory, P. O. Box 500, Batavia, IL 60510, USA}
\author{S.~R.~Hinton}
\affiliation{School of Mathematics and Physics, University of Queensland,  Brisbane, QLD 4072, Australia}
\author{D.~L.~Hollowood}
\affiliation{Santa Cruz Institute for Particle Physics, Santa Cruz, CA 95064, USA}
\author{K.~Honscheid}
\affiliation{Department of Physics, The Ohio State University, Columbus, OH 43210, USA}
\affiliation{Center for Cosmology and Astro-Particle Physics, The Ohio State University, Columbus, OH 43210, USA}
\author{D.~Huterer}
\affiliation{Department of Physics, University of Michigan, Ann Arbor, MI 48109, USA}
\author{B.~Jain}
\affiliation{Department of Physics and Astronomy, University of Pennsylvania, Philadelphia, PA 19104, USA}

\author{D.~J.~James}
\affiliation{Center for Astrophysics $\vert$ Harvard \& Smithsonian, 60 Garden Street, Cambridge, MA 02138, USA}
\author{K.~Kuehn}
\affiliation{Australian Astronomical Optics, Macquarie University, North Ryde, NSW 2113, Australia}
\affiliation{Lowell Observatory, 1400 Mars Hill Rd, Flagstaff, AZ 86001, USA}
\author{O.~Lahav}
\affiliation{Department of Physics \& Astronomy, University College London, Gower Street, London, WC1E 6BT, UK}
\author{S.~Lee}
\affiliation{Jet Propulsion Laboratory, California Institute of Technology, 4800 Oak Grove Dr., Pasadena, CA 91109, USA}
\author{J.~L.~Marshall}
\affiliation{George P. and Cynthia Woods Mitchell Institute for Fundamental Physics and Astronomy, and Department of Physics and Astronomy, Texas A\&M University, College Station, TX 77843,  USA}
\author{J. Mena-Fern{\'a}ndez}
\affiliation{Universit\'e Grenoble Alpes, CNRS, LPSC-IN2P3, 38000 Grenoble, France}
\author{F.~Menanteau}
\affiliation{Center for Astrophysical Surveys, National Center for Supercomputing Applications, 1205 West Clark St., Urbana, IL 61801, USA}
\affiliation{Department of Astronomy, University of Illinois at Urbana-Champaign, 1002 W. Green Street, Urbana, IL 61801, USA}
\author{R.~Miquel}
\affiliation{Instituci\'o Catalana de Recerca i Estudis Avan\c{c}ats, E-08010 Barcelona, Spain}
\affiliation{Institut de F\'{\i}sica d'Altes Energies (IFAE), The Barcelona Institute of Science and Technology, Campus UAB, 08193 Bellaterra (Barcelona) Spain}
\author{J.~Muir}
\affiliation{Department of Physics, University of Cincinnati, Cincinnati, Ohio 45221, USA}
\affiliation{Perimeter Institute for Theoretical Physics, 31 Caroline St. North, Waterloo, ON N2L 2Y5, Canada}
\author{J.~Myles}
\affiliation{Department of Astrophysical Sciences, Princeton University, Peyton Hall, Princeton, NJ 08544, USA}
\author{R.~L.~C.~Ogando}
\affiliation{Centro de Tecnologia da Informa\c{c}\~ao Renato Archer, Campinas, SP, Brazil - 13069-901}
\affiliation{Observat\'orio Nacional, Rio de Janeiro, RJ, Brazil - 20921-400}

\author{A.~Palmese}
\affiliation{Department of Physics, Carnegie Mellon University, Pittsburgh, Pennsylvania 15312, USA}

\author{M.~Paterno}
\affiliation{Fermi National Accelerator Laboratory, P. O. Box 500, Batavia, IL 60510, USA}
\author{P.~Petravick}
\affiliation{Center for Astrophysical Surveys, National Center for Supercomputing Applications, 1205 West Clark St., Urbana, IL 61801, USA}
\author{A.~A.~Plazas~Malag\'on}
\affiliation{SLAC National Accelerator Laboratory, Menlo Park, CA 94025, USA}
\affiliation{Kavli Institute for Particle Astrophysics \& Cosmology, P. O. Box 2450, Stanford University, Stanford, CA 94305, USA}

\author{M.~Raveri}
\affiliation{Department of Physics, University of Genova and INFN, Via Dodecaneso 33, 16146, Genova, Italy}

\author{A.~Roodman}
\affiliation{SLAC National Accelerator Laboratory, Menlo Park, CA 94025, USA}
\affiliation{Kavli Institute for Particle Astrophysics \& Cosmology, P. O. Box 2450, Stanford University, Stanford, CA 94305, USA}
\author{S.~Samuroff}
\affiliation{Institut de F\'{\i}sica d'Altes Energies (IFAE), The Barcelona Institute of Science and Technology, Campus UAB, 08193 Bellaterra (Barcelona) Spain}
\affiliation{Department of Physics, Northeastern University, Boston, MA 02115, USA}
\author{E.~Sanchez}
\affiliation{Centro de Investigaciones Energ\'eticas, Medioambientales y Tecnol\'ogicas (CIEMAT), Madrid, Spain}
\author{E.~Sheldon}
\affiliation{Brookhaven National Laboratory, Bldg 510, Upton, NY 11973, USA}

\author{T.~Shin}
\affiliation{Department of Physics and Astronomy, Stony Brook University, Stony Brook, NY 11794, USA}
\author{M.~Smith}
\affiliation{Physics Department, Lancaster University, Lancaster, LA1 4YB, UK}
\author{E.~Suchyta}
\affiliation{Computer Science and Mathematics Division, Oak Ridge National Laboratory, Oak Ridge, TN 37831}
\author{M.~E.~C.~Swanson}
\affiliation{Center for Astrophysical Surveys, National Center for Supercomputing Applications, 1205 West Clark St., Urbana, IL 61801, USA}
\author{G.~Tarle}
\affiliation{Department of Physics, University of Michigan, Ann Arbor, MI 48109, USA}

\author{D.~Thomas}
\affiliation{Institute of Cosmology and Gravitation, University of Portsmouth, Portsmouth, PO1 3FX, UK}

\author{V.~Vikram}
\affiliation{Department of Physics and Astronomy, University of Pennsylvania, Philadelphia, PA 19104, USA}
\author{A.~R.~Walker}
\affiliation{Cerro Tololo Inter-American Observatory, NSF's National Optical-Infrared Astronomy Research Laboratory, Casilla 603, La Serena, Chile}

\collaboration{DES collaboration}


\date{\today}
\hspace{0.2cm}

\label{firstpage}

\begin{abstract}
\vspace{0.2cm}
This work is part of a series establishing the redshift framework for the $3\times2$pt analysis of the Dark Energy Survey Year 6 (DES Y6). For DES Y6, photometric redshift distributions are estimated using self-organizing maps (SOMs), calibrated with spectroscopic and many-band photometric data. To overcome limitations from color–redshift degeneracies and incomplete spectroscopic coverage, we enhance this approach by incorporating clustering-based redshift constraints (clustering-z, or WZ) from angular cross-correlations with BOSS and eBOSS galaxies, and eBOSS quasar samples. We define a WZ likelihood and apply importance sampling to a large ensemble of SOM-derived $n(z)$ realizations, selecting those consistent with the clustering measurements to produce a posterior sample for each lens and source bin. The analysis uses angular scales \red{corresponding to } 1.5–5 Mpc to optimize signal-to-noise while mitigating modeling uncertainties, and marginalizes over redshift-dependent galaxy bias and other systematics informed by the N-body simulation Cardinal. While a sparser spectroscopic reference sample limits WZ constraining power at $z>1.1$, particularly for source bins, we demonstrate that combining SOMPZ with WZ improves redshift accuracy and enhances the overall cosmological constraining power of DES Y6. We estimate an improvement in $S_8$ of approximately 10$\%$ for cosmic shear and $3\times2$pt analysis, primarily due to the WZ calibration of the source samples.
\end{abstract}


\preprint{DES-2025-0936}
\preprint{FERMILAB-PUB-25-0733-PPD}
\maketitle


\section{Introduction} \label{sec:intro}
Accurately determining the redshift distributions  of lens and source galaxy samples is crucial
for correct interpretation of cosmological measurements of galaxy clustering and weak gravitational lensing \citep[see, e.g.,][]{Hurterer2006,photo-z-perf_cosmo, KIDS_redshift_dis, Stolzner_GaussianFitting}. The cost in time and money obtaining spectroscopic redshifts for every galaxy is impractical for the $>100$~million galaxies in present-day imaging surveys such as the Dark Energy Survey (DES).
\begin{figure*}
\includegraphics[width=\linewidth]{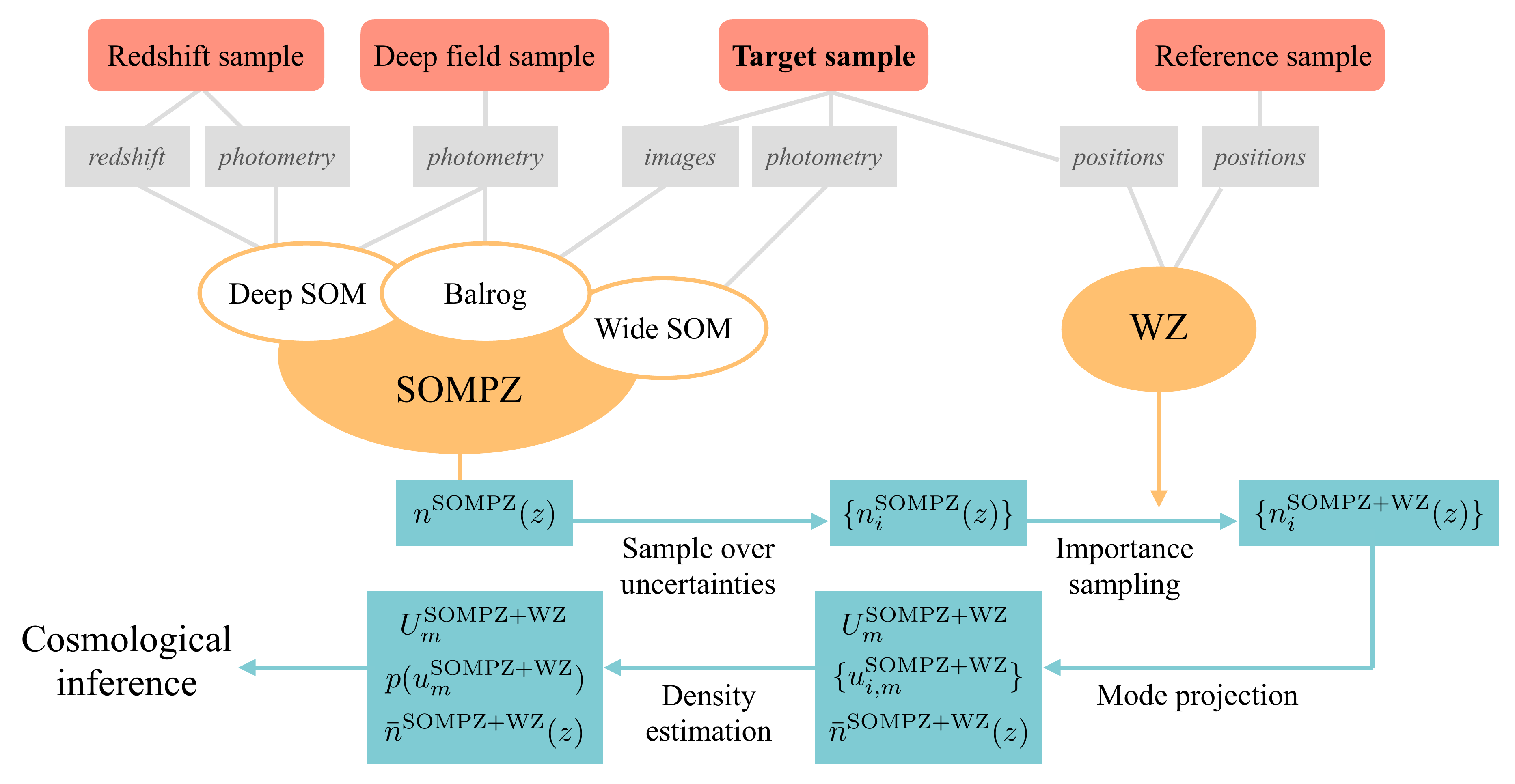}
\caption{Flowchart summarizing the DES Y6 3$\times$2pt redshift calibration pipeline. Photometric data for targeted subsets of the DES Y6 galaxy catalog are placed into Self-Organizing Maps (SOMs), with the redshift distributions of each SOM cell estimated from better-observed galaxies in DES deep fields, along with an observational `transfer function' from deep to wide observational properties derived from the `Balrog' source injection simulations. Sampling over the associated uncertainties produces a set of realizations, $n_i^{\rm SOMPZ}(z)$. The constraints on $n(z)$ implied by clustering information (WZ), described in this work, are realized by 
importance sampling the $n_i^{\rm SOMPZ}(z)$ to yield $n_i^{\rm SOMPZ+WZ}(z)$ realizations.  These are then projected onto a small number modes that capture all of the cosmologically relevant variation, enabling a more accurate marginalization over the redshift distribution in cosmological inference. Note that for weak-lensing source galaxies, there is an additional step of correction for image blending, applied after the mode projection.
 \citet{y6-sompz-metadetect} and \citet{y6-sompz-maglim} contain longer descriptions of the full redshift pipeline.}
\label{fig:flowchart}
\end{figure*}
`Photometric redshifts' (photo-$z$'s, or PZ) can be estimated from multi-band photometry of each galaxy. A wide range of photo-$z$ methods exist \citep[e.g.,][]{photo-z_HSC,Salvato_2019,Ilbert-EP11,Desprez-EP10}. The approach adapted by DES and other surveys \citep[][]{Wright_2020,Wright_2020b,3sdir, DESY3_MAGLIM_z,Campos2024} partitions the photometric space with self-organizing maps (SOMs), and calibrates each SOM `cell' with spectroscopic and/or many-band photometric redshifts, which enables careful quantification of  various sources of uncertainty.  Photometric methods are, however, inherently limited by color-redshift degeneracies, and by the lack of fully representative spectroscopic samples for both training and calibration \citep{Wright_2020,hartley20}.

Clustering-based redshift estimation (WZ, where W refers to $w(\theta)$) provides a complementary method for constraining redshift distributions (see, e.g. \citealt{newman2008, Menard2013,clust_z_mcQuinn_white,theWizz, Scottez}, \citealt*{ Gatti_DESY1}, \citealt{van_den_Busch_2020, KIDS_redshift_dis},  \citealt*{Gatti_Giulia_DESY3}, \citealt{Cawton2022, HSC_clustering_z,Euclid_dassignies}). 
Unlike photometric methods, clustering-based redshift techniques are not affected by photometric noise, the large statistical uncertainties from small deep fields, or the need for fully representative spectroscopic samples. Instead, they rely on angular cross-correlations between the photometric samples and auxiliary spectroscopic samples with secure redshifts. When two samples overlap in redshift, they trace the same underlying dark-matter field, leading to a measurable angular correlation. The amplitude of this correlation thus provides information on the redshift distribution of the photometric sample.
This amplitude is, however, degenerate with the biases of both galaxy samples with respect to the dark matter, making galaxy-bias estimation a crucial step (\citealt{clust_z_mcQuinn_white},  \citealt*{Gatti_DESY1}, \citealt{van_den_Busch_2020}, \citealt*{Gatti_Giulia_DESY3}, \citealt{Cawton2022,DESY3_MAGLIM_z,Naidoo23,Euclid_dassignies}). The bias of the spectroscopic sample, with its known $n(z),$ can be directly \red{inferred} from its auto-correlation functions for a known cosmology.  But estimating the bias of the photometric sample, with unknown $n(z),$ is more challenging.

Since both PZ and WZ approaches to measuring $n(z)$ have their respective strengths and weaknesses, for the analysis of the full 6-year DES catalogs (DES Y6), we combine both types of information to estimate the necessary $n(z)$ functions and their uncertainties. Figure~\ref{fig:flowchart} is a visual summary of the steps in generating the probability distributions of the DES Y6 galaxy samples' $n(z)$ functions.

This article is one of four papers that describe the redshift treatment for the DES Y6 `3$\times$2pt' analysis of angular correlations among weak-lensing and galaxy-density fields: (1) \citet{y6-sompz-metadetect} describe the derivation of the SOMPZ redshift estimates for the source galaxy sample; (2) \citet{y6-sompz-maglim} describe the derivation of the SOMPZ redshift estimates for the lens galaxy sample; (3) this article describes how the clustering information is incorporated into the redshift estimates; and (4) \citet{y6-modes} describe a novel approach to sample over the redshift uncertainty during the cosmological inference. 
Together, the four papers coherently describe the Y6 redshift estimate pipeline. In addition, two related papers play a crucial role in our analysis: (a) \citet{y6-imagesims} describe the source redshift correction due to blending; and (b) \citet{y6-shear_ratio} describe the shear ratio validation for the lens and source redshift.  

In this work, we follow the approach to uncertainties in galaxy biases (and other WZ theory and measurement systematic errors) used for DES Year 3 (DES Y3) analyses by \citet*{Gatti_Giulia_DESY3} and \citet{DESY3_MAGLIM_z}. $N$-body simulated catalogs are used to define priors for the photometric galaxies' bias and its redshift evolution, as well as other clustering-related systematics.  These redshift-dependent systematics are marginalized over in the calibration.

The amplitude of angular clustering, along with its signal-to-noise ratio (SNR), depends on the scale range considered. In earlier studies, very small scales (0.1--1.5 Mpc) were commonly used (\citealt{Menard2013,Schmidt2013}, \citealt*{Gatti_DESY1}, \citealt{Cawton2022,Naidoo23}). These scales offer high-SNR data vectors and are not used for cosmological analyses, ensuring a degree of independence between calibration and cosmological inference.
In DES Y3, a larger scale range (1.5--5 Mpc) was adopted, as it still provided a good SNR while being less affected by non-linearities and 1-halo physics, and remained separate from the scales used in cosmological analyses (\citealt*{Gatti_Giulia_DESY3}, \citealt{DESY3_MAGLIM_z}). More recently, \citet{Euclid_dassignies} tested various scale ranges using state-of-the-art mocks and found 1.5--5 Mpc to be optimal. Based on these findings, we adopt this scale range as the default in this work, while also exploring alternative ranges with our data.

The precision of WZ estimation depends on the redshift and sky overlap between the photometric and the spectroscopic data. In this work, we use spectroscopic data from the BOSS and eBOSS surveys \citep{BOSS_color,eboss_dawson}, which unfortunately overlap only 700--800~deg$^2$ of the $\approx5000$~deg$^2$ of the DES Y6 footprint. Furthermore the dense samples from BOSS and eBOSS surveys only cover the $0<z<1.1 $ range, whereas our DES catalogs extend to $z_{\rm p}=2$. Thus in this work we additionally use the eBOSS QSO sample to extend WZ up to $z=2.2$ to constrain the DES sources bins, but with limited constraining power due to its sparser space density. Unlike DES Y3 WZ, we do not use \textsc{Redmagic} (\citealt{2016rm}, \citealt*{Gatti_Giulia_DESY3}) as a reference sample to calibrate the sources $n(z)$, because the systematic uncertainties arising from its photometric selection and redshift assignment are found to negate any statistical gains from its inclusion.  

The denser and larger DESI DR1 data \citep{DESI_DR1_data} became available in March 2025, too late for inclusion in the DES Y6 $3\times2$ analysis, but a quick examination (also presented in this article) suggests that the DESI DR1 cross-correlations with DES Y6 catalogs are fully consistent with the BOSS/eBOSS values.  Future analyses of DES Y6 that would benefit from more accurate $n(z)$ characterization than given here should include DES$\times$DESI clustering.  

The paper is structured as follows. In Section~\ref{sec:method} we describe the WZ methodology we have adopted to estimate and calibrate the redshift distribution of the \textsc{Maglim} galaxy samples used as the WL lens population and as density tracers;  and the \textsc{Metadetect} galaxy samples that form the WL source populations.  In Section~\ref{sec:data} we describe the data used in this work, including DES data as well as external spectroscopic catalogs. We present the results of our calibration in Section~\ref{sec:results}, and examine the impact of the WZ information on cosmological results. Finally, we conclude in Section~\ref{sec:conclusions}.

\section{Methodology}\label{sec:method}

In this section, we describe the WZ methodology. In Sect. \ref{sec:th_clustering} we introduce the modeling of angular clustering, magnification, and clustering-redshifts systematics. In Sect. \ref{sec:method_data} we briefly present how we  measure 
 in practice the different correlations and covariance  from the data. Finally in Sect. \ref{sec:WZ_likelihood} we explain how the constraints on the redshift distributions are extracted from the data measurements. 

\subsection{Modeling of the angular clustering}\label{sec:th_clustering}

The WZ method can be summarized as follows.
\begin{itemize}
    \item We have one or more unknown samples $\rmu$, each of which has a redshift distribution $n_{\rmu}(z)$ that we wish to determine. 
    \item We have multiple reference samples $\rmr_i,$ each with a secure redshift distribution $n_{\rmr_i}(z)$. In this article, we will use spectroscopic samples.
    \item The intrinsic cross-correlation between two samples is non-zero only for pairs of objects in close proximity in $z$, in which case they trace the same matter field.  This intrinsic signal is contaminated with an apparent angular clustering caused by gravitational lensing magnification.
    \item We evaluate cross-correlations of $\rmu$ and $\{\rmr_i\}$, where the latter are the spectroscopic samples binned into small spectroscopic slices $z_i\pm \Delta z/2$. The amplitudes of the cross-correlations inform $n_{\rmu}(z_i)$.
\end{itemize}
We now detail the modeling of the measured cross-correlations.

\subsubsection{Galaxy clustering}
We use similar conventions to \citet{Y3_DES_Krause} for the two-point statistics. One change relative to previous clustering redshifts works \citep[e.g.][]{Gatti_Giulia_DESY3,Euclid_dassignies} is that the unknowns' distributions $n_\rmu(z)$ are expressed as sums over a set of basis functions $K_j(z)$ rather than assuming top-hat bins (\textit{cf.} Eq. \ref{eq:kernels}), which allows for the inclusion of additional clustering effects. It also improves numerical evaluations of the different WZ elements (in particular the likelihood). The underlying model for the density $\rho_x\propto\diff^3 N/\diff z\,\diff^2\theta$ at redshift $z$ along line of sight $\theta$ for members of a galaxy selection $x$ is
\begin{equation}
\rho_x(z,\theta) = n_x(z) \left(1+B_x\left[z,\delta_{\rm m}(z,\theta)\right]\right) \left[1 + \alpha_x \mu(z,\theta)\right],
\label{eq:rho3d}
\end{equation}
where $\delta_{\rm m}$ is the overdensity of mass at the location $(z,\theta);$ $B_x$ is some biasing function for the galaxies in $x,$ which could depend on redshift; $\mu(z,\theta)$ is the gravitational lensing magnification imposed by the foreground line of sight to $z,\theta;$ and $\alpha_x$ is the magnification coefficient for the galaxies in $x.$  Any correlations between $\delta_{\rm m}(z,\theta)$ and $\mu(z,\theta)$ are weak enough to ignore \citep{magnification_DESY3,y6-magnification}.
\vspace{0.2cm}
\paragraph{Integrated angular clustering correlation:} Let $x$ and $y$ refer to two galaxy samples, with $x=y$ \red{referring to the autocorrelation}. We choose a weighting function\footnote{\red{In this work we use a power law $W(\theta)=\theta^{-1}/\int \diff \theta\, \theta^{-1}$ \citep{Euclid_dassignies}}} $W(\theta)$ to convert the angular cross-clustering $w_{xy}(\theta)$ of the two samples to a scalar value $w_{xy},$ which can be written as
\begin{align}
    w_{xy}
    &=\int \dthet \, {W}(\theta)\,\sum_{\ell}\frac{2\ell +1}{4\pi}C_{\rm gg}^{xy}(\ell) \,\mathcal{L}_{\ell}(\cos{\theta})+w_{xy}^{\mu},
\end{align}
where $C^{xy}_{\rm gg}(\ell)$  is the \red{angular} cross-power power spectrum between the projected intrinsic (unlensed) densities of $x$ and $y$, $\mathcal{L}_\ell$ is the Legendre polynomial of order $\ell$, $w_{xy}^{\mu}$ is the contribution from lensing magnification (which we derive below).
The angular clustering power spectrum can be expressed as
\begin{align}
    C^{xy}_{\rm gg}&(\ell)=
 \int \mathrm{d}\chi\, \frac{\mathcal{W}_{\rm g}^{x}\,\mathcal{W}_{\rm g}^{y}}{\chi^2}\,\,P_{xy}\left(k=\frac{\ell+0.5}{\chi},\,z(\chi)\right)\,\label{eq:linear_bias}
    \\
    &=\int \dz \, n_x(z)\, n_y(z)\, b_x(z)\,b_y(z)\,\frac{H(z)}{c\,\chi^2}P_{\rm m}\left(k(\ell,\chi),\,z\right),\nonumber
\end{align}
where $\mathcal{W}_{\rm g}^x=n_x(z)\frac{\dz}{{\rm d}\chi}$ is the normalized selection function, and $P_{xy}$ and $P_{\rm m}$ are the 3D galaxy cross-spectra and (non-linear) matter power spectra, respectively.  The second line assumes that the biasing functional is linear, $b_x(z,\delta_{\rm m})=b_x(z) \delta_{\rm m}$, for both samples,
as is typically assumed for the WZ method  (\citealt*{Gatti_DESY1,Gatti_Giulia_DESY3}, \citealt{Menard2013,Schmidt2013,van_den_Busch_2020,Naidoo23,HSC_clustering_z}).  
The non-linear power spectrum is numerically evaluated with the default configuration of the \texttt{CCL} library\footnote{\url{https://github.com/LSSTDESC/CCL}} \citep{CCL}, which is based on predictions from \texttt{CLASS} \citep{Diego_Blas_2011_class} and \texttt{halofit} \citep{halofit}. Departures from linear bias are a potential source of systematic errors in this model, which we allow for as described below. For a longer discussion about the validity of this assumption, we refer to \citet{Euclid_dassignies}.\footnote{We refer to Sect. 3.1.3 for a discussion on linear treatment of galaxy bias with non-linear matter power spectrum modeling and its estimation,  Appendix B where the relation between galaxy and matter field is treated at second order, and propagated to the modeling of the observables, and Sect. 4.3 for a measurement of the linear galaxy bias at different scales.}
The results of this latter article\footnote{  \citet{Euclid_dassignies} uses mocks from Euclid Flagship 2 simulation \citep{FS_2024}, and the same numerical power spectrum as in the current work.} is that the assumption of linear bias with non-linear power spectrum is good enough for scales down to 1.5 Mpc. In our context of redshift calibration, we are not trying to extract small-scale cosmological information, merely using the fluctuations as indicators of physical proximity.  In later sections of this work, we perform additional tests on the Y6 data to test this claim.

We introduce the projected matter correlation function
\begin{align}
    w_{\rm m}(z)=&\frac{H(z)}{c\,\chi^2}\int \dthet \, W(\theta)\label{eq:wdm}\\
    \times&\sum_{\ell}\frac{2\ell +1}{4\pi}P_{\rm m}\left(k(\ell,\chi),\,z\right) \,\mathcal{L}_{\ell}(\cos{\theta})\;,\nonumber
\end{align}
and rewrite the galaxy-integrated angular correlation as
\begin{align}
    w_{xy}=\int \dz\,b_x(z)\,b_y(z)\, n_x(z)\,n_y(z)\,w_{\rm m}(z)+w_{xy}^{\mu}\;.\label{eq:wab_th}
\end{align}
\vspace{0.01cm}
\paragraph{Auto-correlation of the reference}
For clustering redshifts, we need to measure the galaxy bias of each of the reference samples ${\rmr_i}$. We neglect the redshift evolution of $b_{\rmr_i}$ in Eq. \eqref{eq:wab_th} as the reference samples are binned in narrow redshift ranges. We neglect as well the lensing magnification, which is nonzero only for galaxy pairs at well-separated redshifts.  Thus the galaxy bias is
\begin{equation}
    b_{\rmr_i}=\sqrt{\frac{w_{\rmr_i\rmr_i}}{\int \dz\, n_{\rmr_i}^2(z)\,w_{\rm m}(z)}}.
    \label{eq:brr}
\end{equation}
As a consequence, if $n_{\rmr_i}$ is known (which is the case for spectroscopic samples), we can use the measurements $w_{\rmr_i\rmr_i}$ to infer the values $b_{\rmr_i}$ of the reference galaxies' biases. 
\vspace{0.2cm}
\paragraph{Cross-correlation} We want to extract information on the redshift distribution of the unknown sample $n_\rmu(z)$ by cross-correlating it with many reference samples $\rmr_i$. First, we decompose the unknown redshift distribution over a basis,
\begin{equation}
    n_\rmu(z)=\sum_j f_\rmu^j\,K_j(z),
\label{eq:kernels}
\end{equation}
with $K_j(z)$ the basis, and $f_\rmu^j$ the coefficients of the decomposition. 
We use sawtooth functions, as they naturally give $n(z)\rightarrow0$ as $z\rightarrow0$, and are continuous, 
\begin{align}
    K_j(z&)= \frac{1}{\Delta z}\,\Lambda \left( \frac{z}{\Delta z} - j \right),
 \\
\text{with}\hspace{.5cm}
    \Lambda(&a) =
    \begin{cases}
        a, & 0 < a < 1 \\
        2 - a, & 1 < a < 2 \\
        0, & \text{otherwise}.
    \end{cases}\nonumber
\end{align}

The modeling of the cross-correlation reduces to 
\begin{align}
    w_{\rmur_i}&=b_{\rmr_i}\int \dz \sum_j b_{\rmu_j}\, f_\rmu^j\,K_j(z)\, n_{\rmr_i}(z)\,w_{\rm m}(z)+w_{\rmur_i}^{\mu}\nonumber\\
    &= \sum_j b_{\rmu_j}\, f_\rmu^j\, A_{ji}+w_{\rmur_i}^{\mu}\label{eq:wur_sumfuA}
\end{align}
where we assume the galaxy bias of the unknown sample is a constant $b_{\rmu_j}$ over the support of each element $j$ of the basis, and where
\begin{equation}
    A_{ji}=b_{\rmr_i}\int \dz\, K_j(z)\, n_{\rmr_i}(z)\,w_{\rm m}(z).
\end{equation}
\vspace{0.01cm}
\subsubsection{Magnification}
\begin{figure*}
    \centering
    \includegraphics[width=1\linewidth]{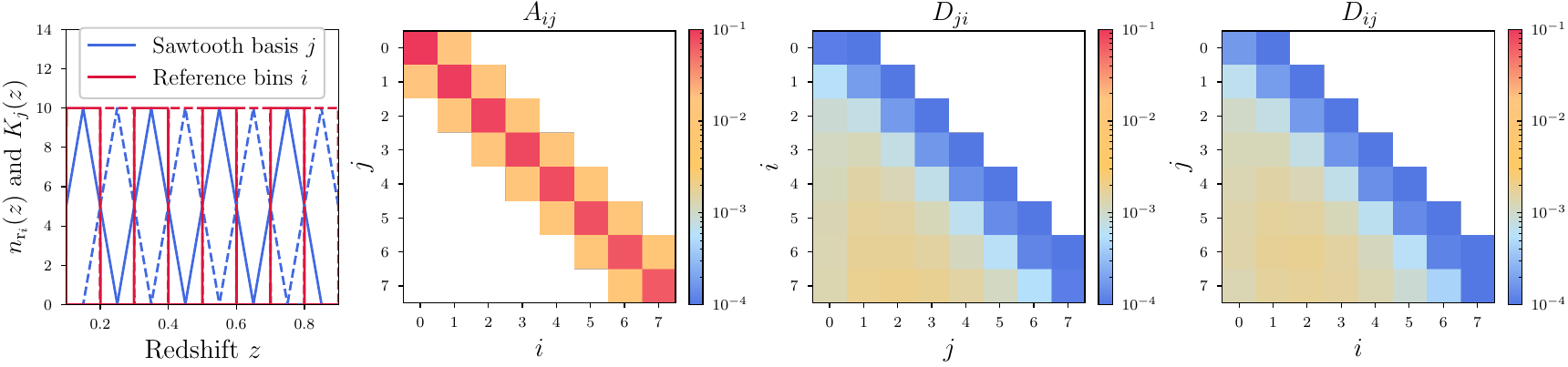}
    \caption{Illustration of the clustering and magnification for 8 sawtooth distributions \red{(on the left panel odd $j$ are represented by solid lines, and even with dashed) } cross-correlated with 8 rectangle reference bins, both with $\Delta z=0.1$. Clustering matrix $A$ and magnification matrices $D$ are then $8\times 8$. The orange values are one order of magnitude smaller than the red, and the blue, two orders of magnitude. Matrix coefficients in white are exactly 0. Please note that $D_{\rmr_i j}$ and $D_{j\rmr_i}$ are very similar but not exactly equal because for this calculation we have taken similar $\Delta z$ binning but with sawtooth kernels for the unknowns $\rmu$ and rectangular (boxcar) kernels for the references $\rmr.$}
    \label{fig:Anr_Dun_Dnr}
\end{figure*}

The angular correlation of $x$ and $y$, caused by magnification is, at first order,
\begin{align}
    w_{xy}^{\mu} =\int \dthet\,&W(\theta)\sum_{ij\in\{{\rm g}\mu,\, \mu {\rm g}\}}c^i_{x}\,c^j_{y}\int \mathrm{d}\chi\, \frac{\mathcal{W}_i^{x}\mathcal{W}_j^{y}}{\chi^2}\,\label{eq:magn_WW}\\\
    \times&\sum_\ell \frac{2\ell+1}{4\pi}\mathcal{L}_\ell(\cos{\theta})\,P_{\rm m}\left(k(\ell,\chi),\,z(\chi)\right)\,,\nonumber
\end{align}
where $c_{x}^{\rm g}=b_{x}$ is the galaxy bias, $c_{x}^\mu=\alpha_x$ is the magnification coefficient,\footnote{Please note there is a large variety of notations in the literature, often in conflict. E.g., what we refer to as $\alpha$ is often written $2(\alpha-1)$  e.g. in \citet{hildebrandt_magn}, $(C_{\rm sample}-C_{\rm area})/2$ e.g. in \citet{magnification_DESY3}.}  and $\mathcal{W}_\mu$ is the lensing efficiency,
\begin{equation}
    \mathcal{W}_\mu^x=\frac{3\, \Omega_{\rm m}\,H_0^2}{2\,c^2}\int_\chi^{\infty}\mathrm{d}\chi^\prime n_x(\chi^{\prime})\frac{\chi}{a(\chi)}\frac{\chi^{\prime}-\chi}{\chi^{\prime}}\,.
\end{equation}
Reordering the terms, and using the integrated matter correlation $w_{\rm m}$ from Equation~\eqref{eq:wdm}, we have
\begin{widetext}
\begin{equation}
w_{xy}^{\mu}=\frac{3\,\Omega_{\rm m}\,H_0 }{c}\int_0^{\infty} \dz\, b_{x}(z)\,n_x(z)\,\frac{H_0\,(1+z)}{H(z)}\, w_{\rm m}(z )\int_{z}^{\infty} \dz'\,\alpha_y(z')\, n_y(z')\,\frac{\chi(z')-\chi(z)}{\chi(z')}\,\chi(z)+\{x\leftrightarrow y\}\,.\label{eq:magnification_wxymu}
\end{equation}
\end{widetext}
where $\{x\leftrightarrow y\}$ represents the same term, but switching the $x$ and $y$ indices.
Decomposing the $\rmu$ distribution over the $K_j$ basis, the two magnification terms can be very compactly written
\begin{align}
    w_{\rmur_i}^{\mu}&=\sum_j f^j_\rmu \left[ b_{\rmu_j}\,\alpha_{ \rmr_i} D_{ji}
+ b_{\rmr_i}\alpha_{ \rmu_j}\, D_{ij}\right],
\end{align}
where we introduce the $D_{ij}$ matrices,  independent of any $b$ or $\mu$ values, that capture the amplitude of the magnification $\mu$ of galaxies in kernel $j$ by the matter in kernel $i$:
\begin{widetext}
\begin{equation}
    D_{ij} = \frac{3\,\Omega_{\rm m}\,H_0 }{c}\int_0^{\infty} \dz_i\,K_i(z_i)\,\frac{H_0\,(1+z_i)}{H(z_i)}\, w_{\rm m}(z_i )\int_{z_i}^{\infty} \dz_j\, K_j(z_j)\,\frac{\chi(z_j)-\chi(z_i)}{\chi(z_j)}\,\chi(z_i). 
\end{equation}
\end{widetext}
In this formula we have assumed that $n_{\rmr_i}(z) = K_i(z)$, but one is free to reintroduce the explicitly known value of $n_{\rmr_i}(z)$ under the appropriate integral---this will not make much difference, as can be seen by comparing the two rightmost panels of Fig.~\ref{fig:Anr_Dun_Dnr}, which swap the rectangular $n_{\rmr_i}$ bins in for the different appearances of the kernel $K_i$ in this equation for $D_{ij}.$
The final theoretical expression for the cross-correlation is then
\begin{equation}
    w_{\rmur_i}=\sum_j f_\rmu^j \left[\,b_{\rmu_j}\, A_{ji}\,+\alpha_{\rmu_j}b_{\rmr_i}D_{ij}+b_{\rmu_j}\alpha_{\rmr_i}D_{ji}\right].
    \label{eq:wuri}
\end{equation}

In Fig. \ref{fig:Anr_Dun_Dnr} we show the clustering and magnification matrices for a set of 8 reference rectangular bins, and 8 sawtooth bins, playing the role of the basis $K_j$.  The clustering matrix is mainly diagonal, as expected. The magnification matrices $D$ are lower-triangular because for tracers in bin $j$ to be magnified by the matter in $i$, the latter has to be localized at lower redshift. We also see that the amplitude of magnification is one order of magnitude smaller than the clustering, but couples more bin pairs. 

\subsubsection{Systematic errors}\label{sec:sys}
The model used to derive Equation~\eqref{eq:wuri} is incomplete in known respects, most importantly that the linear bias model does not hold precisely, especially at smaller separations, and we will fit models in which $b_\rmu, \alpha_\rmu, b_\rmr,$ and $\alpha_\rmr$ are constant across redshift $z.$
There may be other systematic errors as well, which we will detail later. When fitting the data to this model, we will therefore introduce a `systematic error' function $S$ into the model for the intrinsic-clustering term,
 as in \citet*{Gatti_Giulia_DESY3}.   The magnification terms are smaller corrections to the intrinsic-clustering term, so that any deviation from our magnification modeling can be easily absorbed into the bias on intrinsic clustering. 
\begin{align}
    \hat{w}_{\rmur_i}=\sum_j f_\rmu^j\left[ b_{\rmu_j}A_{j\rmr_i}\,(1+S_{\rmu\rmr_i})+\alpha_{\rmu_j}b_{\rmr_i}D_{ij}+b_{\rmu_j}\alpha_{\rmr_i}D_{ji}\right].\label{eq:model:wur_sys}
\end{align}
The $S$ function is chosen to be a function of $z$ whose scale and amplitude of variability can be ``dialed in'' to match the level that the model Equation~\eqref{eq:wuri} departs from the true $w_{\rmur}$ values in simulations.  We choose 
\begin{align}
    &S_{{\rm ur}_i}  =\sum_{k=0}^{M}s_{{\rm u}k}\frac{\sqrt{2k+1}}{0.85}\,\mathcal{L}_k(\tilde z)\label{eq:Sys} \\
    &\text{with }\hspace{0.5cm}\tilde z  \equiv 0.85\times\frac{2\,z-z_{\rm max}-z_{\rm min}}{z_{\rm max}-z_{\rm min}}.\nonumber
\end{align}
$\tilde z$ is a linear remapping of $z$ from the range  $[z_{\rm min},z_{\rm max}]$ to $[-0.85,0.85]$.  The maximum degree $M$ of the Legendre polynomials $\mathcal{L}_k$ controls 
how rapidly $S(z)$ can vary across its range.  Gaussian priors on the $s_{\rmu k}$ coefficients control the 
(scale-dependent) amplitude of the deviations of the model from the simple linear-bias model.  The fit of the model in Equation~\eqref{eq:model:wur_sys} to the data is done with the $\{s_{\rmu k}\}$ allowed to vary.  The choices of the order $M$ and the priors $\sigma(s_{\rmu k})$ are made by examining the deviation between the ``observed'' $w_{\rmur}$ and the model in Eq. \eqref{eq:wuri} in the Cardinal simulations of the DES Y6 catalogs \citep{cardinal}.  This process is described in Section~\ref{sec:cardinalsys}.

We have the option of parameterizing multiple uncorrelated systematic errors independently in the same Legendre basis, with prior widths amplitudes $s_{q{\rmu}k}$ for systematic errors indexed by $q$.  Then, when calculating likelihoods for the data, we can combine the multiple systematic errors' effects by $\sigma^2(s_{\rmu k}) = \sum_q \sigma^2(s_{q\rmu k})$.

\subsection{Measurement of the cross-correlations and their uncertainties}\label{sec:method_data}

In this section, we describe the methods for obtaining estimates of and uncertainties on the observables $\hat{w}_{\rm ur}$ and $\hat{w}_{\rmr\rmr}$ from the catalogs of DES Y6 galaxies and the spectroscopic catalogs.

\subsubsection{Data estimator and scale weighting}

We use the Davis and Peebles \citep[DP, ][]{Davis_Peebles} or the Landy-Szalay \citep[LS,][]{Landy_Szalay} estimator, depending upon whether a simulated catalog of random (unclustered) position sample is available for the unknown data. The estimators for the integrated correlation functions are
\begin{align}
    &w_{xy}^{\rm DP}\approx \sum_{i} \Delta \theta_i\, W(\theta_i)\,\frac{{\rm D}_x{\rm D}_y(\theta_i)-{\rm R}_x{\rm D}_y(\theta_i)}{{\rm R}_x{\rm D}_y(\theta_i)}\label{eq:DP}\,,\\
    &w_{xy}^{\rm LS}\approx \sum_{i} \Delta \theta_i\, W(\theta_i)\,\frac{({\rm D}_x-{\rm R}_x)({\rm D}_y-{\rm R}_y)(\theta_i)}{{\rm R}_x{\rm R}_y(\theta_i)},
\end{align}
where $XY(\theta_i)$ are counts of pairs of catalog members with separations in a bin centered at $\theta_i$ with width $\Delta\theta_i,$ and $X,Y \in \{D,R\}$ are counts of the data or random catalogs. 
We use 10 logarithmically spaced bins between the angles corresponding to comoving transverse scales $r_{\rm p}$ of 1.5 to 5 Mpc, with $W(\theta)=\theta^{-1}/\sum_{i} \Delta \theta_i\, \theta_i^{-1}.$  These estimators are evaluated with \texttt{Treecorr} \citep{Treecorr}, a highly efficient, open-source, user-friendly Python package.

\subsubsection{Covariance}

We estimate the covariance matrix $C_w$ of the $w_{\rmur_i}$ values with the Jackknife method, splitting the overlap of the DES Y6 footprint with the BOSS footprint into 100 sub-regions to capture both shot noise and sample variance. \footnote{\red{The physical size of a 8deg$^2$ area is much larger than the correlation function scale ($\sim 50-100$ Mpc vs. $2$ Mpc)}.}  Inverses of Jackknife covariances are corrected for finite sampling as per \citet{Hartlap}.   Because the spectroscopic reference sample's bins $\rmr_i$ and $\rmr_j$ have no overlap in redshift for $i\ne j,$ we expect ${\rm Cov}(w_{\rmur_i},w_{\rmur_j})$ to be near zero.   This is found to be true to within measurement error both for data and in the much larger area of the Cardinal simulated data, and was also tested with numerical estimates of the covariance. We therefore set to zero all such terms.
As the noise from finite number of patches is greatly suppressed by this
procedure, we do not apply Hartlap corrections when inverting these covariances.  Comparison of the Hartlap-corrected-inverse of the noisy covariance to the inverse of this ``cleaned'' covariance yielded consistent diagonal elements, in both data and Cardinal \red{(cf. Sect. \ref{sec:data})}.

When there are multiple unknown galaxy samples $N_\rmu>1$ under study,  we also set to zero all the covariances involving two different spectroscopic bins, ${\rm Cov}(w_{\rmu_i,\rmr_j}, w_{\rmu_k,\rmr_\ell})$ for $j\ne \ell$, but we can have nonzero ${\rm Cov}(w_{\rmu_i \rmr_k},w_{\rmu_j\rmr_k})$.  If we create a single vector $\vecw$ by concatenating the $w_{\rmur_i}$ values for all samples $\rmu,$ the covariance matrix $C_w$ for this vector is block diagonal, with an $N_\rmu\times N_\rmu$ block for each reference sample $\rmr_k$.  This greatly accelerates the calculation of likelihoods described below and was tested with simulated and analytical estimates of the covariances for different realistic cases.

\subsection{WZ posterior probability}\label{sec:WZ_likelihood}
With the model and measurements for $w_{\rmur}$ both in hand, we wish to evaluate the probability that any given proposal for the $n_\rmu(z)$ functions could have produced the observed $w_{\rmur}$ values. In what follows we will use $\rm u$ and $\rm r$ as indices that run over the samples of unknown and reference galaxies, respectively. The proposed $n_\rmu(z)$ functions are specified by a vector $\vecn \equiv \{f^j_\rmu \}$ of the coefficients of the expansion Equation~\eqref{eq:kernels}, and the data by a vector $\vecw = \{w_{\rmur}\}$ 
with covariance matrix $C_w.$  We have a model $\hat\vecw(\vecn,\vecq)$ 
where there are nuisance parameters $\vecq=\{b_\rmu,\, \alpha_\rmu,\,\alpha_\rmr,\, s_{\rmu k}\}$ and $b_\rmr,$ that appear in the model.  We will assume that the $b_\rmr$ are known from their auto-correlations via Equation~\eqref{eq:brr}, letting any errors be absorbed into the systematic-error functions specified by the $s_{\rmu k}$ coefficients.

We will assume that the likelihood function of the measurements is Gaussian, with a mean at $\hat\vecw$ and covariance matrix of $C_w$.  The posterior probability of the  observations given the  proposed $\vecn$  then becomes 
\begin{widetext}
    
\begin{align} \label{eq:wz_likelihood}
p(\vecw | \vecn) & \propto \int \diff \vecq\,{\mathcal L}(\vecw | \vecn, \vecq) p(\vecq) \\
& \propto \int \diff\vecq\, \exp\left[ -(\vecw-\hat\vecw)^T C_w^{-1} (\vecw-\hat\vecw)/2\right] \nonumber\\ 
& \phantom{\propto} \times \exp\left[ -\left(\alpha_\rmr-\alpha_{\rm r}^{(0)}\right)^2/2\sigma^2(\alpha_\rmr)\right]
\prod_\rmu \exp\left[ -\left(\alpha_\rmu-\alpha_{\rmu}^{(0)}\right)^2/2\sigma^2(\alpha_\rmu)\right]
\prod_{\rmu,k} \exp\left[ -s_{\rmu k}^2/2\sigma^2(s_{ \rmu k})\right].
\end{align}
\end{widetext}
The last line contains the Gaussian priors that we apply to the single value $\alpha_\rmr$ that we assign to all reference bins, the $\alpha_\rmu$ magnification coefficient for each unknown sample, and the (zero-mean) $s_{\rmu k}$ systematic coefficients.  We take flat priors for the $b_\rmu$ values, which are strongly constrained by the WZ data.

Examination of our $\hat\vecw$ model in Equation~\eqref{eq:model:wur_sys} reveals that if we divide the nuisance parameters into $\vecb = \{b_\rmu\}$ and $\vecq^\prime = \{\alpha_\rmu, \alpha_\rmr, s_{\rmu k}\},$ then the posterior probability under the integral is an exponentiated quadratic function of $\vecq^\prime.$ \red{This} means we can write the integrand as a normal distribution over $\vecq^\prime$ with a mean $\vecq_0^\prime$ and a covariance $C_{q}^{\prime}$ that are functions of $\vecn$ and $\vecb:$ 
\begin{equation}
    p(\vecw | \vecn) \propto \int \diff\vecb\, c(\vecn,\vecb) \int \diff\vecq^\prime \, {\mathcal{N}\left[\vecq^\prime | \vecq_0^\prime(\vecn,\vecb), C_q^\prime(\vecn,\vecb)\right]}.
\end{equation}
There is a $\vecb$-dependent normalization factor $c(\vecn,\vecb)$, and 
the integration over $\vecq^\prime$ can then be done analytically, yielding $\left|2\pi C^\prime_q\right|^{1/2}$.  We opt to replace the marginalization over the $b_\rmu$ variables with an evaluation of the maximum of the integrand (i.e. a profile likelihood)---this should still yield a good estimate of the relative compatibility of any $n(z)$ proposal with the WZ observations.\footnote{ We expect the $b_\rmu$ parameter to be uncorrelated with redshift parameters (since it is an amplitude, and we want normalized $n(z)$). In the case of Gaussian likelihood with uninformative flat priors, marginalization and profile likelihood are equal up to a numerical factor.}  This gives
\begin{equation}
    p(\vecw | \vecn) \propto \sup_\vecb \left\{c(\vecn,\vecb)\, \left| C_q^\prime(\vecn,\vecb)\right|^{1/2}\right\}.
\label{eq:wzpost}
\end{equation}
The maximization over $\vecb$ is accomplished by a series of Newton iterations, which are enabled by using automatic differentiation to determine the derivatives of the algebraic expressions for $c(\vecn,\vecb)$ and $C^\prime_q(\vecn,\vecb)$ with respect to $\vecb.$ 

\subsection{Importance sampling and photometric uncertainty marginalization}
\label{sec:importance}
For the cosmological analyses of the DES Y6 data, we need to marginalize the probability of cosmological parameters over the uncertainties in the $n(z)$ distributions of 4 samples of source galaxies for weak gravitational lensing observations (``source bins''), and 6 samples of galaxies used to trace the deflecting matter distributions (``lens bins'').  In each case we wish to combine the constraints on $n(z)$ provided by PZ analyses of the sample members  with the WZ information quantified with the previous section's posterior, to yield our strongest constraints on $n(z).$ 
Using Bayes theorem, 
\begin{equation}
    p(\vecn|{\rm PZ, WZ})\propto p({\rm WZ}|{\rm PZ}, \vecn) \, p(\vecn|{\rm PZ}) = p(\vecw|\vecn) \, p(\vecn|{\rm PZ}),\label{eq:n-wzpz}
\end{equation}  
where $p({\rm WZ}|{\rm PZ}, \vecn)=p(\vecw|\vecn)$ because the WZ likelihood does not depend on the photometry, and $p(\vecn|{\rm PZ})$ is the posterior from photometry described in Sect. \ref{sec:sompz_uncertainty}. 
Thus we sample from  the posterior probability $p(\vecn |{\rm  PZ, WZ})$  by:
\begin{enumerate}
    \item Using the method described in Sect. \ref{sec:sompz_uncertainty} to create a large number ($\approx 10^8$) of realizations $\vecn_i$ from $p(\vecn | {\rm PZ})$.
    \item Calculating $p(\vecw \vert\vecn_i)$ for each PZ sample $\vecn_i$ using Eq.~\eqref{eq:wzpost}.
    \item Retaining each $\vecn_i$ sample with probability $\beta\times p(\vecw\vert\vecn_i) / {\rm max}_j[p(\vecw\vert\vecn_j)],$ with $\beta$ chosen so that the number of samples is around 3,000 \red{\citep[the minimum number needed to get noiseless modes, cf.][]{y6-modes}}.  The samples of $\vecn$ surviving this importance sampling step are drawn from the joint distribution $p(\vecn |{\rm WZ, PZ})$,  \textit{cf.} Eq. \eqref{eq:n-wzpz}. 
    \item We apply the compression techniques described in \citet{y6-modes} to find an encoding matrix $E$ such that $\{\vecu_i\}=\{ E\vecn_i\}$ is a much lower-dimensional representation of $n(z)$ for the unknown population under consideration.\footnote{The notation $\vecu$ for the compressed representation is not related to the index $\rm u$ enumerating the unknown galaxy samples.} A density estimator is then used to find a continuous multidimensional probability distribution $p(\vecu)$ that is plausibly the parent distribution for the $\vecu_i$ samples.  The density estimators are described in \citet{y6-modes}.
    \item Samples of $n(z)$ for use in the cosmological analysis can now be produced by sampling from $p(\vecu),$ setting $\vecn=D \vecu$ with the decoding matrix $D$ found by the modal analysis; and then expressing $n(z)$ from the elements of $\vecn$ using Equation~\eqref{eq:kernels}.  
\end{enumerate}

There are some differences between the application of the method to the source galaxies and the application to the lens galaxies.  The lenses are more straightforward: each of the six bins' $n(z)$ goes through the full WZ process independently. We can do so because the small overlap in redshift between lens bins leaves little covariance between bins' results \citep*{Weaverdyck2025_maglim}, so ${\rm Cov}(w_{\rmu_i \rmr},w_{\rmu_j \rmr})$ can be neglected for $i\ne j.$ This is verified with  both an analytical covariance matrix (generated assuming SOM $n(z)$), and  jackknife estimates of the covariance matrix.
There are six independent SOMs  used for the SOMPZ calibration of each bin \citep*{y6-sompz-maglim}; and six independent applications of the importance sampling; six distinct encoding/decoding matrices. The derived distribution of the compressed $\vecu$ samples are seen to be significantly non-Gaussian in some of the lens bins.
A straightforward ``Gaussianizing'' transformation $\vecu\rightarrow\widetilde\vecu$ yields a set of $\widetilde\vecu$ that are consistent with a multivariate normal distribution with identity-matrix covariance. The cosmology inference pipeline thus needs only to sample over these $\widetilde\vecu$ variables, with statistically independent priors, in order to implement marginalization over $p( \vecn |{\rm  WZ,\,PZ})$ for the lenses.

For the source population, there are cosmologically significant correlations between the $i$th PZ sample $n_{\rmu,i}(z)$ for source bin $\rmu$ with $n_{\rmu^\prime, i}(z)$ from a distinct source bin.  Our WZ sampling process must conserve these correlations, so we concatenate the $\vecn_i$ from all 4 source redshift bins into a single vector, and we multiply together the $p(\vecn | {\rm WZ})$ posteriors for the bins as well when importance sampling. 
Another issue with the source bins' samples is that the PZ samples are too noisy to generate high values of $p(\vecn_i | {\rm WZ}),$ so the importance sampling fails.  To remedy this, we use the mode compression method to project away components of $\vecn_i$ that vary $n(z)$ in ways that do not affect the DES Y6 observables, \textit{e.g.} rapid variation with $z$ \citep{y6-sompz-metadetect}.  The de-noised $\hat\vecn_i=DE\vecn_i$ are used to compute the WZ posterior $p(\hat\vecn_i\vert {\rm WZ})$ for importance sampling.  The mode compression is then recalculated using only the samples that survive the importance sampling, \textit{i.e.} are drawn from the joint PZ$+$WZ probability.  An additional difference between the processing of lens and source galaxies is that the $\vecu_i$ samples for the source galaxies are seen to be consistent with a unit-normal distribution---no Gaussianization is needed.

\subsection{SOMPZ uncertainty}\label{sec:sompz_uncertainty}

As highlighted in sections \ref{sec:WZ_likelihood} and \ref{sec:importance}, we work under the assumption that there exists a set of proposal $\{n_i\}$ realizations of the $n(z)$, representing the distribution $p(\vecn|{\rm PZ})$, which is the range of possible $n(z)$ that is allowed by the information that we have from the photometry of galaxies. In this section we summarize the various sources of uncertainty affecting the distributions of $n(z)$ coming from the photometry of \textsc{Metadetect} and \textsc{Maglim}++ galaxies, and how we generate the $\{n_i\}$ realizations of $n(z)$. Detailed descriptions are provided in \citet{y6-sompz-metadetect} and \citet{y6-sompz-maglim}, respectively. 

The $n(z)$ for \textsc{Maglim}++ and \textsc{Metadetect} galaxies is measured empirically with the SOMPZ methodology \citep{Buchs2019, 3sdir, y3-sompz-sources, highz, DESY3_MAGLIM_z, Campos2024}, which leverages the DES deep fields as a stepping stone between the wide-field galaxy photometry in the target samples and redshift information. The DES deep fields have deeper photometry with additional photometric bands,
and overlap with many-band redshift surveys outside of DES. To establish the probability distribution of wide-field photometry given deep-field ``truth'', \textit{i.e.} the ``transfer function'',
the Balrog software \citep*{y6-balrog} injects simulated galaxies drawn from the DES deep fields into real DES wide-field images. The measured spectroscopic and many-band photometric redshifts of the deep-field galaxies can in this way be transferred to galaxy bins defined using wide-field photometry.
This methodology enables control over all known potential sources of uncertainty impacting the $n(z)$ estimates from galaxy photometry.

To model the uncertainty, \citet{y6-sompz-metadetect} and \citet{y6-sompz-maglim} adopt a methodology similar to \citet*{y3-sompz-sources}, incorporating updates from \citet{highz}. We account for uncertainty arising from: (i) sample variance and shot noise from the finite area covered by the deep fields; (ii) biases in the redshifts assigned to deep-field galaxies lacking spectroscopy and using multi-band photometric assignments instead; (iii) uncertainty in the photometric calibration (zero-point) of deep-field galaxies (ZP). 

To address the statistical uncertainties in item (i) above, 
we use the approximate 3sDir model (a product of three Dirichlet distributions), first presented in \citet{3sdir} and then further developed in \citet*{y3-sompz-sources}. Mathematically the model describes $p(\{f_{zc}\}|\{N_{zc}\})\approx \rm{3sDir}$, where $N_{zc}$  are the number counts of galaxies that have been observed to be at redshift bin $z$ and residing in deep SOM cell $c$, and $\{f_{zc}\}$ are the underlying fractions of the full galaxy population that reside in redshift bin $z$ and cell $c$. We constrain $\sum_{zc}f_{zc}=1$ and $0\leq f_{zc} \leq 1$. We use the SOM algorithm from \citet{3sdir}, which has been further validated by \citet{Campos2024} for use on DES Y6 weak lensing samples.  The 3sDir samples of $\{f_{zc}\}$ incorporate the expected sample variance and shot noise.  Each such sample can be propagated into $n(z)$ functions for the wide bins by using the Balrog transfer function and the observed color distribution of the bin members.

Next the 3sDir process is modified to account for uncertainties in category (ii), namely resulting from 
biases and errors in the deep fields' photometric redshift catalogs: COSMOS2020 \citep{cosmos2020}, PAUS+COSMOS \citep{Alarcon2020} and PAUSW1 \citep{PAUSW1}. As described in \citet{y6-sompz-metadetect} and \citet{y6-sompz-maglim}, a parameterized many-band PZ bias function is constrain separately for distinct magnitude/redshift bin combinations.
This expands on the DES Y3 assumption of biases scaling as $(1+z),$
and also takes a more conservative approach for the uncertainty of these systematic redshift biases.

The 3sDir process is also modified for uncertainties in category (iii), namely the uncertainties from the relative zero-point calibration across different deep fields \citep{hartley20}.  Essentially we need to model the relative zero-point error between the COSMOS field (which contains most of the redshifts) and the other three deep fields.

In total, there are 24 relative zero-point systematic shifts (8 bands and 3 out of 4 fields) from category (iii), plus several parameters of the redshift systematics in category (ii) for COSMOS2020, PAUS+COSMOS, and PAUSW1, each with a Gaussian prior. We draw 100 samples of possible values of this nuisance-parameter vector using a Latin hypercube method. For each of these 100 samples,  we shift the deep fluxes of galaxies with the selected zero-point errors, recalculate their redshift assignments with the selected PZ biases applied, and reassign them to SOM cells to collect new $\{N_{zc}\}$ values. Then, for each of these 100 samples, we generate 1 million $n(z)$ samples using 3sDir and SOMPZ. In total, we generate 100 million $n(z)$ samples for each of the \textsc{Metadetect} and \textsc{Maglim}++ tomographic bins, which have effectively sampled over all three types of PZ uncertainty.

\section{Data and their simulated counterpart}\label{sec:data}
\begin{figure*}
    \centering
    \includegraphics[width=1\linewidth]{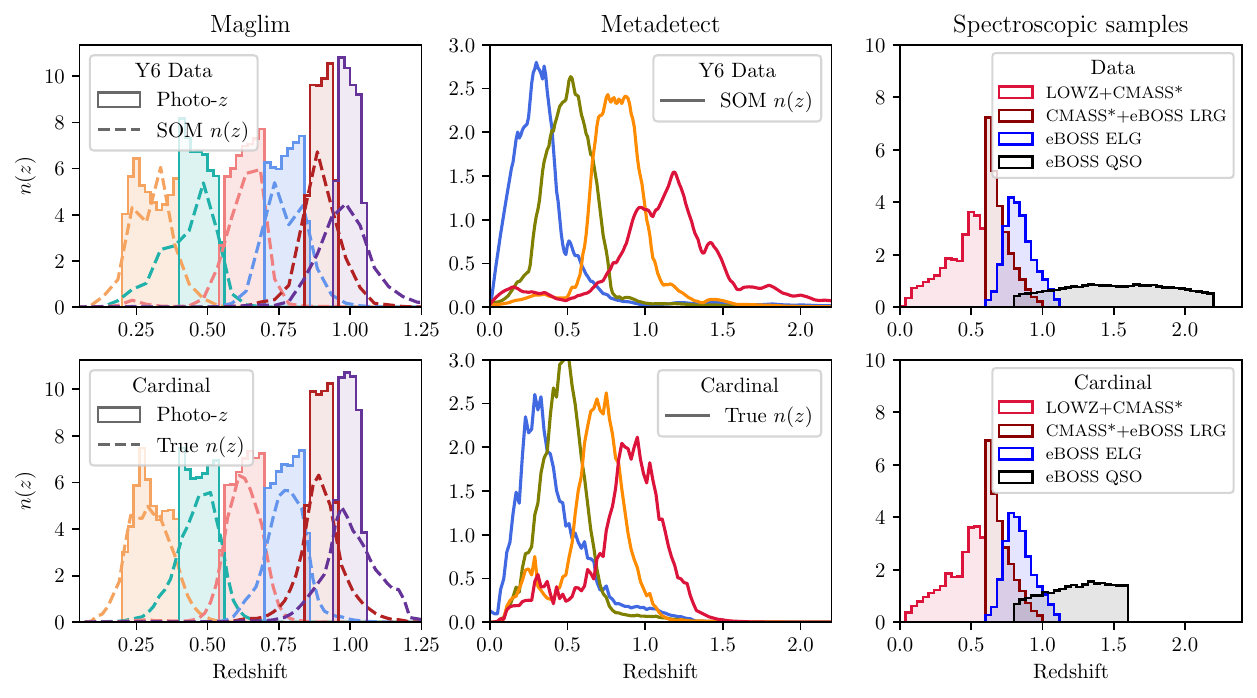}
    \caption{Redshift distributions of the data (\emph{top panel}) and cardinal mocks (\emph{bottom panel}) are shown for \textsc{Maglim}++ (lenses, \emph{left panel}), \textsc{Metadetect} (sources, \emph{middle panel}), and BOSS-eBOSS (reference, \emph{right panel}). Spectroscopic data are presented as a function of spectroscopic redshift. \textsc{Maglim}++ bins are generated with photo-$z$ cuts, and we plot the photo-$z$ histograms of every bin, along with the corresponding SOM estimate of the $n(z)$ for the data. For \textsc{Metadetect} data, only the SOM estimate of the $n(z)$ is shown. The true redshift distributions are provided for all simulated DES samples. Simulated data do not reproduce properly the $z>1.5$ distribution of the data. 
    }
    \label{fig:data_and_mocks_nz}
\end{figure*}

This section describes the various photometric and spectroscopic catalogs used for DES Y6 WZ. The calibration of the nuisance parameters with simulated data is critical for our method, so we need mocks with properties as similar as possible to the real data sets. We use the Cardinal simulation \citep{cardinal} for this purpose. The redshift distributions of the data and mocks are reported in Fig. \ref{fig:data_and_mocks_nz}. 

\subsection{Unknown data}
\subsubsection{\textsc{Metadetect}} 
The source galaxies we calibrate in this work are from the Y6 \textsc{Metadetect} shape catalog presented in \citet*{desy6_metadetect}. The catalog consists of 139.6 million galaxies with the number density of 8.87 galaxies per arcmin$^{2}$ and shape noise of $\sigma_e=0.29$. 
Objects are removed based on star and foreground flags from \citet{y6-gold}, as well as being outliers in properties such as flux and color (see Sect.4.4 of \citet*{desy6_metadetect} and Sect.2 of \citet{y6-sompz-metadetect} for more detail). We also remove objects outside of the joint LSS-shear footprint as described in \citet*{y6-lss_mask}.
The selected \textsc{Metadetect} galaxies are divided into 4 tomographic bins of increasing mean redshift on the basis of which SOM cell they belong to, and 
redshift distributions for each bin are generated under the SOM framework \citep{y6-sompz-metadetect}.

Thanks to the improvements in PSF modeling since DES Y3 described in \citet{y6-piff}, we are now able to include $g$-band flux information for robust redshift calibrations. A statistical weight for each \textsc{Metadetect} object is generated the inverse variance of shear inference from this source
(see Sect.4.5 of \citet*{desy6_metadetect} for more detail). 
All DES Y6 shear measurements incorporate this weighting, hence our $n(z)$ estimations and $\wur$ measurements apply these weights to counts of source galaxies as well. 
Magnification coefficients for the \textsc{Metadetect} bins are evaluated in \citet{y6-magnification}. The properties of the source galaxies in each tomographic bin generated in \citet{y6-sompz-metadetect} are summarized in Table \ref{tab:Metadetect}.

\begin{table}[h!]
    \centering
    \caption{Summary of the \textsc{Metadetect} sample. For each tomographic bin, we give the (SOM-estimated) mean redshift and redshift width,  the number of galaxies,  the number density in arcmin$^{-2}$, and the magnification coefficient. }
    \begin{tabular}{ccccccc}
        \toprule
        \textsc{Metadetect} & $\langle z \rangle_{\rm SOM}$ & $\sigma_z^{\rm SOM}$ & \textbf{$N$ galaxies} & \textbf{$n$ density}& $\alpha$  \\
        \midrule
        Bin 1 &0.415 &0.42 & 33\,707\,071 & 2.555& $-0.186 \pm 0.022$ \\
        Bin 2 & 0.545& 0.33& 34\,580\,888 & 2.252& $-0.248\pm0.026$ \\
        Bin 3 & 0.841& 0.26& 34\,646\,436 & 2.368& $-0.122\pm0.024$ \\
        Bin 4 & 1.175& 0.47& 36\,727\,798  & 1.694&  $\;0.366\pm 0.028$\\
        \bottomrule
    \end{tabular}
    \label{tab:Metadetect}
\end{table}

\subsubsection{\textsc{MagLim}++} 

The lens galaxy catalog we calibrate in this work is the Y6 \textsc{Maglim}++ catalog presented in \citet*{Weaverdyck2025_maglim}. The \textsc{Maglim}++ selection in DES Y6 refines the Y3 \textsc{Maglim} sample to improve purity and mitigate systematics. It consists of a magnitude-redshift cut, near-infrared (NIR) star-galaxy separation optimized for each redshift bin, and a Self-Organizing Map (SOM)-based quality filter.

Galaxies are selected from the Gold catalog using $i$-band limiting magnitude thresholds:
\begin{align}
    &i < 4 \times z_{\text\dnf } + 18, \;\;\;\;\;\;i > 17.5
\end{align}
with $z_{\text\dnf }$ being the mean Directional Neighbourhood Fitting \citep[\dnf,][]{DNF} \ redshift estimates.  This selection has been optimized to balance photometric redshift accuracy and number density, with the goal of to enhance the cosmological constraints on $\Omega_{\rm m}$, $\sigma_8$, and $w$ \citep{Porredon_Maglimy3}. 
Following \citep{Weaverdyck2025stargal}, redshift-bin-optimized cuts in the $(r - z, z - W1)$ color space are used to isolate and remove stars from each bin, where $W1$ is a NIR magnitude from the \textit{unWISE} catalog \citep{unwise2019}.

A SOM trained on \textit{griz} photometry is used to identifies and remove galaxies in regions of color space with highly unreliable photometric redshifts, preferentially excluding quasars and other interlopers.
The final sample is divided into six tomographic bins based on the \dnf\ mean photometric redshift, with redshift bounds that yield similar sky density in each bin. The selection yields 9 186 205 galaxies. These refinements reduce stellar contamination, improve clustering signal fidelity, and enhance suitability for cosmological analyses.

The lens galaxies in each bin are summarized in Table~\ref{tab:maglim}. We apply the LSS weights described in \citet*{Weaverdyck2025_maglim} to each galaxy during all WZ calculations, and use magnification coefficient estimates from \citet{y6-magnification}. The characterization of the $n(z)$ using SOMPZ is described in detail in \citet{y6-sompz-maglim}.

\begin{table}[h!]
    \centering
    \caption{Summary of the \textsc{Maglim}++ sample. For each bin we give the \dnf\ photo-$z$ cuts,  the number of galaxies,  the number density in arcmin$^{-2}$, and the magnification coefficient. }
    \begin{tabular}{cccccc}
        \toprule
        \textsc{Maglim}++ & \textbf{$z$ range} & \textbf{$N$ galaxies} & \textbf{$n$ density}& $ \alpha$  \\
        \midrule
        1 & $0.20 \leq z_{\text{\dnf}} < 0.40$ & 1\,852\,538 & 0.128 & $0.582\pm 0.042$ \\
        2 & $0.40 \leq z_{\text{\dnf}} < 0.55$ & 1\,335\,294 & 0.092 & $0.380\pm 0.105$ \\
        3 & $0.55 \leq z_{\text{\dnf}} < 0.70$ & 1\,413\,738 & 0.097 & $1.043\pm0.078$ \\
        4 & $0.70 \leq z_{\text{\dnf}} < 0.85$ & 1\,783\,834  & 0.123& $1.209\pm 0.081$ \\
        5 & $0.85 \leq z_{\text{\dnf}} < 0.95$ & 1\,391\,521 & 0.096 & $1.451\pm0.144$ \\
        6 & $0.95 \leq z_{\text{\dnf}} < 1.05$ & 1\,409\,280 & 0.097 & $1.416\pm 0.125$ \\
        \bottomrule
    \end{tabular}
    \label{tab:maglim}
\end{table}

\subsection{Reference data}
 

For all unknown samples (sources and lenses), we use as WZ references samples the spectroscopic galaxies from the BOSS and eBOSS surveys \citep{BOSS_color,eboss_dawson}. We also use the final eBOSS QSO sample covering $0.8<z<2.2$ \citep{eboss_qso} to extend WZ coverage to higher redshift.  The DES Y3 analyses did not use the QSO samples.


 The redshift range, number of galaxies, and DES overlap for each spectroscopic reference set are reported in Table~\ref{tab:boss-eboss}, and their redshift distributions are plotted in Fig.~\ref{fig:data_and_mocks_nz}. We make use of the official random (unclustered) catalogs for all the different samples, paying attention that the ratio of random per tracer-sample is constant, equal to 20.  The weight for a BOSS-eBOSS galaxy $i$ is calculated from the catalog weights as in \citet{weight_eboss}:
$ 
w_{\rm tot}^{i}=w^{i}_{\rm noz}\,w^{i}_{\rm sys }\,w^{i}_{\rm CP }.
$
There are no values available for BOSS-eBOSS magnification parameters $\alpha_{\rm r}$, so we assume Gaussian priors centered at $\alpha=0$ with $\sigma=2$.\\

\paragraph{BOSS-eBOSS galaxies} 
As in \citet*{Gatti_Giulia_DESY3}, \citet{DESY3_MAGLIM_z, Cawton2022}, we make use of the LOWZ, CMASS, eBOSS LRG, and ELG samples, covering a common sky area with DES. To avoid replication, we use a $z<0.6$ truncated version of CMASS (that we refer to as CMASS$^\star$), and a $z>0.6$ combination of CMASS (CMASS**) with eBOSS LRG. LOWZ, CMASS, and eBOSS LRG are composed of red galaxies located at $z<1$. In addition, we use the eBOSS ELG sample covering $0.5<z<1.1$.\\

\paragraph{eBOSS QSO}
In this work, in addition to these $z<1.1$ samples, we use all the eBOSS QSOs, covering $0.8<z<2.2$. 
As \citet{clust_z_mcQuinn_white} demonstrate, the reconstructed $n(z)$ uncertainties scale at leading order with the number of reference objects, which is relatively low for the eBOSS QSOs. We thus use broader redshift bins for QSOs than low-$z$ samples to raise signal-to-noise levels. Peculiar velocities of quasars are much larger than standard galaxies', but as we are dealing with angular correlations, the redshift-space-distortion effects are very limited \citep{Euclid_dassignies}. We exclude scales $\rp<1.5$ Mpc. 

\begin{table}[h!]
    \centering
    \caption{ List of the spectroscopic samples from BOSS/eBOSS overlapping with the DES Y6 footprint and covering redshift up to 2.2, used as reference galaxies in this work.}
    \begin{tabular}{ccccc}
        \toprule
        \textbf{ Tracers} & \textbf{$z$ range} & \textbf{$N$ galaxies} & \textbf{Area}  \\
        \midrule
        LOWZ+CMASS*  & $0. \leq z_{\text{spec}} < 0.6$ & 88\,340 & 860 deg$^2$  \\
        CMASS**+LRG & $0.6 \leq z_{\text{spec}} < 1$ & 39\,049 & 860; 700 deg$^2$  \\
        ELG (eBOSS) & $0.5 \leq z_{\text{spec}} < 1.1$ & 86\,055 & 620 deg$^2$  \\
        QSO (eBOSS) & $0.8 \leq z_{\text{spec}} < 2.2$ & 20\,837 & 700 deg$^2$ \\
        \bottomrule
    \end{tabular}
    \label{tab:boss-eboss}
\end{table}

\subsection{Cardinal simulated data}\label{sec:card_mock}
Construction of the Cardinal simulation and the associated data are described in \citet{cardinal,Y6clustermethod}. Here, we briefly summarize important features. Cardinal is developed to produce realistic galaxy populations and lensing signals mimicking the Dark Energy Survey Year 6 dataset (DES Y6). As a successor to the Buzzard simulation \citep{JoeBuzzard}, Cardinal builds on a 10,000~deg$^2$ light-cone from $z=0$ to $z=2.35$ generated by combining three N-body simulations of varying resolutions. Galaxy properties, including positions, velocities, and magnitudes, are assigned to each dark matter particle based on their Lagrangian overdensities \red{calculated in the initial condition of the simulation (inverse distance to the k nearest particles)}. The relation of galaxy properties to the Lagrangian overdensities is obtained from a sub-halo abundance matching (SHAM) model constrained by the galaxy--galaxy correlations and group--galaxy correlations measured in SDSS \citep{SDSS}. An important feature of Cardinal's SHAM model is the inclusion of orphan sub-halos, enhancing the fidelity of galaxy clustering down to small scales. Lensing quantities such as shear, magnification, and deflections are computed using multi-plane ray-tracing \citep{calclens}, with an innovative hybrid correction scheme mitigating resolution effects \citep{cardinal}. The simulation also includes realistic photometric noise, modeled using DES Y6 survey property maps, ensuring that galaxy selection functions accurately mimic survey data. 

The source galaxies in Cardinal are selected based on $g$, $r$, $i$, $z$ photometry, 
and PSF-convolved sizes, ensuring a number density consistent with the DES Y6 \textsc{Metadetect} sample. A key innovation in Cardinal is the incorporation of spatial fluctuations driven by survey depth variations, implemented via a novel abundance-matching algorithm \citep{Y6clustermethod}. The tomographic binning and redshift distribution of source galaxies are derived using the end-to-end DES SOMPZ framework.

To construct the \textsc{Maglim}++ lens galaxy sample in Cardinal, we run a customized \dnf\ pipeline on the entire catalog. As in DES Y6, \textsc{Maglim}++ samples are selected primarily based on 
$i$-band photometry. However, due to differences in photometry between the datasets, we adjust the magnitude cut to ensure that the \textsc{Maglim}++ sample in Cardinal maintains a number density comparable to the DES Y6 \textsc{Maglim}++ sample within each tomographic bin.

We constructed simulated BOSS-eBOSS sample using the simulated \textsc{Maglim}++ sample, selecting subsamples matching each data reference bin's redshift distribution, and using true redshifts as the mock spectroscopic observations, since the impact of peculiar velocities on the spectroscopic samples is negligible for WZ \citep{Euclid_dassignies}.  The BOSS-eBOSS randoms are used to create a BOSS-eBOSS mask in Cardinal, so that the sky distribution is also similar. 

Initially, we built  different  BOSS-eBOSS mocks. The mock ELG catalogs consisted of blue galaxies, while the mock LOWZ-CMASS-LRG catalogs was composed of red galaxies, both selected with a similar selection in Cardinal as with the data (\textit{cf.} App. \ref{app:spec_mocks}). With these color-realistic mocks we measured that  the galaxy biases from cross-correlations of sub-bins did not match the product of the galaxy biases inferred from auto-correlations:
\begin{equation}
w_{{\rm u}_i{\rm r}_i}\neq \sqrt{w_{\rmu_i\rmu_i}\,w_{\rmr_i\rmr_i}}\,.
\end{equation}
A similar behavior was observed in \citet{Euclid_dassignies} using the Flagship 2 simulation, on scales smaller than 1.5 Mpc,  and  interpreted as evidence of 1-halo effects impacting galaxy bias measurements .
In Cardinal, we find the amplitude of this effect to be very large, with biases differing by up to a factor of 3 to 5, and effect persisted up to linear scales of 10 Mpc, whereas a 1-halo origin would suggest it should be confined to scales comparable to the typical 1-halo size ($\sim1$ Mpc). 
Consistency between scales evaluated on data suggests that the behavior observed in App. \ref{app:spec_mocks} was due to the mis-modeling of galaxy-halo properties in Cardinal, rather than a physical process. Furthermore in App. \ref{app:smallscale_blue_red}, we measure the consistency between biases from auto and cross-correlations for some red and blue samples constructed with \textsc{Redmagic} and \textsc{Maglim}++ data samples, and obtain the same results as in \citet{Euclid_dassignies}, namely this effect is limited to $\rp<1.5 $ Mpc.
Constructing the spectroscopic mocks using a subsample of the simulated \textsc{Maglim}++ catalog, we found this discrepancy between biases to be limited and in better agreement with previous works \citep[e.g. ][]{Euclid_dassignies}, and our data test (\textit{cf.} Apps. \ref{app:spec_mocks} and \ref{app:smallscale_blue_red}). Thus, we decided to use the \textsc{Maglim}++-selected BOSS-eBOSS mocks.\\

\section{Results}\label{sec:results}
\subsection{Systematics with Cardinal}
\label{sec:cardinalsys}
\paragraph{Systematics identified }The first step in our analysis is to use simulated samples to define the priors of the systematic parameters $s_{\rmu k}$. This procedure must be carried out for both source and lens samples.
We identify several clustering-redshifts systematics that could impact our measurements:
\begin{itemize} \item Any redshift evolution of the bias for the unknown sample, $b_{\rm u}(z)$, is not incorporated into the model.
\item At our scale range, $\rp \in [1.5, 5]$ Mpc, modeling the galaxies as having linear bias with respect to the non-linear matter power spectrum is not fully accurate \citep{bias_pert_lsst, Martinelli21a}. Additional effects from 1-halo physics may be present, but they are expected to be minor at scales larger than 1.5 Mpc \citep{Euclid_dassignies}.
\item  Simulations show that the correlation of spectroscopic galaxies from narrow redshift bins with photometric galaxies from broader bins can deviate from model prediction, because of the correlation at the border of the spectroscopic bin \citep[\textit{cf.}  ``Limber-1-bin approximation'' in ][]{Euclid_dassignies}. Our new clustering modeling (\textit{cf. } Eq. \ref{eq:wur_sumfuA}) is expected to correct for this effect; the clustering matrix $A$ is not diagonal, and takes into account correlation at the edges of the bins.  
\item The use of the DP estimator instead of LS for source galaxies, in the cases where we lack the associated random catalogs, can cause larger biases in $w_\rmur$ associated with mask shapes and edge effects.
LS is known to better estimator than DP, with  smaller bias and variance \citep{kerscher_comparison_estimator,LSvsDP}.
\end{itemize}

We note that a potential systematic we do not include in Cardinal is the fiber collisions' effect on the spectroscopic samples.\footnote{See \url{https://www.sdss3.org/dr9/algorithms/boss_tiling.php}.} Because of the fibers' diameters, two spectroscopic galaxies cannot be arbitrarily close, and the biases inferred from the auto-correlation might be underestimated with respect to the one from cross-correlation (which is not affected by this problem once weights are applied). We check this in Sect. \ref{sec:test_scale} by comparing the $b_{\rm r}$ obtained from  different ranges of angular separation. \red{Another potential systematic arises from assuming an incorrect cosmology when evaluating $w_{\rm m}$ (see Eq.~\eqref{eq:wdm}), through the Alcock--Paczynski effect. The quantity entering our modeling is $\sqrt{w_{\rm m}}$. Any constant  offset with redshift is absorbed by the free amplitude $b_\rmu$ (see Eq.~\eqref{eq:wzpost}), so only the redshift evolution of this ratio is relevant. We find that the ratio of this quantity computed assuming different cosmologies (DES Y3, KiDS Legacy, and Planck 2018) varies at the $\sim1\%$ level, which is negligible compared to the other systematics considered here.}\\

\paragraph{Evaluating the effect of the systematics on WZ} We evaluate the systematic effects in Cardinal by calculating the ratio of the measured $\wur$ to a model $\hat{w}_{\rmur }$. Systematic effects can enter either the measurement $w$ or the model $\hat{w}$. 
We execute the WZ analysis on Cardinal data under two different cases. 
In the first, ``no systematics'', case we will ``cheat'' by using the truth values for certain quantities that we do not know for the real data, to see if this eliminates the expected systematic errors. 
In the second ``all systematics'' case, we will impose the same level of ignorance on the Cardinal catalog and parameters as we have for the real observations. Please note that we may underestimate cosmic variance in the no-systematics case since the  true redshift distribution of the bins is taken directly from the
full Cardinal simulation, whereas real data are restricted to the 700 deg$^2$ of sky overlapping with BOSS-eBOSS mocks. \red{The $n(z)$ measured in the presence of systematics are no longer expected to be normalised to unity. }

The no-systematics test makes the following alterations to the mock catalogs:
\begin{itemize} 
\item The model $\hat{w}_{\rmu\rmr}$ departs from Equation~\eqref{eq:model:wur_sys} by dropping the $(1+S)$
 systematic correction term and the magnification terms.  We replace $b_\rmu b_{\rmr_i}$ with $b_{\rmu_i} b_{\rmr_i}$,
 \red{where} $b_{\rmu_i}$ estimated directly from a cross-correlation between $\rmr_i$ and the members of $\rmu$ that match the redshift range of $\rmr_i$. Doing so should eliminate systematic errors associated with redshift evolution of bias, and also limit the impact of scale-dependent bias from small scales physics.
\item For each cross-correlation between the $\rmu$ bin and a reference bin $\rmr_i$, we replace the $\rmu$ bin with an idealized bin. This is done by replacing each galaxy whose true redshift falls outside the redshift range of the reference bin with a randomly placed galaxies. This should correct for the ``full-u-bin'' systematic \citep{Euclid_dassignies}. 
\item We use the LS estimator for $\wur$, to have minimal variance and bias from the estimator.
\end{itemize}
Together, these alterations should guarantee that the measured $w_\rmur$ are proportional to $n(z),$ within statistical uncertainties, if our code and methodology are correct.

For all-systematics case, the analysis follows that of the real data:
\begin{itemize}
    \item We use the DP estimator for sources; but we keep  the LS estimator for lenses as we do have lens randoms.
    \item We evaluate the cross-correlation $w_{\rmur_i}$ using the full $\rmu$ bin's catalog instead of one with positions randomized outside the $\rmr_i$ bin redshift range.
    \item We maintain the single unknown $b_\rmu$ value instead of using truth values for $b_{\rmu_i}$ in the model. 
\end{itemize} 

\paragraph{Dominant systematic} We find the redshift evolution of galaxy bias to be the dominant systematic error causing a significant deviation from the model.  
As the other systematics  have limited (if not null) impact, we do not separately analyze the different systematics as mentioned in Sect. \ref{sec:sys}. In particular, we found the full-u-bin systematics to have little impact for our analysis.  Either our modeling of cross-correlations between bins --\textit{cf.} Eq. \eqref{eq:wur_sumfuA}-- has captured the effect, or our precision, using 700 deg$^2$ of BOSS-eBOSS data,  is not good enough to detect these small effects, contrary to \citet{Euclid_dassignies}, for which they were considering cross-correlations between 1,000 deg$^2$ of Euclid and DESI data.\\

\subsubsection{Sys of \textsc{Maglim}++}

\begin{figure*}
    \centering
    \includegraphics[width=0.9\linewidth]{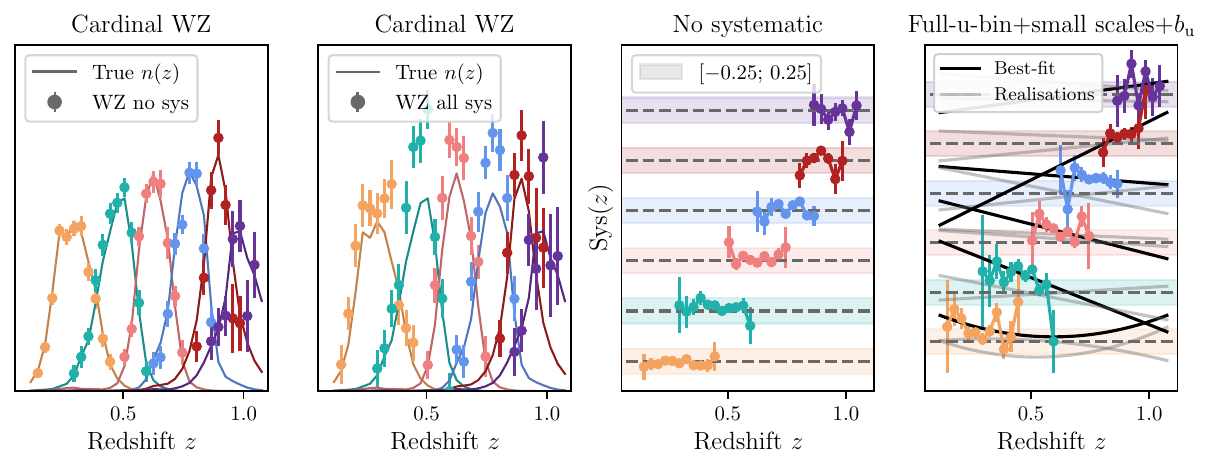}
    \caption{Impact of systematics on WZ for the Cardinal-simulated \textsc{Maglim}++ samples. \emph{Left:} redshift distributions inferred from the no-systematics mock catalog, WZ estimate (points) compared to the true $n(z)$ (solid lines).
\emph{Middle-left:} Same as left but with the all-systematics mock catalog that uses only information available in the real data.
\emph{Middle-right:} Corresponding systematic function $w/\hat w$, for the no-systematic analysis, consistent with no deviation from a normalization constant.
\emph{Right:} Systematic function $w(z)/\hat w(z)$ in the presence of systematics, where deviations from unity $S_{\rmur_i}$ are fitted using a Legendre polynomial basis (black lines).  The gray lines plot some realizations of $S_{\rmur_i}$ from the derived prior on its $s_{\rm uk}$ coefficients.}
    \label{fig:sys_maglim}
\end{figure*}

In Fig. \ref{fig:sys_maglim}, we compare the true redshift distributions, $n(z)$, of the six \textsc{Maglim}++ bins in the Cardinal simulation with the $n(z)$ inferred from WZ, in the no- and all-systematics cases, and the corresponding derived systematic correction function $S_{\rmur_i}$ needed to force agreement between the model and mock data.

In the no-systematics case, the inferred distributions accurately recover the true distributions for all six bins. 
Small deviations at the peak of the Gaussian can be explained by binning effects, as illustrated in Fig. 3 of \citet{Euclid_dassignies}, and these are accounted for in the modeling of $A_{{\rm r}n}$. 

 For the more realistic, all-systematics mock catalog, we observe significant deviations from the truth in the inferred $n(z)$. The corresponding $S_{\rmur_i}$ corrections are shown in the right panel of Fig. \ref{fig:sys_maglim}. We model these systematic functions and find that an order-2 polynomial provides a good fit for bin 1, while an order-1 polynomial is sufficient for bins 2–6. \red{We note that the largest impact of systematics in the first bin could be linked to a broader range of galaxy luminosities selected, leading to more complex redshift-dependent bias evolution.} The priors on the nuisance parameters $s_{\rmu k}$ for the real data are chosen to follow a Gaussian distribution, centered at zero, with the standard deviation set to the best-fit parameter amplitude (i.e., the $1\sigma$ value). An alternative choice could have been to choose a Gaussian distribution centered at the best fit from Cardinal, with same width, so that the no systematic case is at $1 \sigma$. The implicit assumption for such case would be that we trust Cardinal to reproduce the WZ systematics. We decided to rather assume Cardinal was giving a good estimate of the amplitude of the systematics, but was not necessary reproducing them with fidelity.    

\red{Assuming consistency in the galaxy biases inferred from fiducial and large scales (as we will evaluate in Sect. \ref{sec:test_scale}), the low order of the systematic functions is physically motivated. This is because the dominant systematic model error—redshift evolution of the galaxy bias—is expected to be small across the narrow redshift extent of the \textsc{Maglim}++ bins \citep[\textit{cf.} Sect. VI.A for Y3 \textsc{Maglim},][]{Porredon_Maglimy3}.}

\subsubsection{Sys of \textsc{Metadetect}}
\begin{figure*}
    \centering
    \includegraphics[width=0.9\linewidth]{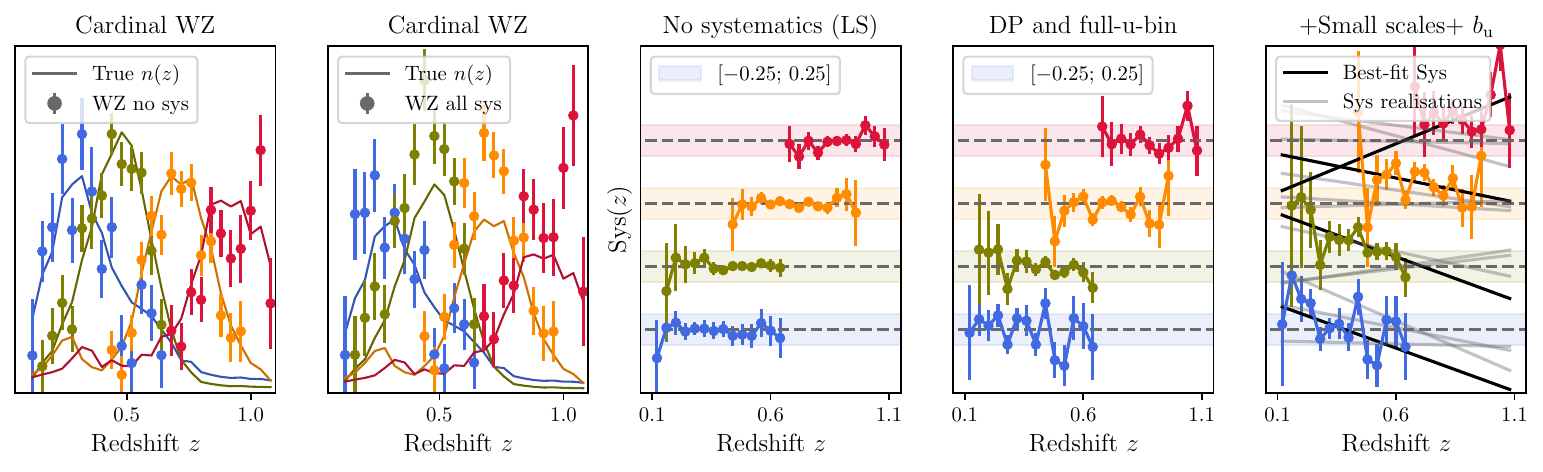}
    \caption{Impact of systematics on WZ for the Cardinal-simulated \textsc{Metadetect} samples. \emph{Left:} redshift distributions inferred from the no-systematics mock catalog, WZ estimate (points) compared to the true $n(z)$ (solid lines). \emph{Middle-left:} Same as left but with the all-systematics mock catalog that uses only information available in the real data. \emph{Middle:} Corresponding systematic function $w/\hat w$, for the no-systematic analysis, consistent with no deviation from a normalization constant. \emph{Middle right and right:} Systematic function $w(z)/\hat w(z)$ in the presence of systematics, where deviations from unity $S_{\rmur_i}$ are fitted using a Legendre polynomial basis (black lines).  The gray lines plot some realizations of $S_{\rmur_i}$ from the derived prior on its $s_{\rm uk}$ coefficients. }
    \label{fig:sys_meta}
\end{figure*}
In Fig. \ref{fig:sys_meta}, we compare the true redshift distributions, $n(z)$, of the four Cardinal-simulated \textsc{Metadetect} bins with the $n(z)$ inferred from WZ,  in the no-systematics and all-systematics scenarios. 

As with the \textsc{Maglim}++ simulations, the no-systematics mock catalog yields $n(z)$ consistent with the truth.
When the realistic all-systematics catalog is used, however, we again observe significant deviations in the inferred $n(z)$. The corresponding $S_{\rmur_i}$ corrections are shown in the right panel of Fig. \ref{fig:sys_meta}. An order-1 polynomial provides an adequate fit for all four bins. As with \textsc{Maglim}++, the priors on the nuisance parameters for the real data are chosen to follow a Gaussian distribution centered at zero, with a standard deviation set to the amplitude of the best-fit Cardinal parameters. A few realizations of the systematic functions drawn from these priors are also shown in the right panel.  Even though the source bins are wider than the lens bins, we still find that linear systematic functions are effective at describing the dominant systematic error---redshift evolution of galaxy bias---across the bins.

\subsection{Data calibration}

\subsubsection{Measurements}\label{sec:measurement}
\begin{figure*}
    \centering
    \includegraphics[width=1.\linewidth]{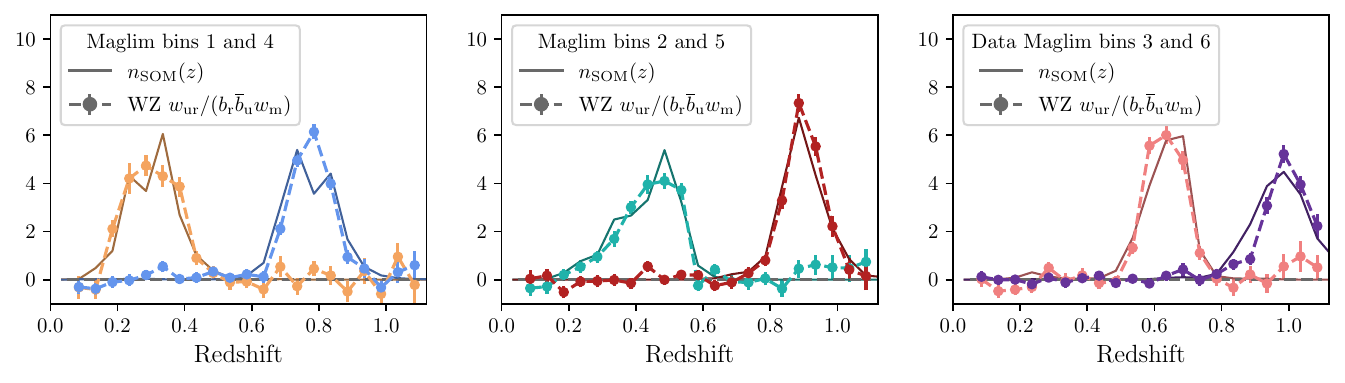}
    \caption{The measured $n(z)$ from WZ ($= w_\rmur/(b_\rmr\,b_\rmu\,w_{\rm m})$ ) for each of the six DES Y6 data \textsc{Maglim}++ bins, without magnification and systematics modeling. BOSS and eBOSS  samples, except QSO, are used as reference samples.  The estimated $n(z)$ from the SOM are reported with solid lines. }
    \label{fig:nz_WZ_maglim}
\end{figure*}

\begin{figure*}
    \centering
    \includegraphics[width=1.\linewidth]{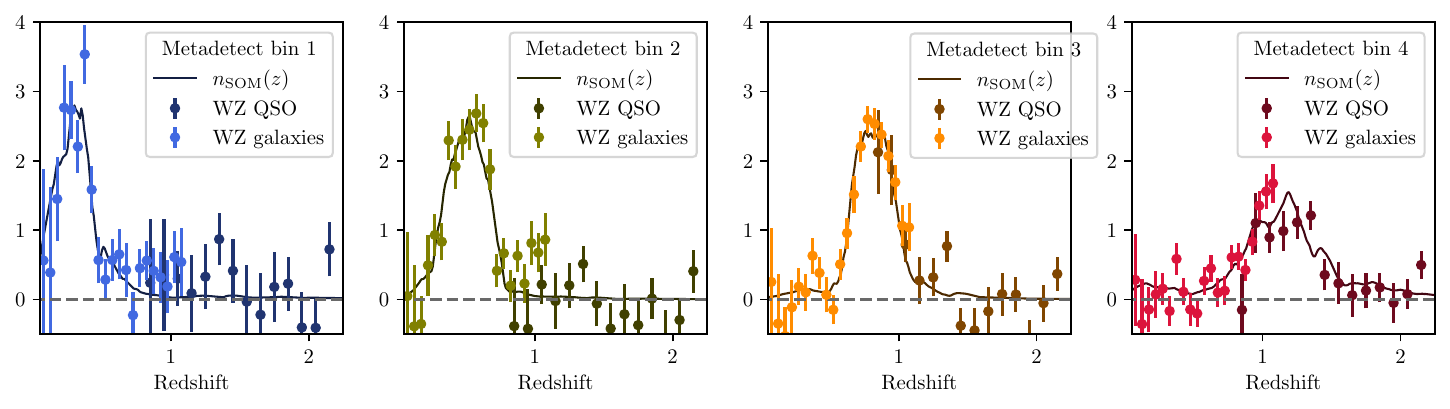}
    \caption{The measured $n(z)$  from WZ ($= w_\rmur(z)/(b_\rmr(z)\,\overline{b}_\rmu\,w_{\rm m}(z))$ ) for each of the four DES Y6 data \textsc{Metadetect} bins, without magnification and systematics modeling. BOSS and eBOSS galaxies and QSO are used as reference samples separately, covering $0<z<2.2$.  The estimated $n(z)$ from the SOM are reported with solid lines. }
    \label{fig:nz_WZ_meta}
\end{figure*}

In Figures \ref{fig:nz_WZ_maglim} and \ref{fig:nz_WZ_meta}, we present the WZ measurements $w_\rmur/(b_\rmr\,b_\rmu\,w_{\rm m})$ for \textsc{Maglim}++ and \textsc{Metadetect} bins, overlaid on the SOM-derived $n(z)$. Here, $b_{\rmr}$ represents the galaxy bias of BOSS-eBOSS, which evolves with redshift, while $b_\rmu$ is the best-fit amplitude of the measurement relative to the SOM $n(z)$. 
We observe a visually reasonable agreement between WZ and the SOM estimates. We report in App. \ref{app:desiWZ} the measurements for \textsc{Metadetect} and \textsc{Maglim}++ using DESI galaxies as reference. We obtain an excellent agreement with BOSS-eBOSS measurements. We retain BOSS-eBOSS as our fiducial WZ measurements rather than slow the Y6KP analysis with a switch to DESI.

Fig. \ref{fig:nz_WZ_meta} includes $z>0.8$ clustering-redshift measurements using eBOSS-QSOs. Due to the sparse density of this tracer, we adopted a binning of $\Delta z=0.1$ to increase the number of spectroscopic redshifts per bin, as the uncertainty on $n(z)$ scales with the number of available spec-$z$ \citep[][]{clust_z_mcQuinn_white}. We detect no significant signal for bins 1 and 2, partially recover the right tail of the distribution for bin 3, and measure most of the distribution for bin 4. Although not highly precise, these measurements serve as a useful validation of the SOM’s performance at high redshift.

As future surveys such as DESI and 4MOST will provide significantly more QSOs up to $z\sim3$, this measurement serves as a promising proof of concept for high-$z$ calibration in upcoming LSST and Euclid \citep{lsst_req,Euclid_overview}. 

\subsubsection{Scales and galaxy biases }\label{sec:test_scale}

\begin{figure*}
    \centering
    \includegraphics[width=1\linewidth]{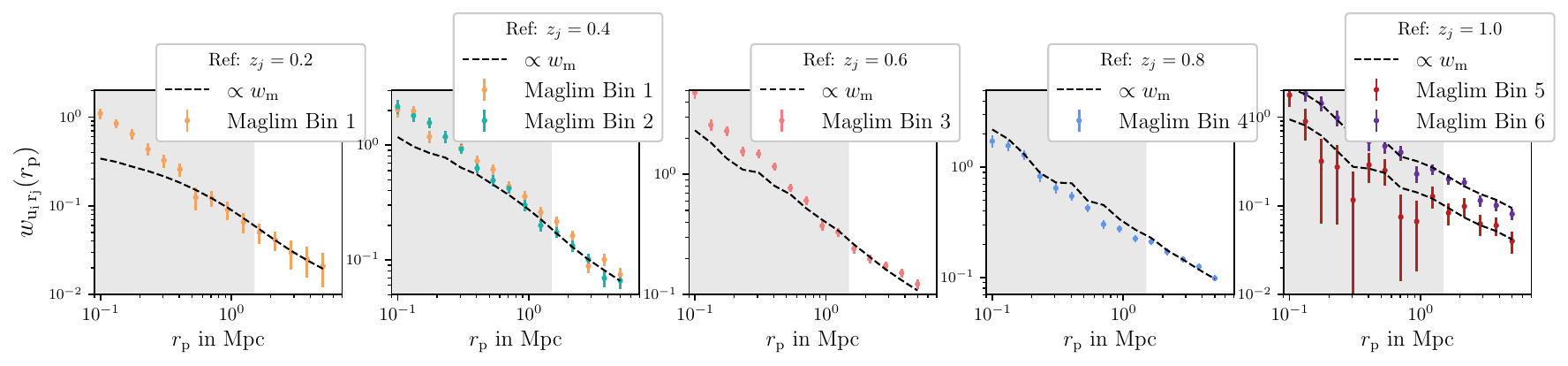}
    \caption{Angular correlation vector for the six \textsc{Maglim}++ bins (colors) and five BOSS-eBOSS reference bins (panels), for angular scales corresponding to $\rp\in[0.1,5]$ Mpc, and reference binning $\Delta z=0.05$. We only plot non-zero correlations (i.e. combinations of u-r bins which overlap in redshift). We also plot a model  $\hat w_{ur}=a\times w_{\rm m}(\rp)$ for the best-fit constant $a.$. The scale discarded for our WZ analysis ($\rp<1.5$ Mpc) are colored in grey.  We observe a good fit for the scale range $[1.5,\,5]$ Mpc, used in this work, despite the simplicity of the modeling.}
    \label{fig:wtheta}
\end{figure*}
\begin{figure*}
    \centering
    \includegraphics[width=0.9\linewidth]{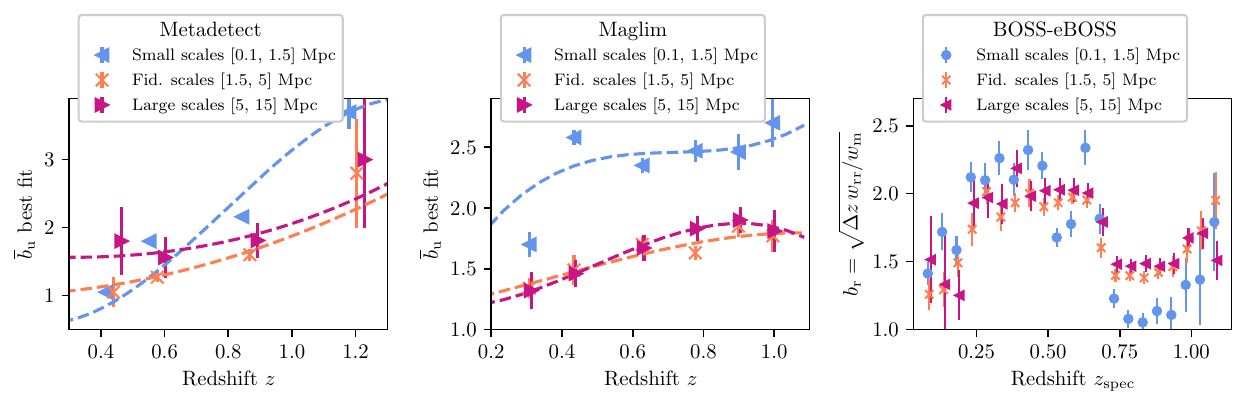}
    \caption{Galaxy biases of the three samples used for DES Y6 WZ are plotted vs redshift.  The values for  \textsc{Metadetect} (\textit{left panel}) and \textsc{Maglim}++(\textit{middle panel}) are derived from a fit of cross-correlations of each tomographic bin with BOSS-eBOSS galaxies (fixing $n_{\rm u}(z)$ to $n_{\rm SOM}(z)$, \textit{cf. } Eq. \ref{eq:fit_bu}), and values for BOSS-eBOSS(\textit{right panel}) from their autocorrelation.  Each panel shows the average bias across three scales range: $[0.1,\, 1.5]$ Mpc (SS, \textit{orange}), $[1.5,\, 5 ]$ Mpc (Fid, \textit{red}), and $[5,\,15]$  Mpc (LS, \textit{purple}).
    In each case, the Fid and LS biases are consistent and well fit by a low order polynomial. The SS cases differ, due to some combination of nonlinear bias and, for BOSS-eBOSS, fiber collisions. }
    \label{fig:galaxy_bias_redshift}
\end{figure*}
For clustering redshifts, we rely on small scales that are not used for cosmological analyses due to the lack of precise modeling. Therefore, it is crucial to test whether these scales can still be reliably used for redshift calibration. A detailed investigation of the impact of small-scale effects on WZ measurements can be found in Sect. 4.3 of \citet{Euclid_dassignies}. The main conclusion of that work is that deviations of $w(z)$ from a model adopting linear, redshift-independent bias for the unknown populations are small enough to be corrected by our systematic-error functions for our chosen scale range of $1.5 \leq r_{\rm p} \leq 5$ Mpc.  The lower bound is a compromise between measurement precision and modeling accuracy, while the upper bound is fixed by the need to maintain statistical independence from the cosmological $w(\theta)$ measurements. As a check, we investigate here whether varying the scale range significantly affects the inferred $n(z)$.  We examine   three different scale ranges: small scales (SS): $[0.1, 1.5]$ Mpc,  fiducial range (Fid): $[1.5, 5]$ Mpc, and large scales (LS): $[5, 15]$ Mpc.\\

\paragraph{Non-linear modeling} As a first sanity check, we examine $\wur(\theta)$ to assess whether our model assumption of linear galaxy bias (\textit{cf}. Eq. \eqref{eq:linear_bias} ) remains valid across the chosen $w(\theta)$ scales. Fig. \ref{fig:wtheta} shows the correlation measurements for \textsc{Maglim}++ bins cross-correlated with different spectroscopic bins, plotted as a function of scale from 0.1 to 5 Mpc.\footnote{We do not show larger scales, as they are used for cosmology, and since we are not using a blinded catalog, we avoid unintentionally unblinding the Y6 data in this figure.} When the \textsc{Maglim}++ bin and the reference bin overlap in redshift so that the $w_\rmur(\theta)$ is appreciable, we observe that the linear-bias fit remains visually good down to $r_p\approx1$~Mpc, across redshifts. \\

\paragraph{Unknown galaxy bias} Two internal consistency checks  on the validity of the linear bias model (\textit{cf.} Eq. \ref{eq:linear_bias}) and the modeling of the bias redshift evolution (\textit{cf.} Eq. \ref{eq:model:wur_sys}), are possible. We examine the value of $\overline b_\rmu$ for every bin that best fit the data with the simplified model
\begin{align}
    &\hat{w}_{\rmu\rmr}(z_i)=\overline{b}_\rmu\,b_\rmr(z_i)\,n_{\rm u}(z_i)\,w_{\rm m}(z_i)\,,\label{eq:fit_bu}
\end{align}
where we use the $n(z)$ derived from the SOMPZ method\footnote{ For the importance sampling, we will get $b_\rmu$ for every realizations, \textit{cf.} Eq. \eqref{eq:wzpost} from a similar fitting, but considering systematics and magnification. For this test, we are assuming the SOM $n(z)$ if a good model, and focus on the galaxy biases. } and measure $w_\rmur$ on the data for three scale ranges (SS, Fid., and LS). We know that the linear bias model is valid for LS.

The robustness of the linear-bias assumption in the full $\hat w_\rmur$ model of Equation~\eqref{eq:model:wur_sys} is supported by the agreement of the bias values between the Fid and LS scale-range choices, as reported in Fig. \ref{fig:galaxy_bias_redshift}.  For
both \textsc{Maglim}++ and \textsc{Metadetect}, the  differences between Fid and LS bias values are easily modeled by our systematic functions $S_{\rmur_i}$.
The galaxy bias measurements using small scales (SS) show strong deviation from the fiducial case. This is not unexpected, as we observe in Fig. \ref{fig:wtheta} that linear-bias model is insufficient at small scales.
The large-scale galaxy bias values for \textsc{Maglim}++ are consistent with those reported in the Y3 results \citep[\textit{cf.} App. B of ][]{Porredon_Maglimy3}.

We can also use the $\overline b_\rmu$ to crudely confirm that our chosen priors on the systematic-functions coefficients $s_{{\rm u}k}$ are introduce sufficient freedom to track the variation of the unknown samples' bias functions with redshift, if we assume that the variation of $\overline b_\rmu$ between different lens or source tomographic bins is a good proxy for the evolution of $b_\rmu(z)$ within each tomographic bin.  
This assumption is not expected to be very accurate, especially for \textsc{Metadetect}, because different color selection criteria are used for distinct bins, so their constituent galaxy populations differ. For \textsc{Maglim}++, as bins are defined with photo-$z$ cuts, this procedure to estimate bias evolution corresponds to the method M3 in \citet{Euclid_dassignies} , and was found to be accurate (\textit{cf.} their Fig. 15).  Proceeding, we plot each the $\overline b_\rmu$ of each bin at the mean of its SOMPZ redshift distribution in Fig. \ref{fig:galaxy_bias_redshift}.  We observe that the redshift evolution of the galaxy bias for both the Fid and LS scale selections are well-described by low-order polynomials, and can be well fit using the priors and degrees for the systematic $s_{\rmu k}$ functions that we derived from the Cardinal simulations in Section~\ref{sec:cardinalsys}. \\

\paragraph{Reference galaxy bias} Next, we examine $w_{\rm rr}(z_i)$ to assess whether the consistency of the reference galaxy biases extracted from the auto-correlations.  The right-hand panel of Fig. \ref{fig:galaxy_bias_redshift} shows the derived biases for BOSS-eBOSS vs redshift of the bin for the three choices of scale range.  The Fid and LS cases agree very well, suggesting that neither scale-dependent bias nor fiber collisions are having substantial impact at the fiducial range.  The SS biases differ substantially, suggesting the presence of one or both effects in the small-scale clustering data.  The substantial changes in bias near $z=0.2$ and $z=0.7$ are expected from the changes in BOSS-eBOSS selection criteria.
\begin{figure*}
    \centering
    \includegraphics[width=0.9\linewidth]{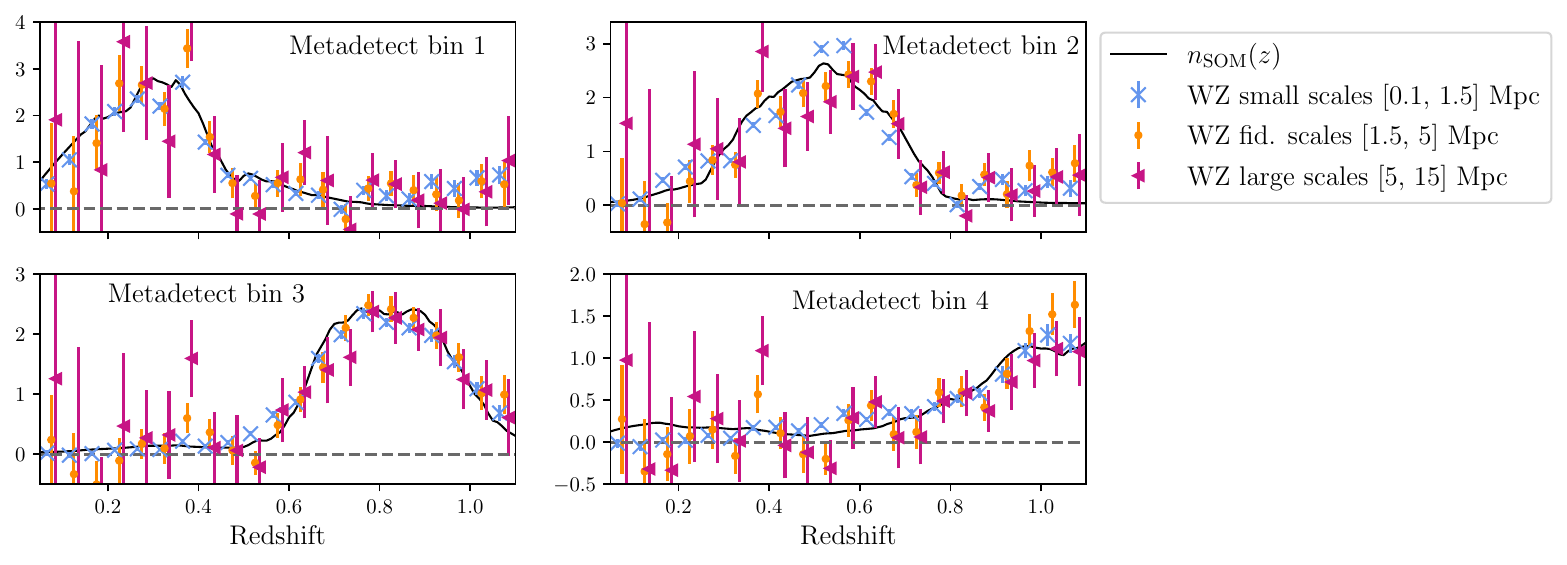}
    \caption{ Redshift distributions of \textsc{Metadetect} four bins,  evaluated as $n(z)=\wur(z)/(\overline{b}_\rmu b_\rmr(z) w_{\rm m}(z))$ (so neglecting magnification and systematics), for three scale ranges. We can observe a very good agreement between the  measurements using fiducial range [1.5, 5] Mpc (\textit{orange}), and the large scale range (\textit{red}) }
    \label{fig:scale_meta}
\end{figure*}
\begin{figure*}
    \centering
    \includegraphics[width=0.9\linewidth]{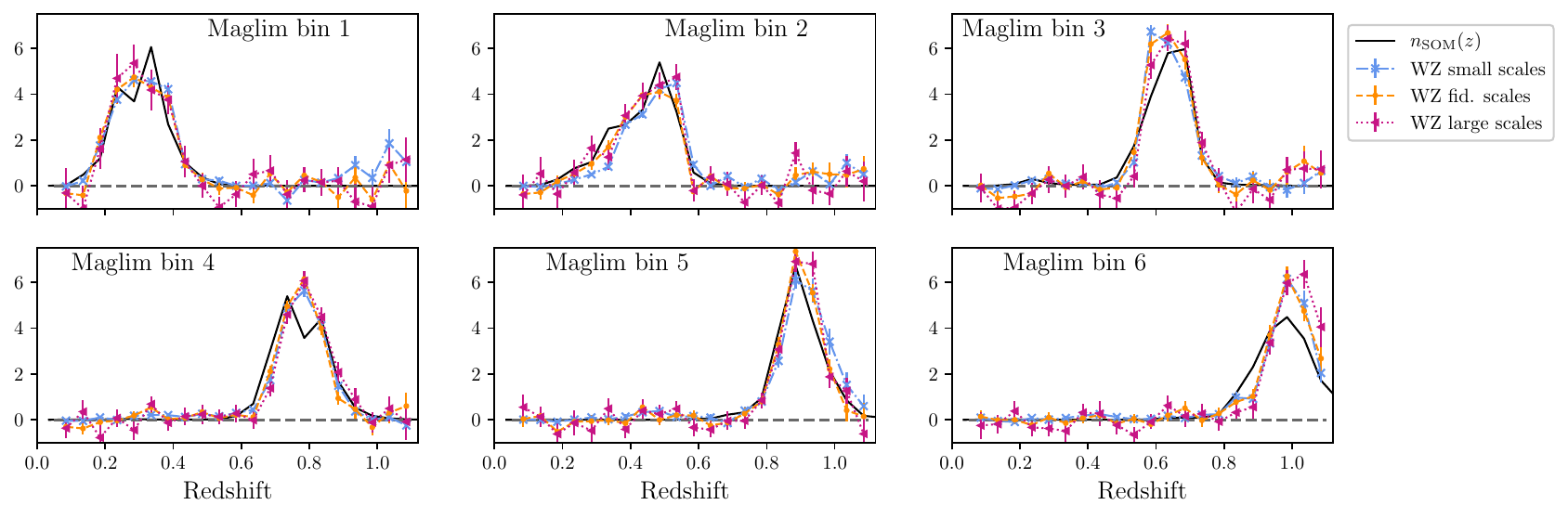}
    \caption{Same as Fig. \ref{fig:scale_meta}, for the six \textsc{Maglim}++ bins. We also observe an excellent agreement between fiducial and large scale ranges. }
    \label{fig:scale_mag}
\end{figure*}

\paragraph{WZ and scales} Figs. \ref{fig:scale_meta} and \ref{fig:scale_mag} compare the $n(z)$ values derived using the SS, Fid, and LS scale ranges for each \textsc{Metadetect} and \textsc{Maglim}++ bin, respectively, under the simplified model $n(z)=\wur(z)/(\overline{b}_\rmu b_\rmr(z) w_{\rm m}(z)).$  The LS and Fid cases yield to two WZ results consistent within their statistical uncertainties.  The SS WZ result often differs to the two others WZ results by substantially more than its (very small) statistical uncertainties.
To quantify any scale-dependent deviations, we compute the ratio of the inferred redshift distributions between different scale ranges: $
 \eta_{\rm scales}(z) = {n(z\vert {\rm scales}_1)}/\;{n(z\vert {\rm scales}_2)}, $
The results are shown are reported in App. \ref{app:scales}.   For both \textsc{Maglim}++ and  \textsc{Metadetect}, and for Fid vs LS, we find that any differences are well within the freedom that we have allotted to the systematic-error function. \red{We find no significant evidence that fiber collisions affect our results. Fiber collisions should impact auto-correlations more strongly than cross-correlations and primarily on small (fiducial) scales, however we obtain consistent galaxy biases from auto-correlations across both fiducial and large-scale ranges (cf. Fig. \ref{fig:galaxy_bias_redshift}), as well as consistent $n(z)$ between these two scale choices (cf. Figs. \ref{fig:scale_meta} and \ref{fig:scale_mag}). In addition, the $n(z)$ inferred from BOSS and eBOSS agrees with that obtained from DESI on the same scales, despite their different fiber-collision properties (cf. App. \ref{app:desiWZ}). This is consistent with \cite{choppin_2025}, where they showed that fiber collisions have limited impact even on 1–5 Mpc scales. Because of this, and taking into account that any residual effect can be absorbed by our systematic function, provided it does not vary too rapidly with redshift, we are not explicitly modeling fiber collision in this paper. .}

\subsection{Importance sampling}\label{sec:fullshape}

We now turn to the results of applying the full WZ importance sampling process (\textit{cf.} Sect. \ref{sec:importance}) to the samples of $n(z)$ produced by the SOMPZ$+$3sDir photometric analyses.
\begin{figure*}
    \centering
    \includegraphics[width=1\linewidth]{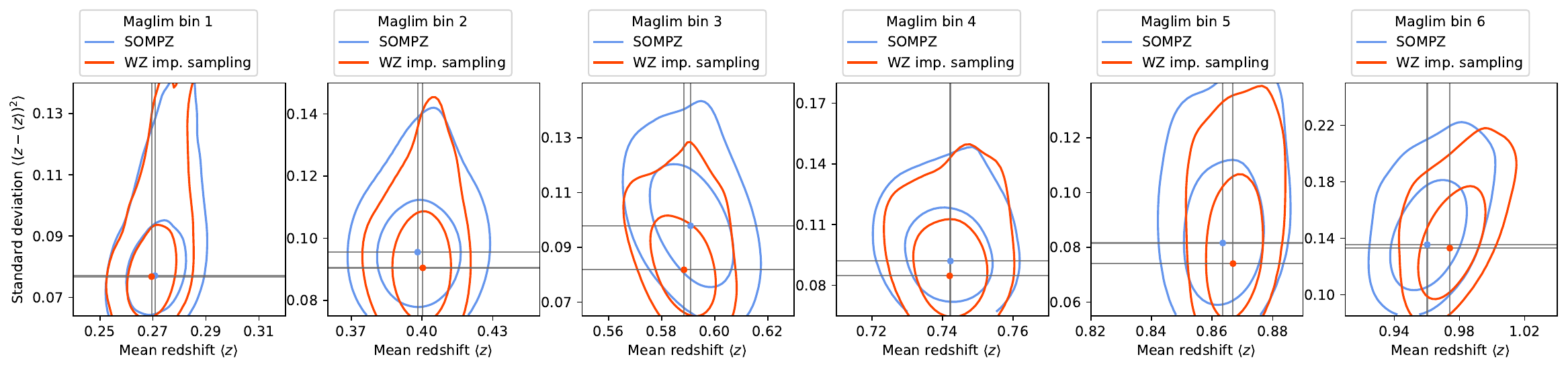}
    \caption{ Distributions of the mean redshift and standard deviation of $n(z)$ samples drawn from $p(n|{\rm PZ})$ using SOMPZ (blue), and samples drawn from $p(n|{\rm WZ,\,PZ})$ using WZ importance sampling (red), for each of the six \textsc{Maglim}++ tomographic bins. The contours represents 68$\%$ and 95$\%$ of the data-points. The lines report the position of the peaks. The  WZ-selected $n(z)$ samples tend to have slightly lower standard deviations, and slightly less variation in the mean. Only Bin 6 shows a substantial shift in the derived mean redshift.} 
    \label{fig:hist_WZ_maglim}
\end{figure*}

\begin{figure*}
    \centering
    \includegraphics[width=1\linewidth]{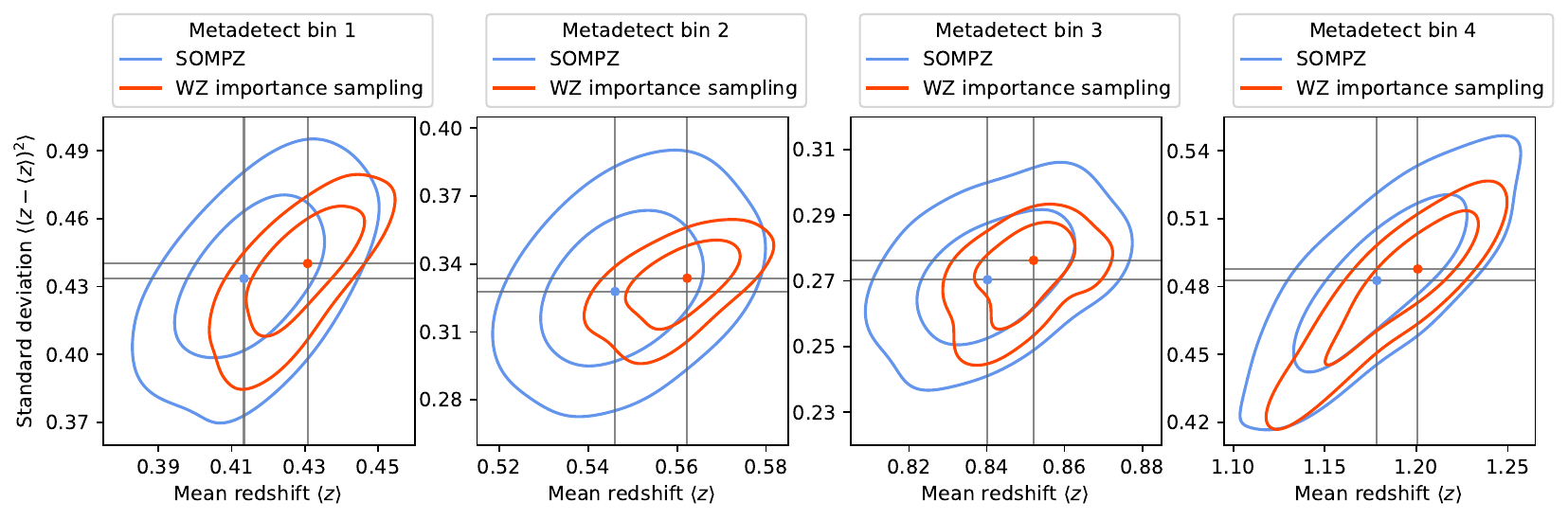}
    \caption{Same as Fig. \ref{fig:hist_WZ_maglim} for \textsc{Metadetect} four bins. Samples drawn from $p(n|PZ)$ are reported in blue, and samples drawn from the importance sampling on the compressed PZ are reported in red. Contrary to \textsc{Maglim}++, here WZ has an impact on the mean redshift of the selected samples, with preference for higher mean redshifts. The variations in standard deviation is also reduced with a preference for larger bins.}\label{fig:hist_WZ_meta}
\end{figure*}

\begin{figure*}
    \centering
    \includegraphics[width=1\linewidth]{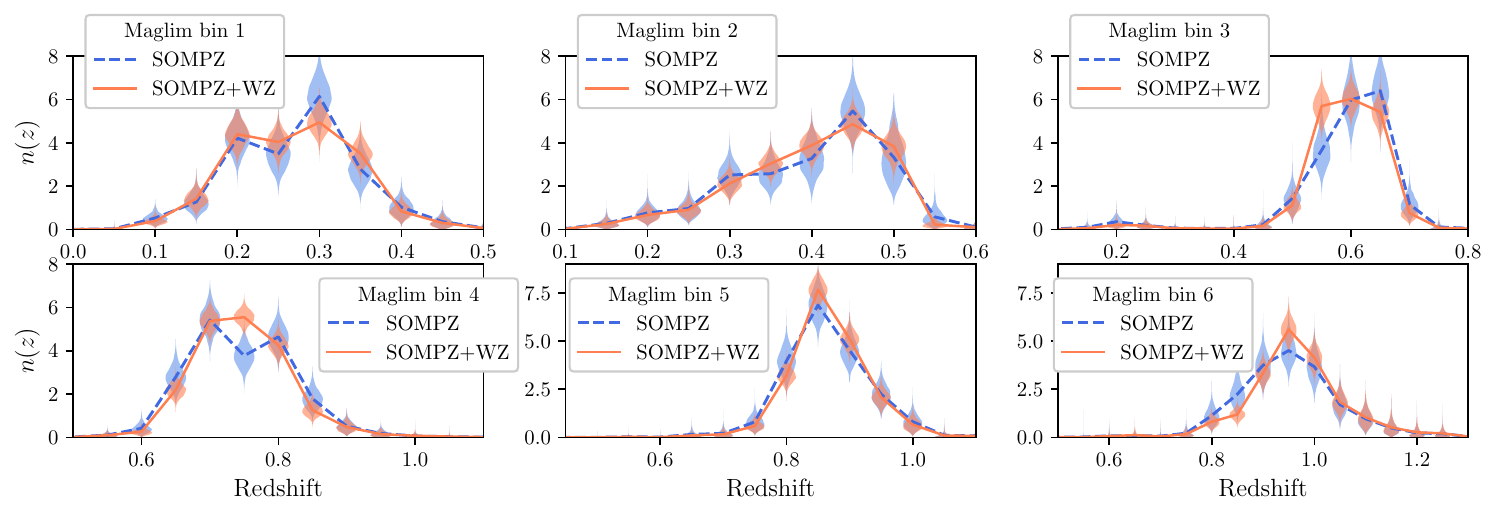}
    \caption{\emph{Left:} Violin plots showing the distribution of $n(z)$ functions in each of the six \textsc{Maglim}++ bins.  Blue violins and lines are the distributions and means of $n(z)$ taken directly from the 3sDir sampling.  Red  lines and violins represent similar quantities from the WZ selected subsample. }
    \label{fig:violin_maglim}
\end{figure*}
\begin{figure*}
    \centering
    \includegraphics[width=1\linewidth]{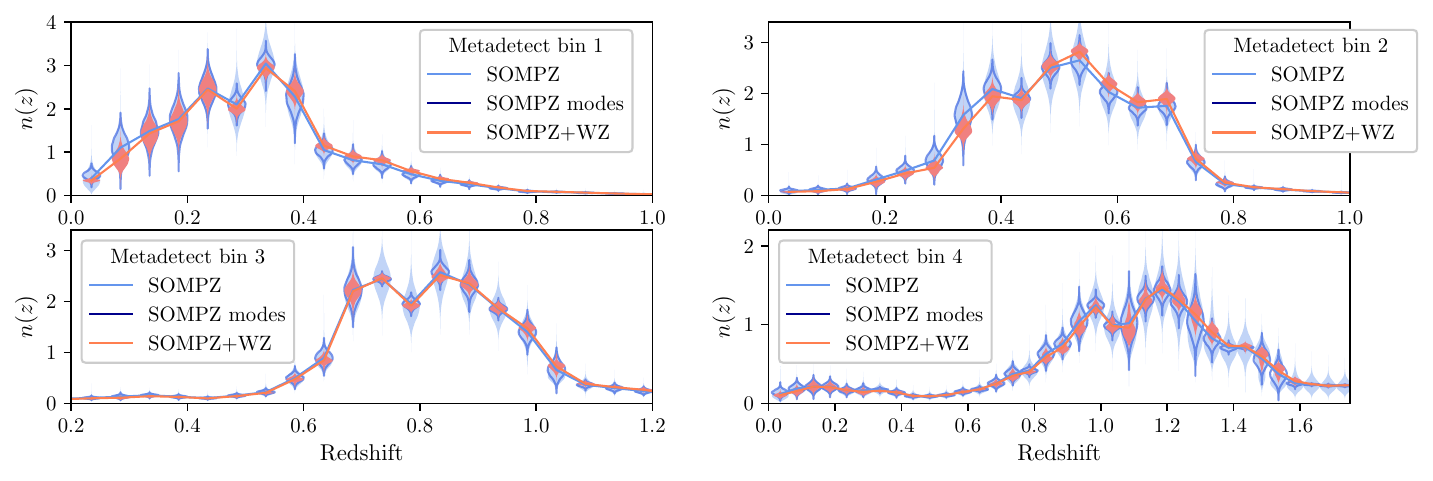}
    \caption{ Violin plots showing the distribution of $n(z)$ functions in each of the four \textsc{Metadetect} redshift bins.  Blue violins and lines are the distributions and means of $n(z)$ taken directly from the 3sDir sampling.  Dark blue contour violins have used mode projection to remove fluctuations with no significant cosmological impact (the mean is unchanged by this process).  Orange are the distributions after importance sampling with the WZ likelihood. }
    \label{fig:violin_meta}
\end{figure*}

For \textsc{Maglim}++,
importance sampling of the SOMPZ $n(z)$ samples using the WZ likelihood is performed independently for the six lens bins. For each $n(z)$ sample, we compute the mean $z$ and standard deviation of $z.$  Distributions of these quantities computed from 100 million input PZ samples, and from the few thousand selected via WZ, are reported in Fig. \ref{fig:hist_WZ_maglim}. The WZ information has led to negligible shifts in the mean redshift in bins 1, 2, 3, and 4, small shifts for bins 5, and a shift $\Delta \langle z \rangle\approx0.015$ in bin 6.
As seen in Fig. \ref{fig:nz_WZ_maglim}, this shift in bin 6 is primarily driven by a comparatively weak WZ signal at 
$z\sim0.85$ relative to the SOM mean distribution. Bringing the WZ$+$PZ point at this $z$ into agreement with the mean of the PZ values would require a multiplicative error of factor $\approx 3,$ restricted only to this point, which is not within the realm of WZ systematic errors expected nor encountered in Cardinal. As $z\sim 0.85$ corresponds to the transition in the reference sample between an LRG dominated sample and an ELG dominated sample, we rerun WZ for both LRG and ELG separately and report the results in App. \ref{app:maglimb6_red_blue}. Both reference samples, composed of the same galaxy tracer, are in good agreement. We thus consider this shift of the bin 6 leading edge to be a definitive implication of the WZ measurement. 
Regarding the standard deviation of $n(z)$, WZ information alters the results of PZ sampling noticeably only by requiring a narrower $n(z)$ for bin 3.

Figure \ref{fig:violin_maglim} presents violin plots of the distributions of \textsc{Maglim}++ $n(z)$ before and after WZ importance sampling. We observe that incorporating WZ information yields smoother $n(z)$ distributions, removing the ``divots'' in $n(z)$ in bins 1, 2, and 4. Additionally, the sizes of the violins are reduced, as WZ sampling eliminates $n(z)$ samples from PZ that have nonphysical fluctuations, \textit{i.e.} the typical $n(z)$ samples is less noisy after WZ importance sampling.

For \textsc{Metadetect} source redshifts, as explained in Sect.~\ref{sec:importance}, we first remove cosmologically irrelevant fluctuations from the input SOMPZ $n(z)$ samples to ensure that importance sampling yields a sufficiently large output sample. Figure~\ref{fig:violin_meta} compares the uncertainties in the PZ$+$WZ $n(z)$ elements with those in the PZ-only case, both before and after mode compression of the PZ samples. As expected, mode compression significantly reduces fluctuations in $n(z)$ at each redshift, \textit{i.e.} the violins are shorter. The addition of WZ selection further suppresses fluctuations, in line with expectations.

For the four source redshift bins, Fig.~\ref{fig:hist_WZ_meta} shows the distributions of the mean redshift \(\langle z \rangle\) and standard deviation \(\mathrm{std}(z)\) across two cases: the original 3sDir samples, and compressed PZ$+$WZ samples selected with importance sampling. The most noticeable feature is an upward shift of approximately \(1\sigma\) in \(\langle z \rangle\)  bin 1 and 2, and a moderate upward shift for bin 3 and 4, due to WZ data, along with a increase of the standard deviation of $n(z)$ (larger bin widths). 
We performed importance sampling with larger priors on the systematics function, to check whether the priors were not too small to describe the  galaxy bias evolution $b_\rmu(z)$, but we got similar shifts for all the bins. 

\begin{figure*}
    \centering
    \includegraphics[width=1\linewidth]{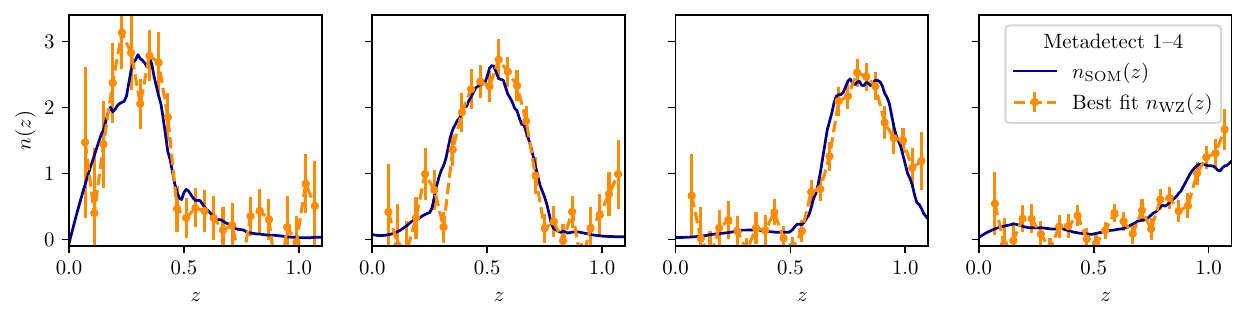}
    \caption{The $n(z)$ from the SOM, and the best fit distribution from the WZ data for \textsc{Metadetect} four bins (including magnification, bias, and systematic parameters). One can see that for the first two bins, the best model oscillates around the SOM distribution which does not capture some of the high-$z$ features in the WZ best fit (e.g. for the second bin, the excess at $z\sim 0.55$). }
    \label{fig:gary_test}
\end{figure*}

In Fig.~\ref{fig:gary_test}, we further investigate the cause of the shifts in mean $z$. We start with the mean $n(z)$ from the SOM and determine the best-fit model by varying the $b_{\rm u}$, magnification, and systematic parameters of $\hat{w}_\rmur$ with respect to the BOSS WZ measurements. Using these best-fit parameters, we transform $w_\rmur(z)$ into a redshift distribution:
\begin{equation}
    n_{\rm WZ}(z) = (w_\rmur(z) - w^{\mu}_{\rm bf}) / (b_\rmr b_{\rm u, bf}).
\end{equation}
where the magnification contribution $ w^\mu$ to the covariance and the bias $b_u$ are given their values in the best-fit model to the data.
The resulting $n(z)$ distributions are shown in Fig.~\ref{fig:gary_test} and compared with the mean SOMPZ $n(z).$  For bin 3, the WZ-derived model is significantly narrower than the SOMPZ $n(z)$ but remains relatively centered. 
In contrast, for bins 1 and 2, the WZ-derived model fluctuates around the SOMPZ $n(z)$, possibly
leading to the observed shift in mean $z$.
Achieving full consistency between the SOM $n(z)$ and the best-fit model would require a higher-order systematic function, allowing for these oscillatory in the model, which is not physically motivated.  Focusing on Fig. \ref{fig:gary_test}, it seems WZ is predicting an excess for the first two bins at $z\sim 1.05$, which could hint for a potential systematic, such as wrong magnification priors, leading to an underestimation of this effect.  One can then naturally think these $z\sim 1$ WZ points are responsible of  the mean-$z$ shifts observed in Fig. \ref{fig:hist_WZ_meta}.  If we now look at Fig. \ref{fig:violin_meta}, we see no excess on the WZ selected $n(z)$ at $z\sim 1$ with respect to PZ ones.  For bin 1 and 2, the reason is straightforward:  the PZ samples are only marginally deviating from 0 at $z=1$, and thus the corresponding WZ data points have no effect for the importance sampling. 
For bin 3 and 4,  we there is no significant mismatch between the best fit WZ $n(z)$ and the SOM one (which does not correspond tot the true $n(z)$), showing that if there was a localized systematic effect, it is corrected by the Sys function. 
 A closer investigation reveals that this shift primarily arises because the WZ data favor 3sDir samples from regions in the Latin hypercube where COSMOS and PAUS redshifts have been perturbed upwards, though still within their estimated uncertainties.

\subsection{Impact on cosmology}
\label{sec:chains}
\begin{figure*}
    \centering
    \includegraphics[width=1\linewidth]{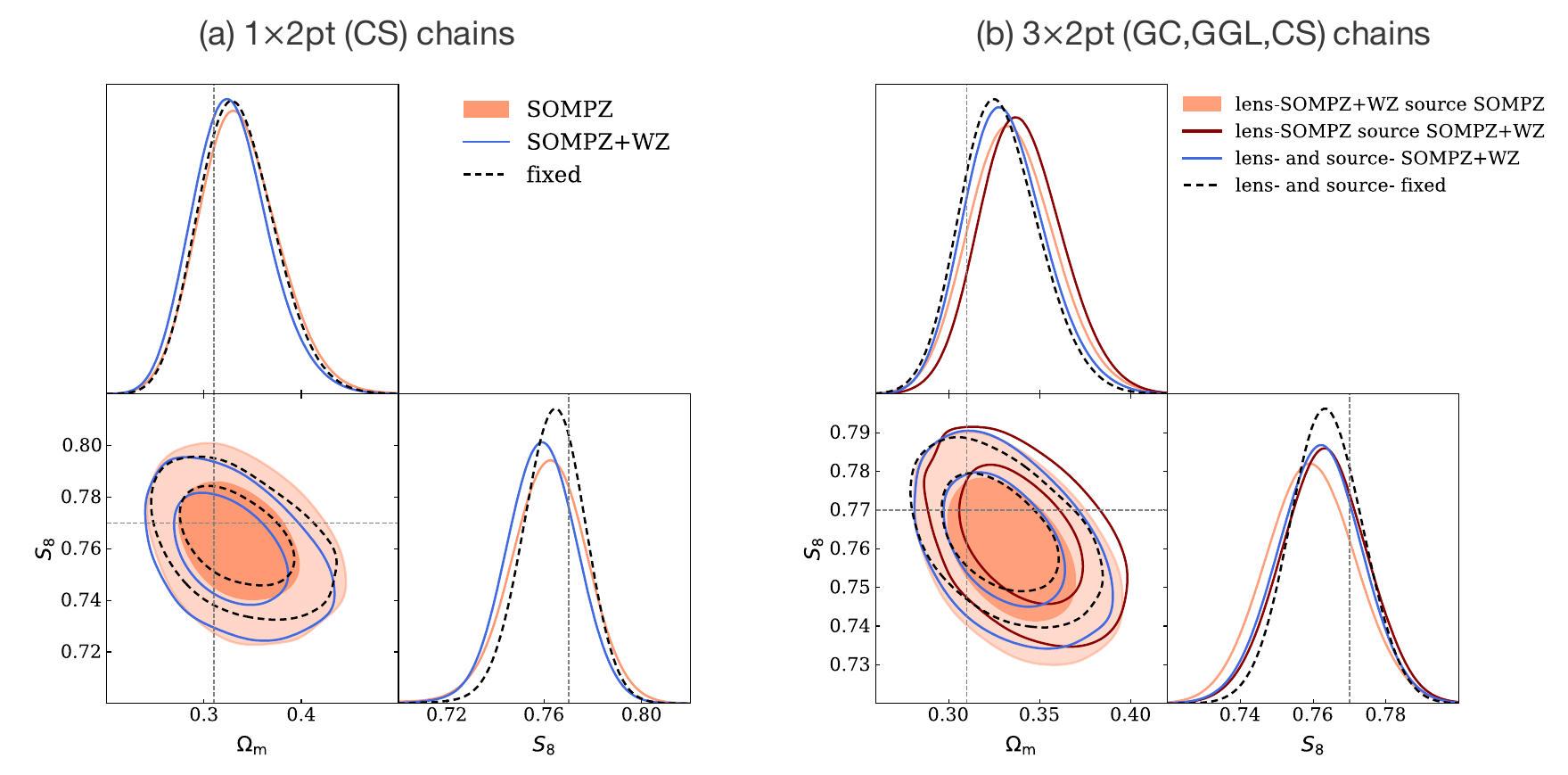}
    \caption{\emph{Left} Cosmic-shear (1$\times$2pt) chains for three setups: SOMPZ uncertainty in orange, SOMPZ+WZ uncertainty in blue, and fixed (i.e. no uncertainty) in black. We see a significant gain on $S_8$ adding WZ uncertainty, but still smaller constraining power in comparison to fixed. All the three chains give consistent results. \emph{Right: } $3\times 2$pt chains for the four uncertainty setups: lens SOMPZ+WZ and source SOMPZ in orange, lens SOMPZ and source SOMPZ+WZ in red, both lens and source with SOMPZ+WZ in blue, and both lens and source fixed (i.e. no uncertainty) in black. All the chains give consistent results, with a clear gain adding WZ information, mainly for the source. }
    \label{fig:chains_Boyan}
\end{figure*}

In this section, we assess the impact of including WZ information on cosmological constraints. Specifically, we aim to answer two key questions: (1) does incorporating WZ information shift the peak of the posterior distributions of cosmological parameters?  (2) Does it reduce the size of the posterior contours?
To investigate these effects, we analyze two sets of Markov Chain Monte Carlo (MCMC) runs: one considering cosmic shear (CS, $1\times 2$ pt) alone, and one with galaxy clustering, galaxy-galaxy lensing, and cosmic shear ($3\times2$ pt). The marginalization over photometric uncertainties follows the mode-based approach described in Sect. \ref{sec:importance}, replacing the shift-and-stretch model used in Y3. The data vector is generated assuming a flat $\Lambda$CDM cosmology (\red{$\Omega_{\rm m}=0.31$, $A_{\rm s}10^9=1.831$, $n_{\rm s}=0.965$, $\Omega_{\rm b}=0.051$, $h=0.69$}) and for both lenses and sources, assumes a true $n(z)$ that is the mean of our derived sampling of $p(n|{\rm WZ,\,PZ}).$

\subsubsection{Cosmic shear}
We test three setups for the source uncertainties assumed for the  cosmological inference: 
\begin{itemize}
    \item SOMPZ: the  $n(z)$ modes and priors are derived from compression of the full SOMPZ samples from $p(n|{\rm PZ})$,
    \item SOMPZ+WZ: the $n(z)$ modes and priors are derived from the importance-sampled $p(n|{\rm WZ,\,PZ})$,
    \item fixed: no photo-$z$ uncertainty are marginalized over. 
\end{itemize}
The posterior distributions of CS chains are shown on the left of Fig. \ref{fig:chains_Boyan}. Incorporating WZ calibration of sources using the BOSS-selected sample improves the constraints in $S_8$ by $10\%$ in the cosmic shear analysis. The fixed posteriors improves SOMPZ+WZ constraints in $S_8$ by 10$\%$, meaning we are still losing some constraining power because of the redshift uncertainties.
We observe consistency between the three results. SOMPZ and WZ are expected to provide consistent best-fit cosmological parameters, with the inclusion of WZ leading to improved constraints.

\subsubsection{ 3$\times2 $pt}
For lenses and sources, we have modes and priors derived from $p(n|{\rm PZ})$ referred as SOMPZ, and from the WZ-importance-sampled $p(n|{\rm WZ,\,PZ})$ referred as SOMPZ+WZ. We then test four setups for the uncertainties: 
\begin{itemize}
    \item source SOMPZ and lens SOMPZ+WZ,
    \item source SOMPZ+WZ and lens SOMPZ,
   \item source SOMPZ+WZ and lens SOMPZ+WZ,
    \item fixed: no photo-$z$ uncertainty are marginalized over for both lens and source. 
\end{itemize}
The posterior distributions of $3\times2$pt chains are shown on the right of Fig. \ref{fig:chains_Boyan}. Adding the WZ information, we observe a gain of around $10\%$ in $S_8$. Comparing the combinations of SOMPZ and SOMPZ+WZ uncertainty, we see that this gain is mainly coming from the uncertainties of sources (orange vs red), and adding WZ for lenses has a negligible impact in $S_8$ (red vs blue), but results in a small gain in $\Omega_{\rm m}$. The fixed posteriors improves SOMPZ+WZ constraints in $S_8$ by 15$\%$, meaning we are still losing significant constraining power because of the redshift uncertainties. We observe consistency between the results. SOMPZ and WZ are expected to provide consistent best-fit cosmological parameters, with the inclusion of WZ leading to improved constraints.

\section{Conclusion}\label{sec:conclusions}

This work is part of a series establishing the redshift calibration  for the Dark Energy Survey Year 6 (DES Y6) $3\times2$pt cosmology analysis. Photometric redshifts in DES Y6 are estimated using self-organizing maps (SOMs), calibrated with spectroscopic and many-band photometric data, as described in \citet{y6-sompz-metadetect}, \citet{y6-sompz-maglim}. In this paper, we complement these estimates with clustering-based redshift (WZ) constraints by cross-correlating DES galaxies with BOSS, eBOSS, and eBOSS quasar samples. These angular cross-correlations define a WZ likelihood, which is then used to perform importance sampling on a large ensemble of SOM-derived $n(z)$ realizations, retaining those consistent with the WZ measurements. The resulting posterior samples of $n(z)$ for each lens and source bin are then compressed using a PCA-like decomposition introduced in \citet{y6-modes}, enabling efficient exploration within cosmological parameter inference.

Our main contributions include:
\begin{itemize}
\item Introducing the WZ methodology for DES Y6, including a redshift decomposition--\textit{cf.} Eq. \eqref{eq:kernels}, magnification--\textit{cf.} Eq. \eqref{eq:magnification_wxymu} , and a systematics model--\textit{cf.} Eq. \eqref{eq:Sys}. All combined, we can express our WZ data vectors as  linear combinations of analytical quantities--\textit{cf.} Eq. \eqref{eq:model:wur_sys}.
\item Presenting an analytically marginalizable WZ likelihood that incorporates systematics and magnification--\textit{cf.} Eq. \eqref{eq:wz_likelihood}, allowing direct sampling of $n(z)$ realizations consistent with both SOMPZ and WZ constraints--\textit{cf.} Sect. \ref{sec:importance}.
\item Investigating the impact of modeling systematics using Cardinal mocks and showing that a low-order polynomial bias evolution model (linear or quadratic) is sufficient for DES Y6 and significantly simpler than the fifth-order model used in DES Y3--\textit{cf.} Sect. \ref{sec:cardinalsys}.
\item Performing cross-correlation measurements using BOSS and eBOSS galaxy and quasar samples--\textit{cf.} Sect. \ref{sec:measurement}. 
\item Comparing our redshift measurements with DESI, showing consistency and confirming the reliability of the method--\textit{cf.} App. \ref{app:desiWZ}.
\item Validating the scale range used in the analysis, demonstrating robustness to nonlinear and small-scale effects--\textit{cf.} Sect. \ref{sec:test_scale} and App. \ref{app:scales}.
\item Applying importance sampling, significantly reducing redshift uncertainties by selecting realizations consistent with clustering data--\textit{cf.} Sect. \ref{sec:fullshape}.
\item Demonstrating that incorporating WZ leads to improved constraints on cosmological parameters in test chains--\textit{cf.} Sect. \ref{sec:chains}.
\end{itemize}

This work highlights the importance of clustering-based redshift constraints as a complement to photometric methods. While SOM-based methods remain a powerful baseline, combining them with WZ constraints improves robustness and reduces uncertainties. This hybrid approach will be crucial for future Stage-IV surveys such as Euclid, LSST, and Roman, which will benefit from deeper imaging and much greater spectroscopic overlap. For instance, Euclid will obtain a spectroscopic redshift sample up to $z\sim1.8$ with H$\alpha$ emitters across its entire footprint. Even higher redshifts, up to $z\sim 3$, will be accessible through quasars with DESI. The methodology developed here is readily extensible to these surveys, paving the way for precise and reliable redshift calibration in next-generation cosmological analyses.

\section*{Acknowledgments}

We thank Cristobal Padilla for his support and advices, Jonas Chaves Montero for his comments on the small scales clustering and mock realism, Andreu Font-Ribera for his comments relative to fiber collision. 

W. d’A. acknowledges support from the  MICINN projects PID2019-111317GB-C32, PID2022-141079NB-C32 as well as predoctoral program AGAUR-FI ajuts (2024 FI-1 00692) Joan Oró. IFAE is partially funded by the CERCA program of the Generalitat de Catalunya. GMB acknowledges support from DOE grant DE-SC0007901 and NSF grant AST-2009210 during this work. MM acknowledges support from the MINCINN
grant CNS2023-144328 (CARTEU) and PID2022-141079NB-C32. The project that gave rise to these results received the support of a fellowship from "la Caixa" Foundation (ID 100010434). The fellowship code is LCF/BQ/PI23/11970028.

\textbf{Author Contributions:} All authors contributed to this paper and/or to the infrastructure work that made this analysis possible. WdA extended the Y3 WZ methodology, developed the WZ code, carried out the measurements, performed the validation tests requested by the collaborations, and prepared the manuscript. GB contributed to the methodological development with innovative ideas, implemented the importance sampling, and assisted in manuscript preparation. BY and GG implemented the SOM pipelines for the source and lens samples and contributed to the manuscript preparation. AAl helped with the statistical tests and their interpretation during the development of this work, and contributed to the manuscript preparation. MM supervised WdA throughout the development of this work. RC and MG made significant contributions to the manuscript preparation as internal reviewers for the collaboration. KB and IS contributed to the development of the Y6 Gold sample. MY, MB, TS,  NW, AP, JV and MR contributed to the building of the \textsc{Metadetect} and \textsc{Maglim}++ samples, and JP produced their catalog versions. CT contributed to the development of Cardinal. DA contributed to the development of Balrog for DES Y6. RC and JF contributed to the design of the BOSS–eBOSS mocks. EL evaluated the magnification coefficients for the different samples, and helped with the galaxy color splitting. AAm, DG, CS, and MT contributed to the interpretation of the analysis as members of the DES Y6 redshift WG.  AP, SA, MC, DS, AF, and CC contributed to the development of this work as members of the DES Y6 SWG.
The remaining authors have made contributions to this paper that include, but are not limited to, the construction of DECam and other aspects of collecting the data; data processing and calibration; catalog creation; developing
broadly used methods, codes, and simulations; running the pipelines
and validation tests; and promoting the science analysis.

Funding for the DES Projects has been provided by the U.S. Department of Energy, the U.S. National Science Foundation, the Ministry of Science and Education of Spain, 
the Science and Technology Facilities Council of the United Kingdom, the Higher Education Funding Council for England, the National Center for Supercomputing 
Applications at the University of Illinois at Urbana-Champaign, the Kavli Institute of Cosmological Physics at the University of Chicago, 
the Center for Cosmology and Astro-Particle Physics at the Ohio State University,
the Mitchell Institute for Fundamental Physics and Astronomy at Texas A\&M University, Financiadora de Estudos e Projetos, 
Funda{\c c}{\~a}o Carlos Chagas Filho de Amparo {\`a} Pesquisa do Estado do Rio de Janeiro, Conselho Nacional de Desenvolvimento Cient{\'i}fico e Tecnol{\'o}gico and 
the Minist{\'e}rio da Ci{\^e}ncia, Tecnologia e Inova{\c c}{\~a}o, the Deutsche Forschungsgemeinschaft and the Collaborating Institutions in the Dark Energy Survey. 

The Collaborating Institutions are Argonne National Laboratory, the University of California at Santa Cruz, the University of Cambridge, Centro de Investigaciones Energ{\'e}ticas, 
Medioambientales y Tecnol{\'o}gicas-Madrid, the University of Chicago, University College London, the DES-Brazil Consortium, the University of Edinburgh, 
the Eidgen{\"o}ssische Technische Hochschule (ETH) Z{\"u}rich, 
Fermi National Accelerator Laboratory, the University of Illinois at Urbana-Champaign, the Institut de Ci{\`e}ncies de l'Espai (IEEC/CSIC), 
the Institut de F{\'i}sica d'Altes Energies, Lawrence Berkeley National Laboratory, the Ludwig-Maximilians Universit{\"a}t M{\"u}nchen and the associated Excellence Cluster Universe, 
the University of Michigan, NSF NOIRLab, the University of Nottingham, The Ohio State University, the University of Pennsylvania, the University of Portsmouth, 
SLAC National Accelerator Laboratory, Stanford University, the University of Sussex, Texas A\&M University, and the OzDES Membership Consortium.

Based in part on observations at NSF Cerro Tololo Inter-American Observatory at NSF NOIRLab (NOIRLab Prop. ID 2012B-0001; PI: J. Frieman), which is managed by the Association of Universities for Research in Astronomy (AURA) under a cooperative agreement with the National Science Foundation.

The DES data management system is supported by the National Science Foundation under Grant Numbers AST-1138766 and AST-1536171.
The DES participants from Spanish institutions are partially supported by MICINN under grants PID2021-123012, PID2021-128989 PID2022-141079, SEV-2016-0588, CEX2020-001058-M and CEX2020-001007-S, some of which include ERDF funds from the European Union. IFAE is partially funded by the CERCA program of the Generalitat de Catalunya.

We  acknowledge support from the Brazilian Instituto Nacional de Ci\^encia
e Tecnologia (INCT) do e-Universo (CNPq grant 465376/2014-2).

This document was prepared by the DES Collaboration using the resources of the Fermi National Accelerator Laboratory (Fermilab), a U.S. Department of Energy, Office of Science, Office of High Energy Physics HEP User Facility. Fermilab is managed by Fermi Forward Discovery Group, LLC, acting under Contract No. 89243024CSC000002.

This research used resources of the National Energy Research Scientific Computing Center (NERSC), a Department of Energy User Facility using NERSC award HEP-ERCAP0035553.

\section*{Data availability}

  The DES Y6 data products used in this work, as well as the full ensemble of DES Y6 \textsc{Maglim}++ \citep{Weaverdyck2025_maglim} and \textsc{Metadetect}  \citep{desy6_metadetect} samples described by this work, are publicly available at \url{https://des.ncsa.illinois.edu/releases}. 
The reference samples are available at:  \url{https://data.sdss.org/sas/dr12/boss/lss/} for BOSS \citep{BOSS_color},  \url{https://data.sdss.org/sas/dr17/eboss/lss/catalogs/DR16/} for eBOSS \citep{eboss_dawson}, and \url{https://data.desi.lbl.gov/public/dr1/survey/catalogs/dr1/LSS/iron/LSScats/v1.5/} for DESI DR1 \citep{DESI_DR1_data}, as well as \url{https://cosmohub.pic.es/home}.
As cosmology likelihood sampling software we use \texttt{cosmosis}, available at https://github.com/joezuntz/cosmosis. Simple examples of WZ codes are available at \url{https://github.com/wdassignies/Clustering_z}. 

\bibliographystyle{mnras_2author}
\bibliography{refs}

\providecommand{\noopsort}[1]{}\providecommand{\singleletter}[1]{#1}%
\begin{thebibliography}{}
\makeatletter
\relax
\def\mn@urlcharsother{\let\do\@makeother \do\$\do\&\do\#\do\^\do\_\do\%\do\~}
\def\mn@doi{\begingroup\mn@urlcharsother \@ifnextchar [ {\mn@doi@} {\mn@doi@[]}}
\def\mn@doi@[#1]#2{\def\@tempa{#1}\ifx\@tempa\@empty \href {http://dx.doi.org/#2} {doi:#2}\else \href {http://dx.doi.org/#2} {#1}\fi \endgroup}
\def\mn@eprint#1#2{\mn@eprint@#1:#2::\@nil}
\def\mn@eprint@arXiv#1{\href {http://arxiv.org/abs/#1} {{\tt arXiv:#1}}}
\def\mn@eprint@dblp#1{\href {http://dblp.uni-trier.de/rec/bibtex/#1.xml} {dblp:#1}}
\def\mn@eprint@#1:#2:#3:#4\@nil{\def\@tempa {#1}\def\@tempb {#2}\def\@tempc {#3}\ifx \@tempc \@empty \let \@tempc \@tempb \let \@tempb \@tempa \fi \ifx \@tempb \@empty \def\@tempb {arXiv}\fi \@ifundefined {mn@eprint@\@tempb}{\@tempb:\@tempc}{\expandafter \expandafter \csname mn@eprint@\@tempb\endcsname \expandafter{\@tempc}}}

\bibitem[\protect\citeauthoryear{{Adame} \& {Aguilar} et~al.,}{{Adame} et~al.}{2025}]{DESI_DR1_data}
{Adame} A.~G.,  et~al. 2025, \mn@doi [\jcap] {10.1088/1475-7516/2025/07/017}, \href {https://ui.adsabs.harvard.edu/abs/2025JCAP...07..017A} {2025, 017}

\bibitem[\protect\citeauthoryear{{Alarcon} \& {Gaztanaga} et~al.,}{{Alarcon} et~al.}{2021}]{Alarcon2020}
{Alarcon} A.,  et~al. 2021, \mn@doi [\mnras] {10.1093/mnras/staa3659}, \href {https://ui.adsabs.harvard.edu/abs/2021MNRAS.501.6103A} {501, 6103}

\bibitem[\protect\citeauthoryear{{Anbajagane}, {Tabbutt}  et~al.}{{Anbajagane} et~al.}{2025}]{y6-balrog}
{Anbajagane} D.,  {Tabbutt} M.,   et~al., 2025

\bibitem[\protect\citeauthoryear{{Bechtol} et~al.}{{Bechtol} et~al.}{2025}]{y6-gold}
{Bechtol} K.,  et~al., 2025

\bibitem[\protect\citeauthoryear{{Becker}}{{Becker}}{2013}]{calclens}
{Becker} M.~R.,  2013, PhD thesis, University of Chicago

\bibitem[\protect\citeauthoryear{{Bernstein} \& {Assignies Doumerg} et~al.,}{{Bernstein} et~al.}{2025}]{y6-modes}
{Bernstein} G.,  et~al. 2025, \mn@doi [arXiv e-prints] {10.48550/arXiv.2506.00758}, \href {https://ui.adsabs.harvard.edu/abs/2025arXiv250600758B} {p. arXiv:2506.00758}

\bibitem[\protect\citeauthoryear{{Blanton} \& {Schlegel} et~al.,}{{Blanton} et~al.}{2005}]{SDSS}
{Blanton} M.~R.,  et~al. 2005, \mn@doi [\aj] {10.1086/429803}, \href {https://ui.adsabs.harvard.edu/abs/2005AJ....129.2562B} {129, 2562}

\bibitem[\protect\citeauthoryear{Blas, Lesgourgues  \& Tram}{Blas et~al.}{2011}]{Diego_Blas_2011_class}
Blas D.,  Lesgourgues J.,   Tram T.,  2011, \mn@doi [JCAP] {10.1088/1475-7516/2011/07/034}, 2011, 034–034

\bibitem[\protect\citeauthoryear{{Bordoloi}, {Lilly}  \& {Amara}}{{Bordoloi} et~al.}{2010}]{photo-z-perf_cosmo}
{Bordoloi} R.,  {Lilly} S.~J.,   {Amara} A.,  2010, \mn@doi [\mnras] {10.1111/j.1365-2966.2010.16765.x}, \href {https://ui.adsabs.harvard.edu/abs/2010MNRAS.406..881B} {406, 881}

\bibitem[\protect\citeauthoryear{{Buchs} \& {Davis} et~al.,}{{Buchs} et~al.}{2019}]{Buchs2019}
{Buchs} R.,  et~al. 2019, \mn@doi [\mnras] {10.1093/mnras/stz2162}, \href {https://ui.adsabs.harvard.edu/abs/2019MNRAS.489..820B} {489, 820}

\bibitem[\protect\citeauthoryear{{Campos} \& {Yin} et~al.,}{{Campos} et~al.}{2024}]{Campos2024}
{Campos} A.,  et~al. 2024, \href {https://ui.adsabs.harvard.edu/abs/2024arXiv240800922C} {p. arXiv:2408.00922}

\bibitem[\protect\citeauthoryear{{Cawthon} \& {Elvin-Poole} et~al.,}{{Cawthon} et~al.}{2022}]{Cawton2022}
{Cawthon} R.,  et~al. 2022, \mn@doi [\mnras] {10.1093/mnras/stac1160}, \href {https://ui.adsabs.harvard.edu/abs/2022MNRAS.513.5517C} {513, 5517}

\bibitem[\protect\citeauthoryear{{Chisari} \& {Alonso} et~al.,}{{Chisari} et~al.}{2019}]{CCL}
{Chisari} N.~E.,  et~al. 2019, \mn@doi [\apjs] {10.3847/1538-4365/ab1658}, \href {https://ui.adsabs.harvard.edu/abs/2019ApJS..242....2C} {242, 2}

\bibitem[\protect\citeauthoryear{{Choppin de Janvry} \& {Gontcho} et~al.,}{{Choppin de Janvry} et~al.}{2025}]{choppin_2025}
{Choppin de Janvry} J.,  et~al. 2025, \mn@doi [arXiv e-prints] {10.48550/arXiv.2511.18133}, \href {https://ui.adsabs.harvard.edu/abs/2025arXiv251118133C} {p. arXiv:2511.18133}

\bibitem[\protect\citeauthoryear{{Davis} \& {Peebles}}{{Davis} \& {Peebles}}{1983}]{Davis_Peebles}
{Davis} M.,  {Peebles} P.~J.~E.,  1983, \mn@doi [\apj] {10.1086/160884}, \href {https://ui.adsabs.harvard.edu/abs/1983ApJ...267..465D} {267, 465}

\bibitem[\protect\citeauthoryear{{Dawson} \& {Schlegel} et~al.,}{{Dawson} et~al.}{2013}]{BOSS_color}
{Dawson} K.~S.,  et~al. 2013, \mn@doi [\aj] {10.1088/0004-6256/145/1/10}, \href {https://ui.adsabs.harvard.edu/abs/2013AJ....145...10D} {145, 10}

\bibitem[\protect\citeauthoryear{{Dawson} \& {Kneib} et~al.,}{{Dawson} et~al.}{2016}]{eboss_dawson}
{Dawson} K.~S.,  et~al. 2016, \mn@doi [\aj] {10.3847/0004-6256/151/2/44}, \href {https://ui.adsabs.harvard.edu/abs/2016AJ....151...44D} {151, 44}

\bibitem[\protect\citeauthoryear{{De Vicente}, {S{\'a}nchez}  \& {Sevilla-Noarbe}}{{De Vicente} et~al.}{2016}]{DNF}
{De Vicente} J.,  {S{\'a}nchez} E.,   {Sevilla-Noarbe} I.,  2016, \mn@doi [\mnras] {10.1093/mnras/stw857}, \href {https://ui.adsabs.harvard.edu/abs/2016MNRAS.459.3078D} {459, 3078}

\bibitem[\protect\citeauthoryear{{DeRose} \& {Wechsler} et~al.,}{{DeRose} et~al.}{2019}]{JoeBuzzard}
{DeRose} J.,  et~al. 2019, \href {https://ui.adsabs.harvard.edu/abs/2019arXiv190102401D} {p. arXiv:1901.02401}

\bibitem[\protect\citeauthoryear{{Elvin-Poole} \& {MacCrann} et~al.,}{{Elvin-Poole} et~al.}{2023}]{magnification_DESY3}
{Elvin-Poole} J.,  et~al. 2023, \mn@doi [\mnras] {10.1093/mnras/stad1594}, \href {https://ui.adsabs.harvard.edu/abs/2023MNRAS.523.3649E} {523, 3649}

\bibitem[\protect\citeauthoryear{{Euclid Collaboration} \& {Mellier} et~al.,}{{Euclid Collaboration}}{2025a}]{Euclid_overview}
{Euclid Collaboration} 2025a, \mn@doi [\aap] {10.1051/0004-6361/202450810}, \href {https://ui.adsabs.harvard.edu/abs/2025A&A...697A...1E} {697, A1}

\bibitem[\protect\citeauthoryear{{Euclid Collaboration} \& {Castander} et~al.,}{{Euclid Collaboration}}{2025b}]{FS_2024}
{Euclid Collaboration} 2025b, \mn@doi [\aap] {10.1051/0004-6361/202450853}, \href {https://ui.adsabs.harvard.edu/abs/2025A&A...697A...5E} {697, A5}

\bibitem[\protect\citeauthoryear{{Euclid Collaboration: Desprez} \& {Paltani} et~al.,}{{Euclid Collaboration: Desprez} et~al.}{2020}]{Desprez-EP10}
{Euclid Collaboration: Desprez} G.,  et~al. 2020, \mn@doi [\aap] {10.1051/0004-6361/202039403}, \href {https://ui.adsabs.harvard.edu/abs/2020A&A...644A..31E} {644, A31}

\bibitem[\protect\citeauthoryear{{Euclid Collaboration: Ilbert} \& {de la Torre} et~al.,}{{Euclid Collaboration: Ilbert} et~al.}{2021}]{Ilbert-EP11}
{Euclid Collaboration: Ilbert} O.,  et~al. 2021, \mn@doi [\aap] {10.1051/0004-6361/202040237}, \href {https://ui.adsabs.harvard.edu/abs/2021A&A...647A.117E} {647, A117}

\bibitem[\protect\citeauthoryear{{Gatti} \& {Vielzeuf} et~al.,}{{Gatti} et~al.}{2018}]{Gatti_DESY1}
{Gatti} M.,  et~al. 2018, \mn@doi [\mnras] {10.1093/mnras/sty466}, \href {https://ui.adsabs.harvard.edu/abs/2018MNRAS.477.1664G} {477, 1664}

\bibitem[\protect\citeauthoryear{{Gatti} \& {Giannini} et~al.,}{{Gatti} et~al.}{2022}]{Gatti_Giulia_DESY3}
{Gatti} M.,  et~al. 2022, \mn@doi [\mnras] {10.1093/mnras/stab3311}, \href {https://ui.adsabs.harvard.edu/abs/2022MNRAS.510.1223G} {510, 1223}

\bibitem[\protect\citeauthoryear{{Giannini} \& {Alarcon} et~al.,}{{Giannini} et~al.}{2024}]{DESY3_MAGLIM_z}
{Giannini} G.,  et~al. 2024, \mn@doi [\mnras] {10.1093/mnras/stad2945}, \href {https://ui.adsabs.harvard.edu/abs/2024MNRAS.527.2010G} {527, 2010}

\bibitem[\protect\citeauthoryear{{Giannini}, {Alarcon}  et~al.}{{Giannini} et~al.}{2025}]{y6-sompz-maglim}
{Giannini} G.,  {Alarcon} A.,   et~al., 2025, \href {https://ui.adsabs.harvard.edu/abs/2025arXiv250907964G} {p. arXiv:2509.07964}

\bibitem[\protect\citeauthoryear{{Giannini} \& {Camacho-Ciurana} et~al.,}{{Giannini} et~al.}{2026}]{y6-shear_ratio}
{Giannini} G.,  et~al. 2026, \mn@doi [arXiv e-prints] {10.48550/arXiv.2601.15175}, \href {https://ui.adsabs.harvard.edu/abs/2026arXiv260115175G} {p. arXiv:2601.15175}

\bibitem[\protect\citeauthoryear{{Hartlap}, {Simon}  \& {Schneider}}{{Hartlap} et~al.}{2007}]{Hartlap}
{Hartlap} J.,  {Simon} P.,   {Schneider} P.,  2007, \mn@doi [\aap] {10.1051/0004-6361:20066170}, \href {https://ui.adsabs.harvard.edu/abs/2007A&A...464..399H} {464, 399}

\bibitem[\protect\citeauthoryear{{Hartley} \& {Chang} et~al.,}{{Hartley} et~al.}{2020}]{hartley20}
{Hartley} W.~G.,  et~al. 2020, \mn@doi [\mnras] {10.1093/mnras/staa1812}, \href {https://ui.adsabs.harvard.edu/abs/2020MNRAS.496.4769H} {496, 4769}

\bibitem[\protect\citeauthoryear{{Hildebrandt}}{{Hildebrandt}}{2016}]{hildebrandt_magn}
{Hildebrandt} H.,  2016, \mn@doi [\mnras] {10.1093/mnras/stv2575}, \href {https://ui.adsabs.harvard.edu/abs/2016MNRAS.455.3943H} {455, 3943}

\bibitem[\protect\citeauthoryear{{Hildebrandt} \& {van den Busch} et~al.,}{{Hildebrandt} et~al.}{2021}]{KIDS_redshift_dis}
{Hildebrandt} H.,  et~al. 2021, \mn@doi [\aap] {10.1051/0004-6361/202039018}, \href {https://ui.adsabs.harvard.edu/abs/2021A&A...647A.124H} {647, A124}

\bibitem[\protect\citeauthoryear{{Huterer} \& {Takada} et~al.,}{{Huterer} et~al.}{2006}]{Hurterer2006}
{Huterer} D.,  et~al. 2006, \mn@doi [\mnras] {10.1111/j.1365-2966.2005.09782.x}, \href {https://ui.adsabs.harvard.edu/abs/2006MNRAS.366..101H} {366, 101}

\bibitem[\protect\citeauthoryear{{Jarvis}}{{Jarvis}}{2015}]{Treecorr}
{Jarvis} M.,  2015, {TreeCorr: Two-point correlation functions}, Astrophysics Source Code Library, record ascl:1508.007

\bibitem[\protect\citeauthoryear{{Kerscher}, {Szapudi}  \& {Szalay}}{{Kerscher} et~al.}{2000}]{kerscher_comparison_estimator}
{Kerscher} M.,  {Szapudi} I.,   {Szalay} A.~S.,  2000, \mn@doi [\apjl] {10.1086/312702}, \href {https://ui.adsabs.harvard.edu/abs/2000ApJ...535L..13K} {535, L13}

\bibitem[\protect\citeauthoryear{{Krause} \& {Fang} et~al.,}{{Krause} et~al.}{2021}]{Y3_DES_Krause}
{Krause} E.,  et~al. 2021, \href {https://ui.adsabs.harvard.edu/abs/2021arXiv210513548K} {p. arXiv:2105.13548}

\bibitem[\protect\citeauthoryear{{Landy} \& {Szalay}}{{Landy} \& {Szalay}}{1993}]{Landy_Szalay}
{Landy} S.~D.,  {Szalay} A.~S.,  1993, \mn@doi [\apj] {10.1086/172900}, \href {https://ui.adsabs.harvard.edu/abs/1993ApJ...412...64L} {412, 64}

\bibitem[\protect\citeauthoryear{{Legnani} et~al.}{{Legnani} et~al.}{2026}]{y6-magnification}
{Legnani} E.,  et~al., 2026

\bibitem[\protect\citeauthoryear{{Martinelli} \& {Tutusaus} et~al.,}{{Martinelli} et~al.}{2021}]{Martinelli21a}
{Martinelli} M.,  et~al. 2021, \mn@doi [\aap] {10.1051/0004-6361/202039835}, \href {https://ui.adsabs.harvard.edu/abs/2021A&A...649A.100M} {649, A100}

\bibitem[\protect\citeauthoryear{{Mau } et~al.}{{Mau } et~al.}{prep}]{y6-imagesims}
{Mau } S.,  et~al., in prep., To be submitted

\bibitem[\protect\citeauthoryear{{McCullough} \& {Amon} et~al.,}{{McCullough} et~al.}{2024}]{blue_shear}
{McCullough} J.,  et~al. 2024, \href {https://ui.adsabs.harvard.edu/abs/2024arXiv241022272M} {p. arXiv:2410.22272}

\bibitem[\protect\citeauthoryear{{McQuinn} \& {White}}{{McQuinn} \& {White}}{2013}]{clust_z_mcQuinn_white}
{McQuinn} M.,  {White} M.,  2013, \mn@doi [\mnras] {10.1093/mnras/stt914}, \href {https://ui.adsabs.harvard.edu/abs/2013MNRAS.433.2857M} {433, 2857}

\bibitem[\protect\citeauthoryear{{M{\'e}nard} \& {Scranton} et~al.,}{{M{\'e}nard} et~al.}{2013}]{Menard2013}
{M{\'e}nard} B.,  et~al. 2013, \href {https://ui.adsabs.harvard.edu/abs/2013arXiv1303.4722M} {p. arXiv:1303.4722}

\bibitem[\protect\citeauthoryear{{Morrison} \& {Hildebrandt} et~al.,}{{Morrison} et~al.}{2017}]{theWizz}
{Morrison} C.~B.,  et~al. 2017, \mn@doi [\mnras] {10.1093/mnras/stx342}, \href {https://ui.adsabs.harvard.edu/abs/2017MNRAS.467.3576M} {467, 3576}

\bibitem[\protect\citeauthoryear{{Myles} \& {Alarcon} et~al.,}{{Myles} et~al.}{2021}]{y3-sompz-sources}
{Myles} J.,  et~al. 2021, \mn@doi [\mnras] {10.1093/mnras/stab1515}, \href {https://ui.adsabs.harvard.edu/abs/2021MNRAS.505.4249M} {505, 4249}

\bibitem[\protect\citeauthoryear{{Naidoo} \& {Johnston} et~al.,}{{Naidoo} et~al.}{2023}]{Naidoo23}
{Naidoo} K.,  et~al. 2023, \mn@doi [\aap] {10.1051/0004-6361/202244795}, \href {https://ui.adsabs.harvard.edu/abs/2023A&A...670A.149N} {670, A149}

\bibitem[\protect\citeauthoryear{{Navarro-Giron{\'e}s} \& {Gazta{\~n}aga} et~al.,}{{Navarro-Giron{\'e}s} et~al.}{2024}]{PAUSW1}
{Navarro-Giron{\'e}s} D.,  et~al. 2024, \mn@doi [\mnras] {10.1093/mnras/stae1686}, \href {https://ui.adsabs.harvard.edu/abs/2024MNRAS.534.1504N} {534, 1504}

\bibitem[\protect\citeauthoryear{{Newman}}{{Newman}}{2008}]{newman2008}
{Newman} J.~A.,  2008, \mn@doi [\apj] {10.1086/589982}, \href {https://ui.adsabs.harvard.edu/abs/2008ApJ...684...88N} {684, 88}

\bibitem[\protect\citeauthoryear{{Nicola} \& {Hadzhiyska} et~al.,}{{Nicola} et~al.}{2024}]{bias_pert_lsst}
{Nicola} A.,  et~al. 2024, \mn@doi [JCAP] {10.1088/1475-7516/2024/02/015}, \href {https://ui.adsabs.harvard.edu/abs/2024JCAP...02..015N} {p.~015}

\bibitem[\protect\citeauthoryear{{Palanque-Delabrouille} \& {Magneville} et~al.,}{{Palanque-Delabrouille} et~al.}{2016}]{eboss_qso}
{Palanque-Delabrouille} N.,  et~al. 2016, \mn@doi [\aap] {10.1051/0004-6361/201527392}, \href {https://ui.adsabs.harvard.edu/abs/2016A&A...587A..41P} {587, A41}

\bibitem[\protect\citeauthoryear{{Porredon} \& {Crocce} et~al.,}{{Porredon} et~al.}{2022}]{Porredon_Maglimy3}
{Porredon} A.,  et~al. 2022, \mn@doi [\prd] {10.1103/PhysRevD.106.103530}, \href {https://ui.adsabs.harvard.edu/abs/2022PhRvD.106j3530P} {106, 103530}

\bibitem[\protect\citeauthoryear{{Raichoor} \& {Comparat} et~al.,}{{Raichoor} et~al.}{2017}]{raichoor17}
{Raichoor} A.,  et~al. 2017, \mn@doi [\mnras] {10.1093/mnras/stx1790}, \href {https://ui.adsabs.harvard.edu/abs/2017MNRAS.471.3955R} {471, 3955}

\bibitem[\protect\citeauthoryear{{Rau} \& {Dalal} et~al.,}{{Rau} et~al.}{2023}]{HSC_clustering_z}
{Rau} M.~M.,  et~al. 2023, \mn@doi [\mnras] {10.1093/mnras/stad1962}, \href {https://ui.adsabs.harvard.edu/abs/2023MNRAS.524.5109R} {524, 5109}

\bibitem[\protect\citeauthoryear{{Rodr{\'\i}guez-Monroy}, {Weaverdyck}  et~al.}{{Rodr{\'\i}guez-Monroy} et~al.}{2025}]{y6-lss_mask}
{Rodr{\'\i}guez-Monroy} M.,  {Weaverdyck} N.,   et~al., 2025, \href {https://ui.adsabs.harvard.edu/abs/2025arXiv250907943R} {p. 2509.07943}

\bibitem[\protect\citeauthoryear{{Ross} \& {Bautista} et~al.,}{{Ross} et~al.}{2020}]{weight_eboss}
{Ross} A.~J.,  et~al. 2020, \mn@doi [\mnras] {10.1093/mnras/staa2416}, \href {https://ui.adsabs.harvard.edu/abs/2020MNRAS.498.2354R} {498, 2354}

\bibitem[\protect\citeauthoryear{{Rozo} \& {Rykoff} et~al.,}{{Rozo} et~al.}{2016}]{2016rm}
{Rozo} E.,  et~al. 2016, \mn@doi [\mnras] {10.1093/mnras/stw1281}, \href {https://ui.adsabs.harvard.edu/abs/2016MNRAS.461.1431R} {461, 1431}

\bibitem[\protect\citeauthoryear{{Salvato}, {Ilbert}  \& {Hoyle}}{{Salvato} et~al.}{2019}]{Salvato_2019}
{Salvato} M.,  {Ilbert} O.,   {Hoyle} B.,  2019, \mn@doi [Nature Astronomy] {10.1038/s41550-018-0478-0}, \href {https://ui.adsabs.harvard.edu/abs/2019NatAs...3..212S} {3, 212}

\bibitem[\protect\citeauthoryear{{S{\'a}nchez} \& {Raveri} et~al.,}{{S{\'a}nchez} et~al.}{2020}]{3sdir}
{S{\'a}nchez} C.,  et~al. 2020, \mn@doi [\mnras] {10.1093/mnras/staa2542}, \href {https://ui.adsabs.harvard.edu/abs/2020MNRAS.498.2984S} {498, 2984}

\bibitem[\protect\citeauthoryear{{S{\'a}nchez} \& {Alarcon} et~al.,}{{S{\'a}nchez} et~al.}{2023}]{highz}
{S{\'a}nchez} C.,  et~al. 2023, \mn@doi [\mnras] {10.1093/mnras/stad2402}, \href {https://ui.adsabs.harvard.edu/abs/2023MNRAS.525.3896S} {525, 3896}

\bibitem[\protect\citeauthoryear{{Schlafly}, {Meisner}  \& {Green}}{{Schlafly} et~al.}{2019}]{unwise2019}
{Schlafly} E.~F.,  {Meisner} A.~M.,   {Green} G.~M.,  2019, \mn@doi [\apjs] {10.3847/1538-4365/aafbea}, \href {https://ui.adsabs.harvard.edu/abs/2019ApJS..240...30S} {240, 30}

\bibitem[\protect\citeauthoryear{{Schmidt} \& {M{\'e}nard} et~al.,}{{Schmidt} et~al.}{2013}]{Schmidt2013}
{Schmidt} S.~J.,  et~al. 2013, \mn@doi [\mnras] {10.1093/mnras/stt410}, \href {https://ui.adsabs.harvard.edu/abs/2013MNRAS.431.3307S} {431, 3307}

\bibitem[\protect\citeauthoryear{{Schutt} \& {Jarvis} et~al.,}{{Schutt} et~al.}{2025}]{y6-piff}
{Schutt} T.,  et~al. 2025, \mn@doi [The Open Journal of Astrophysics] {10.33232/001c.132299}, \href {https://ui.adsabs.harvard.edu/abs/2025OJAp....8E..26S} {8, 26}

\bibitem[\protect\citeauthoryear{{Scottez} \& {Benoit-L{\'e}vy} et~al.,}{{Scottez} et~al.}{2018}]{Scottez}
{Scottez} V.,  et~al. 2018, \mn@doi [\mnras] {10.1093/mnras/stx3056}, \href {https://ui.adsabs.harvard.edu/abs/2018MNRAS.474.3921S} {474, 3921}

\bibitem[\protect\citeauthoryear{{St{\"o}lzner} \& {Joachimi} et~al.,}{{St{\"o}lzner} et~al.}{2021}]{Stolzner_GaussianFitting}
{St{\"o}lzner} B.,  et~al. 2021, \mn@doi [\aap] {10.1051/0004-6361/202040130}, \href {https://ui.adsabs.harvard.edu/abs/2021A&A...650A.148S} {650, A148}

\bibitem[\protect\citeauthoryear{{Takahashi} \& {Sato} et~al.,}{{Takahashi} et~al.}{2012}]{halofit}
{Takahashi} R.,  et~al. 2012, \mn@doi [\apj] {10.1088/0004-637X/761/2/152}, \href {https://ui.adsabs.harvard.edu/abs/2012ApJ...761..152T} {761, 152}

\bibitem[\protect\citeauthoryear{{Tanaka} \& {Coupon} et~al.,}{{Tanaka} et~al.}{2018}]{photo-z_HSC}
{Tanaka} M.,  et~al. 2018, \mn@doi [PASJ] {10.1093/pasj/psx077}, \href {https://ui.adsabs.harvard.edu/abs/2018PASJ...70S...9T} {70, S9}

\bibitem[\protect\citeauthoryear{{The LSST Dark Energy Science Collaboration} \& {Mandelbaum} et~al.,}{{The LSST Dark Energy Science Collaboration}}{2018}]{lsst_req}
{The LSST Dark Energy Science Collaboration} 2018, \href {https://ui.adsabs.harvard.edu/abs/2018arXiv180901669T} {p. arXiv:1809.01669}

\bibitem[\protect\citeauthoryear{{To} \& {DeRose} et~al.,}{{To} et~al.}{2024}]{cardinal}
{To} C.-H.,  et~al. 2024, \mn@doi [\apj] {10.3847/1538-4357/ad0e61}, \href {https://ui.adsabs.harvard.edu/abs/2024ApJ...961...59T} {961, 59}

\bibitem[\protect\citeauthoryear{{To} et~al.}{{To} et~al.}{prep}]{Y6clustermethod}
{To} C.-H.,  et~al., in prep., To be submitted

\bibitem[\protect\citeauthoryear{{Vargas-Maga{\~n}a} \& {Bautista} et~al.,}{{Vargas-Maga{\~n}a} et~al.}{2013}]{LSvsDP}
{Vargas-Maga{\~n}a} M.,  et~al. 2013, \mn@doi [\aap] {10.1051/0004-6361/201220790}, \href {https://ui.adsabs.harvard.edu/abs/2013A&A...554A.131V} {554, A131}

\bibitem[\protect\citeauthoryear{{Weaver} \& {Kauffmann} et~al.,}{{Weaver} et~al.}{2022}]{cosmos2020}
{Weaver} J.~R.,  et~al. 2022, \mn@doi [\apjs] {10.3847/1538-4365/ac3078}, \href {https://ui.adsabs.harvard.edu/abs/2022ApJS..258...11W} {258, 11}

\bibitem[\protect\citeauthoryear{{Weaverdyck}, {Rodr{\'\i}guez-Monroy}  et~al.}{{Weaverdyck} et~al.}{2026}]{Weaverdyck2025_maglim}
{Weaverdyck} N.,  {Rodr{\'\i}guez-Monroy} M.,   et~al., 2026

\bibitem[\protect\citeauthoryear{{Weaverdyck} et~al.}{{Weaverdyck} et~al.}{prep}]{Weaverdyck2025stargal}
{Weaverdyck} N.,  et~al., in prep., \prd

\bibitem[\protect\citeauthoryear{{Wright} \& {Hildebrandt} et~al.,}{{Wright} et~al.}{2020a}]{Wright_2020}
{Wright} A.~H.,  et~al. 2020a, \mn@doi [\aap] {10.1051/0004-6361/201936782}, \href {https://ui.adsabs.harvard.edu/abs/2020A&A...637A.100W} {637, A100}

\bibitem[\protect\citeauthoryear{{Wright} \& {Hildebrandt} et~al.,}{{Wright} et~al.}{2020b}]{Wright_2020b}
{Wright} A.~H.,  et~al. 2020b, \mn@doi [\aap] {10.1051/0004-6361/202038389}, \href {https://ui.adsabs.harvard.edu/abs/2020A&A...640L..14W} {640, L14}

\bibitem[\protect\citeauthoryear{{Yamamoto} \& {Becker} et~al.,}{{Yamamoto} et~al.}{2025}]{desy6_metadetect}
{Yamamoto} M.,  et~al. 2025, \mn@doi [\mnras] {10.1093/mnras/staf1661}, \href {https://ui.adsabs.harvard.edu/abs/2025MNRAS.tmp.1569Y} {}

\bibitem[\protect\citeauthoryear{{Yin} \& {Amon} et~al.,}{{Yin} et~al.}{2025}]{y6-sompz-metadetect}
{Yin} B.,  et~al. 2025, \href {https://ui.adsabs.harvard.edu/abs/2025arXiv251023566Y} {p. arXiv:2510.23566}

\bibitem[\protect\citeauthoryear{{d'Assignies} \& {Manera} et~al.,}{{d'Assignies} et~al.}{2025}]{Euclid_dassignies}
{d'Assignies} W.,  et~al. 2025, \mn@doi [\aap] {10.1051/0004-6361/202555551}, \href {https://ui.adsabs.harvard.edu/abs/2025A&A...702A.155D} {702, A155}

\bibitem[\protect\citeauthoryear{{van den Busch} \& {Hildebrandt} et~al.,}{{van den Busch} et~al.}{2020}]{van_den_Busch_2020}
{van den Busch} J.~L.,  et~al. 2020, \mn@doi [\aap] {10.1051/0004-6361/202038835}, \href {https://ui.adsabs.harvard.edu/abs/2020A&A...642A.200V} {642, A200}

\makeatother
\end{thebibliography}

\appendix

\section{Realistic spectroscopic mocks and small scale systematics in Cardinal}

\label{app:spec_mocks}
\begin{figure*}
    \centering
    \includegraphics[width=1\linewidth]{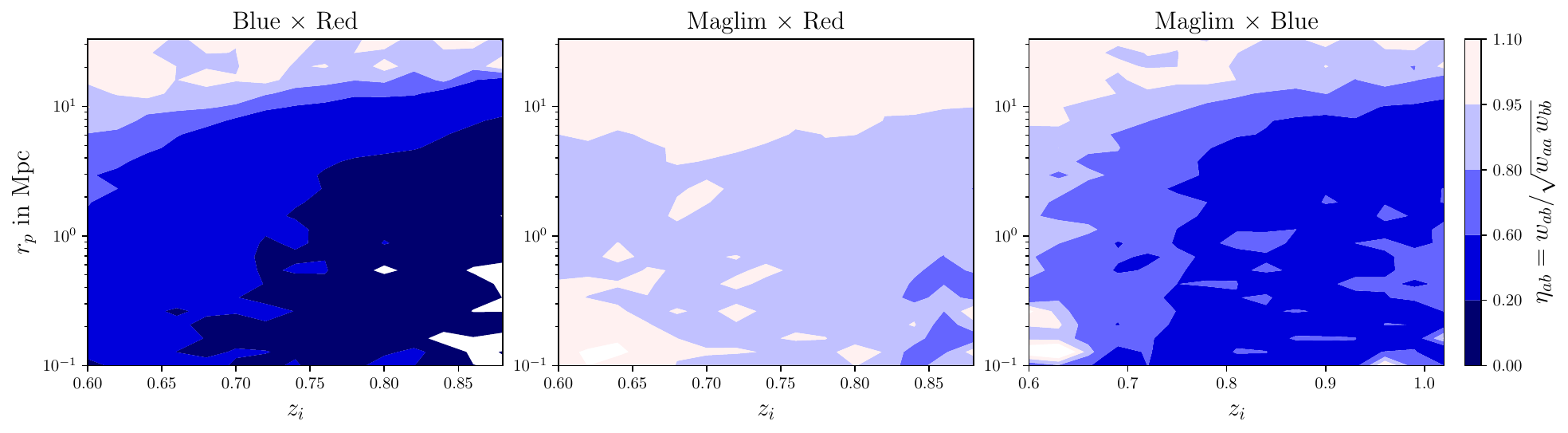}
    \caption{Difference of the biases in the cross-correlations and the biases of the auto-correlations, quantified by a Pearson-like coefficients. We evaluate this quantity for different combinations of Mocks.  }
    \label{fig:Pears}
\end{figure*}

For our fiducial analysis, we use BOSS-eBOSS mocks constructed from the \textsc{Maglim}++ mock sample, which reproduces the redshift distribution of the BOSS-eBOSS data. In an earlier version of this work, we relied on a different set of mocks. Specifically, we initially used a ``Blue'' Cardinal eBOSS ELG sample, built using the same color and magnitude cuts as the eBOSS ELG sample in the South Galactic Cap, and a ``Red'' Cardinal LRG sample, also reproducing BOSS/eBOSS selection \citep{raichoor17}.
In principle, realistic mocks can be used to assess the impact of 1-halo physics, as in \citet{Euclid_dassignies}. On very small scales, effects such as halo assembly history, quenching, stochasticity, and conformity can cause deviations in the cross-correlation galaxy bias from the product of the individual auto-correlation biases. This deviation is quantified using a Pearson-like coefficient:
\begin{equation}
    \eta_{ab}(\rp)=\frac{w_{ab}(\rp)}{\sqrt{w_{aa}(\rp)\,w_{bb}(\rp)}},
\end{equation}
where $a$ and $b$ denote two galaxy samples with similar redshift distributions, $n(z)$. In the linear bias regime, $\eta_{ab}$ is expected to approach unity on large scales.
Figure~\ref{fig:Pears} shows the Pearson coefficients measured in Cardinal as a function of scale and redshift, for different combinations of $a$ and $b$. As expected, $\eta_{ab} \rightarrow 1$ for $\rp > 10$ Mpc. However, we also observe a significant suppression of $\eta_{ab}$ at $\rp < 1.5$ Mpc, with a clear redshift dependence. Unlike \citet{Euclid_dassignies}, we find that this suppression extends to intermediate scales ($1.5 < \rp < 10$ Mpc), beyond the typical 1-halo regime.

Tests on the DES Y6 data (see Figs. \ref{fig:scale_meta} and \ref{fig:scale_mag}) show limited differences across redshifts in the WZ results when using smaller ($<1.5$~Mpc) and larger scales ($>5$~Mpc). For these two figures, we use $b_\rmr^{w_{\rm rr}}$ as a proxi to  $b_\rmr^{w_{\rm ur}}$, showing that any potential deviation is limited for the reference galaxy bias. 
For the unknown galaxy bias, we know (i) from linear theory that the large scale $b_{\rm u}^{w_\rmur}$ is equal to the large scale $b_{\rm u}^{w_{\rm uu}}$,  (ii)  $b_{\rm u}^{w_{\rm ur}}$ values are consistent between $1.5$ Mpc to 15 Mpc ( \textit{cf.} Fig. \ref{fig:galaxy_bias_redshift})
(iii) we tested galaxy bias from $w_{\rm uu}$ are consistent between  between $1.5$ Mpc to 15 Mpc; thus 
we can deduce (iv) at scales $\rp \sim 1.5$ Mpc, the $b_{\rm u}^{w_{\rm ur}}$ would be consistent with $b_{\rm u}^{w_{\rm uu}}$.
In addition in App. \ref{app:smallscale_blue_red}, we measure the $\eta_{ab}$ functions for different red and blue tracer samples generated from \textsc{Maglim}++ and \textsc{Redmagic}. We observe some deviations from unity, but limited to the range $\rp<1.5 $ Mpc, and far from the factor 5 and redshift evolution observed in Fig. \ref{fig:Pears}. Thus the variations of $\eta$ observed in Cardinal do not appear to be physical.

In Cardinal, the auto-correlation properties of red and blue galaxies, as well as their mixtures, were validated against SDSS data (see Fig. 5 of \citealt{cardinal}). However, the clustering of faint galaxies already shows significant discrepancies with the data, even at large scales. More importantly, the cross-correlation between red and blue galaxies was not validated. Therefore, the type of small-scale cross-correlation analysis we aim to perform (at $\sim 1$ Mpc) was not reliably tested in Cardinal.
We found that the \textsc{Maglim}++-based mocks exhibit fewer of these issues. Although we cannot produce Figure~\ref{fig:Pears} directly from photometric data, the cross-correlations in the \textsc{Maglim}++ mocks appear consistent with our data measurements, as seen in Figures~\ref{fig:galaxy_bias_redshift} and \ref{fig:eta_mag_meta}. Consequently, we adopted the \textsc{Maglim}++-based mocks for our fiducial analysis.
We emphasize that mocks are used to define priors on systematics—not to parameterize them precisely with informative priors. Our methodology is deliberately conservative and flexible, allowing for systematic amplitudes larger than those observed in Cardinal. Therefore, despite concerns about mock realism, we are confident in the robustness of our results.

\section{Impact of the colors of the reference sample}{\label{app:color}}
\begin{figure*}
    \centering
    \includegraphics[width=1\linewidth]{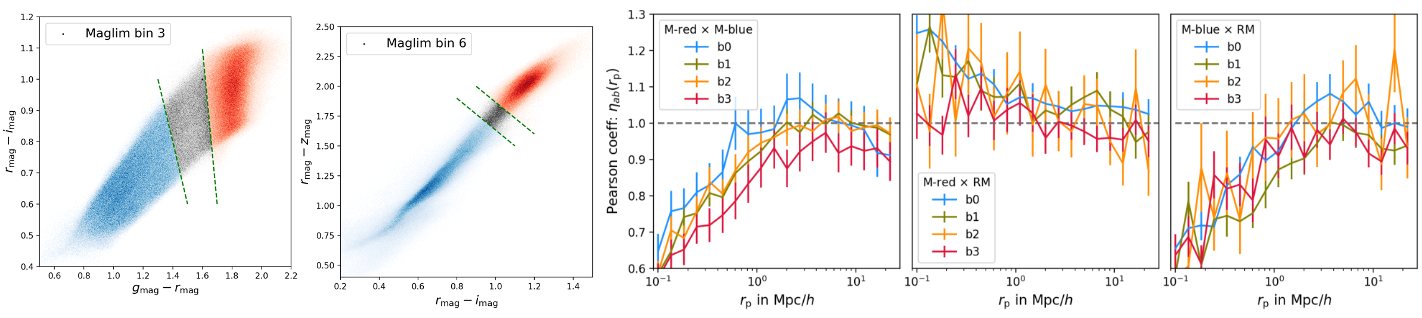}
    \caption{\emph{Left:} Separation of the blue (in blue) and red (in red) samples in color space for two of the 4 \textsc{Maglim}++ bins we considered. \emph{Right} Pearson-like coefficient for the cross-correlations of the \textsc{Maglim}++ blue and red samples, the \textsc{Maglim}++ red and \textsc{Redmagic} samples, and the \textsc{Maglim}++ blue and \textsc{Redmagic} samples for four bin pairs b0--b3. We measure deviations from unity at $\rp<1.5$ Mpc, but a rather constant equal to unity trend for larger scales. The cross-correlations of red and blue samples results in an under-correlations, while the cross-correlations of the two red samples seems to result in an excess. }
    \label{fig:red_blue_mag_rm}
\end{figure*}

\subsection{On small scale systematics}\label{app:smallscale_blue_red}

The goal here is to measure $\eta_{ab}(\rp, z)$ for combinations of blue and red galaxy samples. We use the last four \textsc{Maglim}++ bins and split each into blue and red subsamples \citep[one can find color sample split for source sample in ][ with SOM method]{blue_shear}, isolating the color bimodality in color--color space. Two examples of this bimodality are shown in Fig.~\ref{fig:red_blue_mag_rm}.  
In addition, we consider an extra red sample using \textit{Redmagic}, but whose galaxies in common with \textsc{Maglim}++ have been removed to ensure independence. We then create four \textsc{Redmagic} bins whose redshift ranges closely match those of the four \textsc{Maglim}++ bins. Unlike the approach used for \textit{Cardinal} in App.~\ref{app:spec_mocks}, here we cannot measure $w_{ab}$, $w_{aa}$, and $w_{bb}$ for $a$ and $b$ within the same narrow redshift range. Therefore, we must account for differences in the redshift distributions.  
We estimate the $n(z)$ of each sample using BOSS--eBOSS cross-correlations on significantly larger scales ($5 < \rp < 30$~Mpc) than usual, to avoid small-scale effects, and we neglect galaxy bias evolution.  
To disentangle the impact of galaxy bias from that of the redshift distribution on $\eta$, we define a modified correlation function:  

\begin{align}
    &\tilde{w}_{xy}(\rp)=\frac{\int \mathrm{d}z \, n_x(z)\,n_y(z)\,\xi_{xy}(z,\,\rp)}{\int \mathrm{d}z \, n_x(z)\,n_y(z)\,\xi_{\rm m}(z,\,\rp)} 
    \hspace{0.5cm },\\ 
    &\eta_{ab}(\rp)=\frac{\tilde{w}_{ab}(\rp)}{\sqrt{\tilde{w}_{aa}(\rp)\,\tilde{w}_{bb}(\rp)}}.
\end{align}
We obtain $\tilde{w}$ by dividing the measured correlations $w^{\rm meas}_{xy}(\rp)$ by the denominator of the right term in the Eq. below. 
We only consider cross-correlations between color bins that overlap significantly in redshift. The results of the three cross-correlations (M-red $\times $ M-blue, M-red $\times $ RM, M-blue $\times $ RM ) for the four bin pairs (referred as b0--b3) are shown in Fig.~\ref{fig:red_blue_mag_rm}.  
For scales larger than $1.5$~Mpc, $\eta_{ab}$ is generally consistent with unity, except for bin 4 of \textsc{Maglim}++ red $\times$ \textsc{Maglim}++ blue, which shows a constant offset even at large scales, where linear theory predicts $\eta_{ab} = 1$. This behavior likely reflects a miscalibration of the individual $n(z)$ used in computing $\tilde{w}$, rather than a physical effect.  
For blue--red cross-correlations, we generally observe a decrease in $\eta$ at small scales, as expected since blue and red galaxies tend to occupy different halos \citep[see discussion in Sect.~4.3 of][]{Euclid_dassignies}. Interestingly, for the two distinct red samples, we measure an increase in $\eta$ at small scales.

All the trends we observe are far from the extreme behaviors seen in \textit{Cardinal} (App.~\ref{app:spec_mocks}), where $\eta \sim 0.2$ even at $\rp \sim 5$~Mpc. This validates our choice to switch to the second set of BOSS--eBOSS mocks, which ended up mitigating this effect in Cardinal.  

\subsection{On  \textsc{Maglim}++ bin 6 WZ}\label{app:maglimb6_red_blue}
\begin{figure}
    \includegraphics[width=0.45\textwidth]{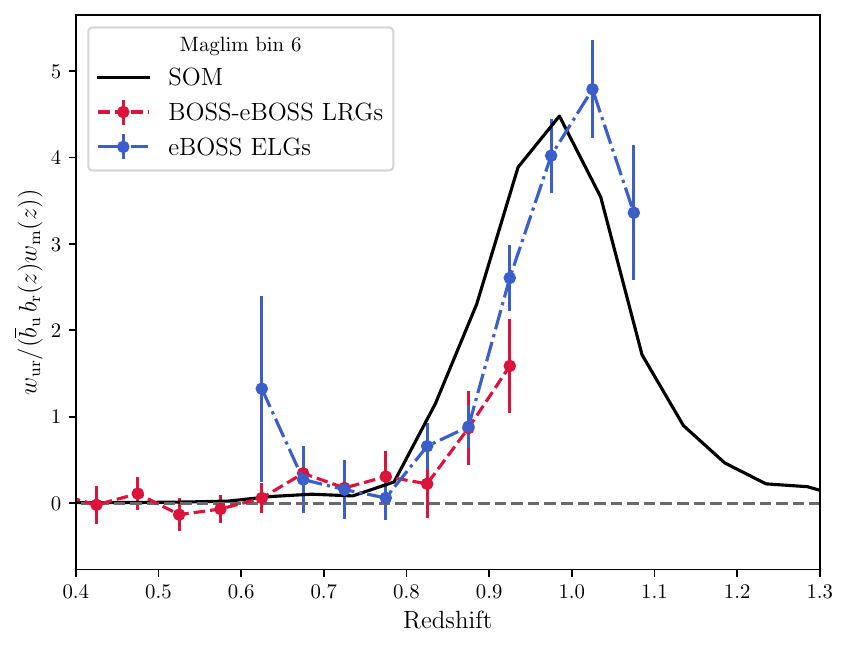}
    \caption{WZ for the \textsc{Maglim}++ lens bin 6, using separately the reference red (LOWZ-CMASS-eBOSS LRG) and blue (eBOSS ELG) samples. We observe both reference sample are consistent and predict a smaller $n(z)$ at $z\sim 0.85$.}\label{fig:bin6 red vs blue}
\end{figure}

For the last \textsc{Maglim}++ bin, WZ predicts a significantly smaller $n(z)$ than the SOM estimate around $z \sim 0.85$. This redshift corresponds to the transition in the reference sample $\rm r$ from LRGs to ELGs. We evaluate correlations on small scales, where environmental effects can play an important role. In particular, if $b^{w_{\rm ur}}_{\rm u}$ differs at small scales when $\rmr$ is composed solely of LRGs versus solely of ELGs, this could introduce a non-physical jump in the inferred $n_{\rm u}(z)$ near the LRG--ELG transition \citep{Euclid_dassignies}.  
To test this, we perform WZ measurements separately for BOSS+eBOSS LOWZ--CMASS--LRGs and for eBOSS ELGs, as recommended by \citet{Euclid_dassignies}, assuming a fixed value of $b_{\rm u} = 2$. The results are shown in Fig.~\ref{fig:bin6 red vs blue}. We find good agreement between the two measurements, ruling out the reference tracer type as the cause of the low $n(z)$ observed in this bin.

\section{DESI WZ}\label{app:desiWZ}
\begin{figure*}
    \centering
    \includegraphics[width=1\linewidth]{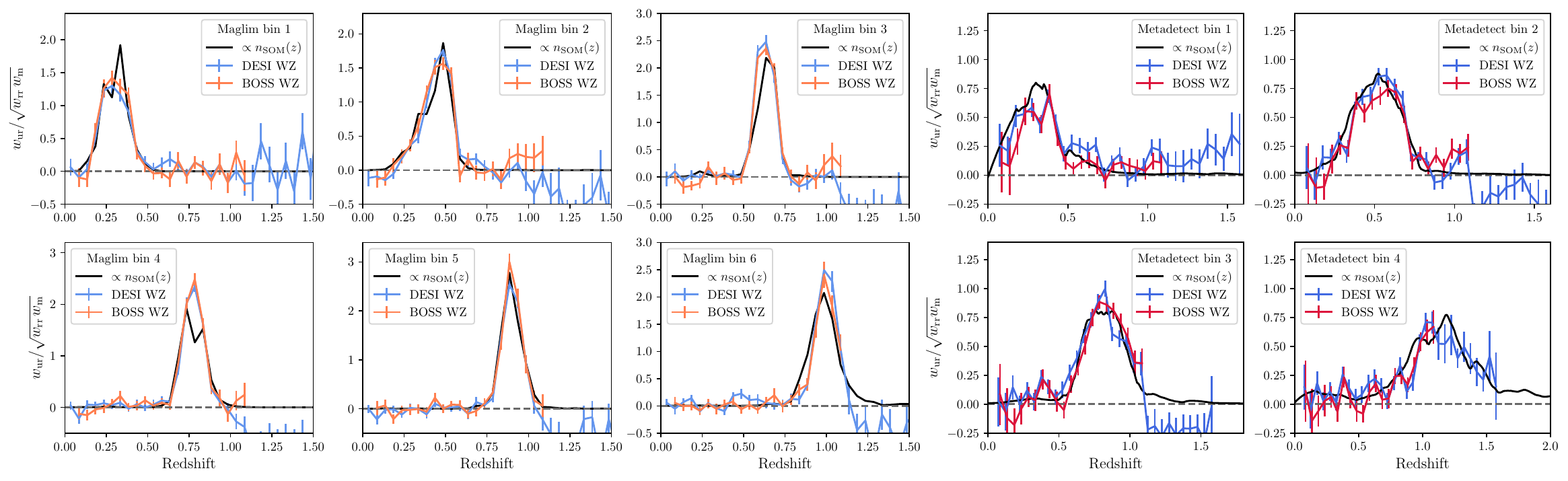}
    \caption{WZ measurements using DESI and BOSS-eBOSS galaxy samples, for the six \textsc{Maglim}++ bins and the four \textsc{Metadetect} bins. We observe an excellent agreement.  }
    \label{fig:DESI_WZ}
\end{figure*}

\begin{figure*}
    \centering
    \includegraphics[width=1\linewidth]{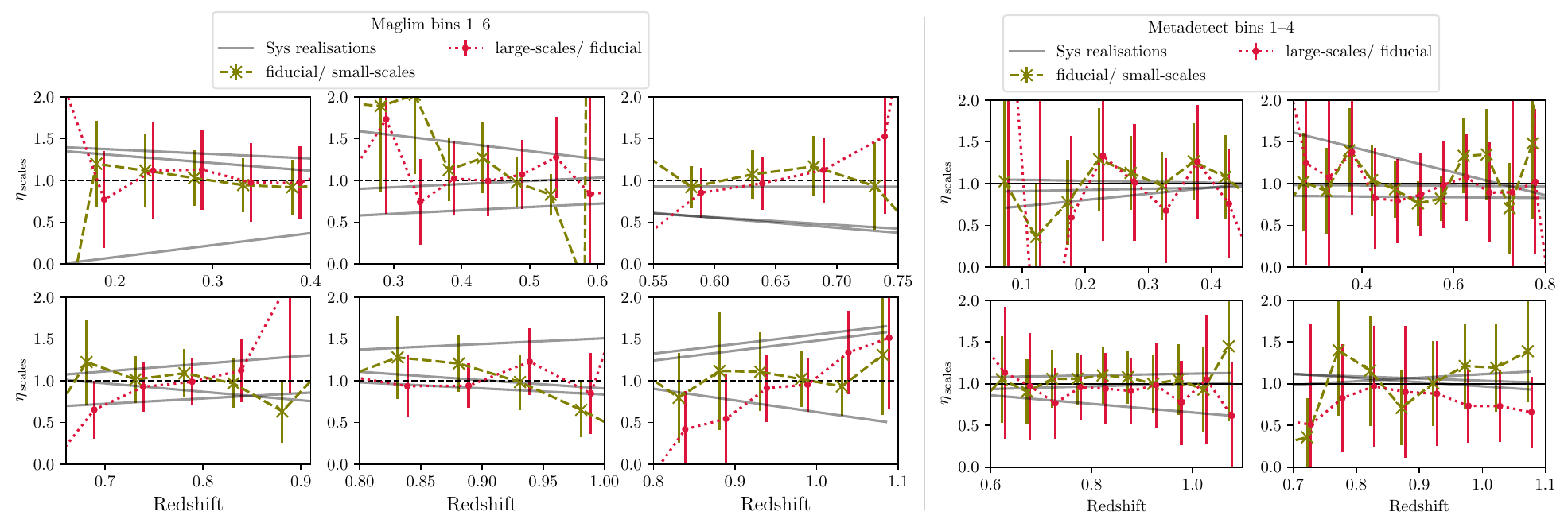}
    \caption{Ratio of the $n(z)$ inferred from different scale ranges for the six \textsc{Maglim}++ bins (right) and the four \textsc{Metadetect} bins (left). The redshift range for every panel was chosen to remove all the points corresponding to small $n(z)$, as  evaluating ratios for those is not relevant. The  fiducial scales/small-scales  is reported in green, and the large-scale/fiducial in red. We report as well some realizations of the Sys functions given our priors measured in Figs. \ref{fig:sys_meta} and \ref{fig:sys_maglim}.}
    \label{fig:eta_mag_meta}
\end{figure*}
In this section, we report the $n(z)$ using DESI DR1 galaxies as reference samples (for a DESI-DR2 WZ analysis of HSC data see \citep{choppin_2025}. In short, we use DESI BGS, LRGs, ELGs, up to $z=1.6$ \citep{DESI_DR1_data}. We obtain a sky overlap with DES of 800 deg$^2$, with a slightly different footprint than BOSS-eBOSS. The WZ measurements are performed over $1.5$ to $5$ Mpc, and reported in Fig. \ref{fig:DESI_WZ}. For the six \textsc{Maglim}++ bins, the agreement with BOSS-eBOSS is excellent. For \textsc{Metadetect} the agreement also appear very good, but with more deviations, such as the excess of $n(z)$ for the first bin at $z=0.55$. With a simple shift and stretch model, we found the mean-$z$ inferred from DESI to be consistent with the one from BOSS-eBOSS for every bin.
We also observe in Fig. \ref{fig:DESI_WZ} that a magnification-like seems to be large for ELG, at $z>1$, as we measured $\wur<0$. Fibre collisions, which are particularly relevant for ELGs in DESI-DR1, could also contribute to this behavior. Including DESI WZ would require estimate of the magnification coefficients, test of the validity of the scale range, and new sys priors, as our measurement now extend up to $z=1.6$. The photometric redshift uncertainty is not the most limiting systematics for DES Y6, as explained in Sect. \ref{sec:chains}. Thus, in order to avoid delaying the DES Y6 analysis, we did not include these DESI measurement to the official pipeline.  

\section{Impact of the scale ranges on WZ}{\label{app:scales}}

In Sect. \ref{sec:test_scale}, we evaluated the WZ measurements using three different scale ranges: small scales (SS): $[0.1, 1.5]$ Mpc,  fiducial range: $[1.5, 5]$ Mpc, large scales: $[5, 15]$ Mpc, and we estimate the redshift distributions, neglecting magnification and systematics. 
 To quantify any scale-dependent deviations, we compute the ratio of the inferred redshift distributions between different scale ranges: $ \eta_{\rm scales}(z) = {n(z\vert {\rm scales}_1)}/\;{n(z\vert {\rm scales}_2)}$.
The results are shown in Fig. \ref{fig:eta_mag_meta}. We observe no statistical deviation from unity for the large-scale/fiducial ratio, validating our scale range choice. More surprisingly, we also do not observe significant deviation using the smaller scale range, this measurement being associated with higher SNR. We also report few realizations of the systematic functions, randomly sampling over our Gaussian priors, and observe that our systematic function can describe these kind of (non-significant) variations.


\end{document}